\documentclass[graybox]{svmult}

\usepackage{mathptmx}       % selects Times Roman as basic font
\usepackage{helvet}         % selects Helvetica as sans-serif font
\usepackage{courier}        % selects Courier as typewriter font
\usepackage{type1cm}        % activate if the above 3 fonts are
\usepackage{makeidx}         % allows index generation
\usepackage{graphicx}        % standard LaTeX graphics tool
\usepackage{multicol}        % used for the two-column index
\usepackage[bottom]{footmisc}% places footnotes at page bottom

\makeindex             % used for the subject index
\usepackage{natbib}
\usepackage[caption=false]{subfig}
\newcommand{\me}{$M_{\oplus}$}
\newcommand{\re}{$R_{\oplus}$}

\newcommand{\teff}{$T_{\rm eff}$}
\newcommand{\tint}{$T_{\rm int}$}
\newcommand{\teq}{$T_{\rm eq}$}
\newcommand{\tb}{$T_{\mathrm B}$}
\newcommand{\cp}{\citep}
\newcommand{\ct}{\citet}

\newcommand{\apj}{ApJ}
\newcommand{\apjl}{ApJ Letters}
\newcommand{\araa}{ARA\&A}

\newcommand{\apjs}{Astrophysical Journal Supplement} 
\newcommand{\aj}{AJ}

\newcommand{\aap}{A \& A}
\newcommand{\pasp}{Publications of the Astronomical Society of the Pacific}

\newcommand{\nat}{Nature}
\newcommand{\mnras}{MNRAS}

\begin{document}

\title*{Modeling Exoplanetary Atmospheres: \\ An Overview}
\author{Jonathan J. Fortney}
\institute{Jonathan J. Fortney \at Department of Astronomy and Astrophysics, Other Worlds Laboratory (OWL), University of California, Santa Cruz, \email{jfortney@ucsc.edu}
}
\maketitle

\abstract{We review several aspects of the calculation of exoplanet model atmospheres in the current era, with a focus on understanding the temperature-pressure profiles of atmospheres and their emitted spectra.  Most of the focus is on gas giant planets, both under strong stellar irradiation and in isolation.  The roles of stellar irradiation, metallicity, surface gravity, C/O ratio, interior fluxes, and cloud opacity are discussed.  Connections are made to the well-studied atmospheres of brown dwarfs as well as sub-Neptunes and terrestrial planets, where appropriate.  Illustrative examples of model atmosphere retrievals on a thermal emission spectrum are given and connections are made between atmospheric abundances and the predictions of planet formation models.}

%\abstract{We review several aspects of the calculation of exoplanet model atmospheres in the current era, with a focus on understanding the temperature-pressure profiles of atmospheres and their emitted spectra.  Most of the focus is on gas giant planets, both under strong stellar irradiation and in isolation.  The role of stellar irradiation, metallicity, surface gravity, C/O ratio, interior fluxes, and cloud opacity are discussed.  Connections are made to the well-studied atmospheres of brown dwarfs as well as sub-Neptunes and terrestrial planets, where appropriate.  Illustrative examples of model atmosphere retrievals on a thermal emission spectrum are given and connections are made between atmospheric abundances and the predictions of planet formation models.}

\section{Why Study Atmospheres?}
\label{sec:1}
While atmospheres often make up only a tiny fraction of a planet's mass, they have an out-sized importance in determining a number of physical properties of planets, how they evolve with time, and their physical appearance.  Atmospheres dramatically influence a planet's energy balance, as the relative importance of gaseous absorption or scattering from clouds or gasses dictate a planet's albedo.  Atmospheres can impact cooling, as interior convection or conduction must give way to a radiative atmosphere to lose energy out to space.  Atmospheres, by their composition, can tell us a rich story of the gain and loss of volatiles, since atmospheres can be accreted from the nebula, outgassed from the interior, lost to space by escape processes, or regained by the interior.

We tend to think of two broad reasons for studying planetary atmospheres. One is that atmospheres are inherently interesting, with a diverse array of physical and chemical processes at work.  What sets the temperature structure of an atmosphere?  Why do some have thermal inversions and others do not?  What sets the chemical abundances in an atmosphere?  Why are some atmospheres dominated by clouds, and why are others mostly cloud-free?  What determines the day-night temperature contrast on the planet?  How fast can winds blow?  Can planets lose their entire atmosphere, never to attain one again?

An entirely other set of questions focuses more on what an atmosphere can tell us about the formation and evolution of the planet.  Atmospheric composition can tell us a lot about the integrated history of a planet.  The metal-enrichment of a giant planet, compared to its star, can help us to understand aspects of planet formation.  The comparative planetology of rocky worlds, like Earth and Venus, one with water vapor in the atmosphere, and one without, informs our understanding of divergent evolution.  Noble gas abundances teach us about the accretion of primordial volatiles.

\begin{figure}[htp]  
\includegraphics[clip,width=1.0\columnwidth]{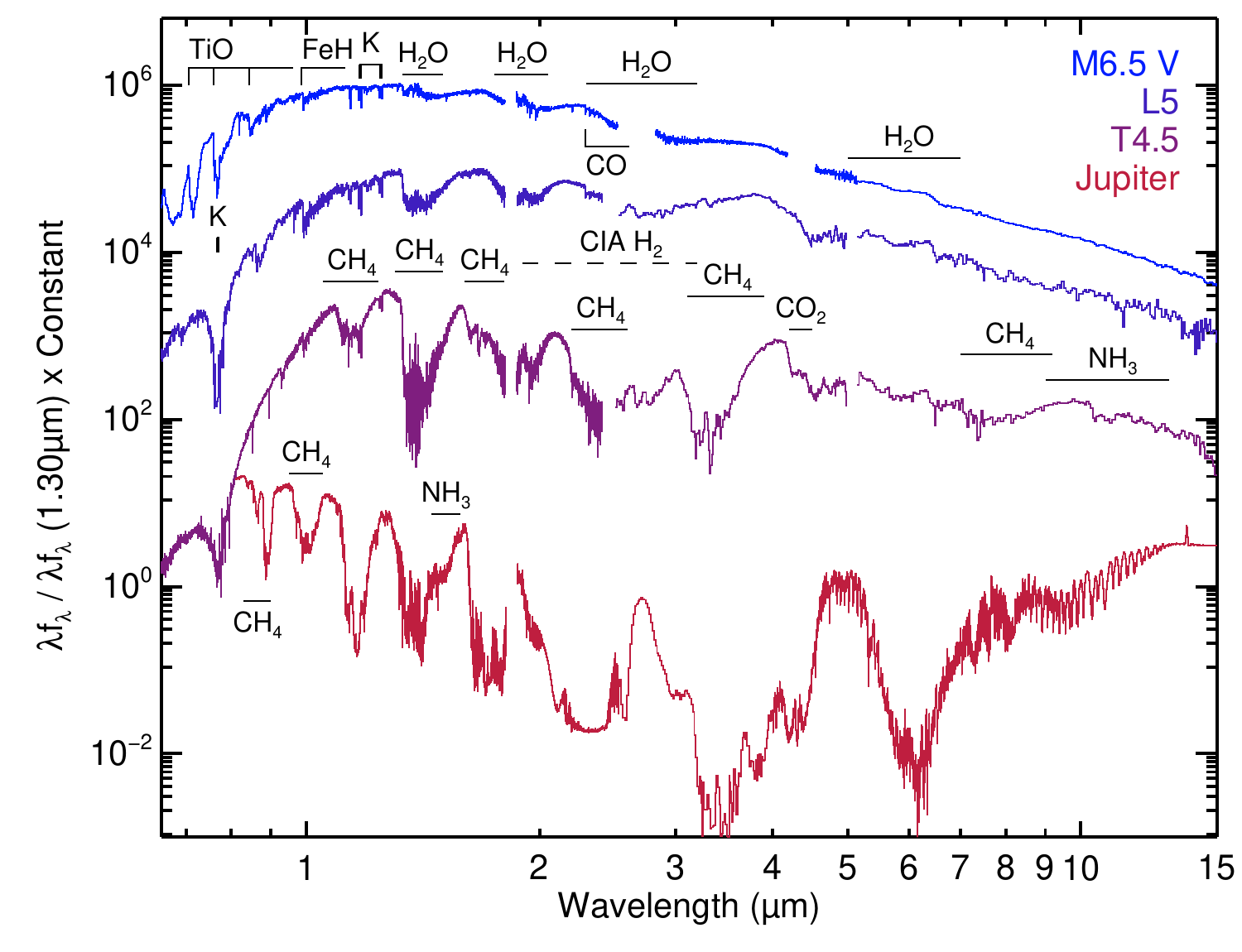}  
\caption{An entirely empirical showcase of the continuum in atmospheres from the coolest stars, to brown dwarfs, to the spectrum of Jupiter.  Dominant infrared molecular absorption features of H$_2$O, CO, CO$_2$, CH$_4$, and NH$_3$ are shown, as well as TiO and K at optical wavelengths. This sequence is \teff\ values of 2700 K, 1550 K, 1075 K, and 125 K.  Note that  nearly all flux from Jupiter short-ward of 4.5 $\mu$m is reflected solar flux, not thermal emission.  Figure courtesy of Mike Cushing.\label{cushing}} 
\end{figure}

The tools we use to model exoplanetary atmospheres are often the very same tools, or descendents of the tools, that we used to model the atmospheres of solar system planets.  Other such tools were used to study cool stellar atmospheres or brown dwarfs.  In that way, exoplanetary atmospheres can be thought of as a meeting of the minds, tools, and prejudice  of the models and methods of planetary atmospheres and stellar atmospheres.  The continuum from the coolest stars, to brown dwarfs and hot planets, to cool planets is real, and can be readily seen in Figure \ref{cushing}.

The field of exoplanetary atmospheres has exploded in the past decade. With the 2019 launch of the \emph{James Webb Space Telescope}, the field is poised for dramatic advances.  We are lucky that a number of recent texts have emerged that discuss the physics and chemistry of exoplanetary and solar system atmospheres.  All are worth a detailed reading, including \ct{Seager10b}, \ct{Pierrehumbert10}, \ct{Heng17}, and \ct{Catling17}, while classic solar system texts like \ct{ChambHunt} are also still essential reading.

\section{Energy Balance and Albedos}  \label{sec:2}
For any planet with an atmosphere, the atmosphere will help to set the energy balance of the planet with that of its parent star.  Let's begin by striving to be clear about the albedo (reflectivity) of a planet, and how that enables estimates of planetary fluxes and temperatures.  Often the descriptions of various albedos are actually much clearer in words than in mathematics, which is somewhat unusual.  Excellent references on this topic exist from the ``early days'' of exoplanetary atmospheres, including \ct{Marley99} and \ct{Sudar00}.

\subsection{Geometric Albedo}
Geometric albedo is the reflectivity of the planet when seen at full illumination, called ``at full phase.''  (Think of the full moon.)  Within the solar system, this is basically how we always see the giant planets, since they are on orbits at much larger separations than the Earth.  In the exoplanet context, we can determine the geometric albedo of a planet at secondary eclipse (or ``occultation'') when its flux disappears as it passes behind its parent star.  The geometric albedo, $A_{\mathrm G}$, is always specified at a particular wavelength or in a given bandpass.

An oddity of $A_{\mathrm G}$ is that it can sometimes \emph{fail} as a true measure of planetary reflectivity in the exoplanet context for hot planets \cp[e.g.][]{Burrows08b,Fortney08a}.  This is because hot Jupiters can have appreciable thermal emission at visible wavelengths.  Thus, $A_{\mathrm G}$ can be higher than naively expected, due to some (or even most!) of the ``planetary'' flux coming from the planet being due to thermal emission, rather than reflected light.

\subsection{Spherical Albedo}
The spherical albedo, $A_{\mathrm S}$, is similar to the geometric albedo.  Again, it is specified at a given wavelength or a bandpass.  However, here we are interested in the reflectivity over all angles -- just the total reflectivity of the stellar flux, not caring about scattering angle.  Recall that $A_{\mathrm G}$ is the stellar light that we get back when viewing the planet at full phase.

This means that $A_{\mathrm S}$ can only be determined with some care.  Within the solar system, the most straightforward way is to send a spacecraft to the planet to observe scattering from the planet at all phase angles.  In practice, the spherical albedo is not discussed much in the literature.  However, if the spherical albedo is integrated over all \emph{stellar} wavelengths, then we have a rather interesting quantity: the Bond albedo.

\subsection{Bond Albedo}
The Bond albedo, $A_{\mathrm B}$, is typically the most important albedo for planets.  It is the ratio of the total reflected stellar power (in say, erg s$^{-1}$) to the total power incident upon the planet.  Within a modeling framework, it is $A_{\mathrm S}$ integrated over the stellar spectrum.  The value of $A_{\mathrm B}$ is important because it determines how much total power is absorbed or scattered by a planet. 

The single most important thing to recall about $A_{\mathrm B}$ is that, unlike $A_{\mathrm G}$ and $A_{\mathrm S}$, \emph{it is not a quantity that is inherent to the planet alone.}  The value of $A_{\mathrm B}$, for a given planet, \emph{strongly depends on the incident spectrum from the parent star.}  Meaning, the same planet, around two different stars, will have two different Bond albedos.  Typically, around an M star, more flux is emitted in the infrared. There is less scattering, more absorption, and lower $A_{\mathrm B}$, compared to illumination by a Sunlike star, where there is more short-wavelength incident flux that is Rayleigh scattered away \cp[e.g.,][]{Marley99}.

While $A_{\mathrm B}$ is straightforward to discuss, it is difficult to measure in practice.  Within the solar system, it can be determined by observing light scattered from planetary atmospheres in all directions ($A_{\mathrm S}$) over a broad wavelength range that samples from the near UV to mid IR, where the Sun is brightest.  In the exoplanet context, such a measurement is much more difficult.  At least for strongly irradiated planets, $A_{\mathrm B}$ is probably best determined by just observing how hot a planet actually is, by measuring its total thermal emission.

\subsection{Temperatures of Interest}
Planets that do not have an intrinsic energy source will be in energy balance with the input from their parent star.  That is, the power absorbed by the planet will be re-radiated back to space.  For a planet like the Earth, the intrinsic energy due to secular cooling of the interior, along with radiative decay, is negligible in terms of energy balance; thus, absorbed power from the Sun entirely dominates the atmospheric energy balance.  However, for very young rocky planets \cp{Lupu14}, and for giant planets at essentially any age \cp{Burrows97,Baraffe03,Marley07}, the flux from the planet's interior is appreciable and affects the atmospheric temperature structure and energy balance. 
If a planet is in energy balance with its star, the equilibrium temperature, \teq, can be written:
\begin{equation}
T_{\rm eq}^4 =  f(1-A_{\rm B})L_* / (16 \pi \sigma d^2),
\end{equation}
where $f$ is 1 if the absorbed radiation is able to be radiated
away over the entire planet (4$\pi$ steradians) or 2 if it only radiates
on the dayside (2$\pi$ sr), which is then hotter. $A_{\rm B}$ is the planet’s Bond albedo, $L_*$ is the luminosity of the star, $\sigma$ is the Stefan-Boltzmann constant,
and $d$ is the planet's orbital distance.

If the effective temperature, \teff, is defined as the temperature of a blackbody of the same radius that would emit the equivalent flux as the real planet, \teff\ and \teq\ can be simply related.  This relation requires the inclusion of a third temperature, \tint, the ``intrinsic effective temperature," that describes the flux from the planet's interior.  These temperatures are related by:
\begin{equation}
T_{\rm eff}^4 = T_{\rm eq}^4 + T_{\rm int}^4 
\end{equation}
We then recover our limiting cases: if a planet is self-luminous (like
a young giant planet) and far from its parent star, $T_eff \approx T_int$; for most
rocky planets, or any planets under extreme stellar irradiation,
$T_eff \approx T_eq$.

\subsection{Absorption and Emission of Flux}
\begin{figure}[htp]
\includegraphics[width=1.0\columnwidth]{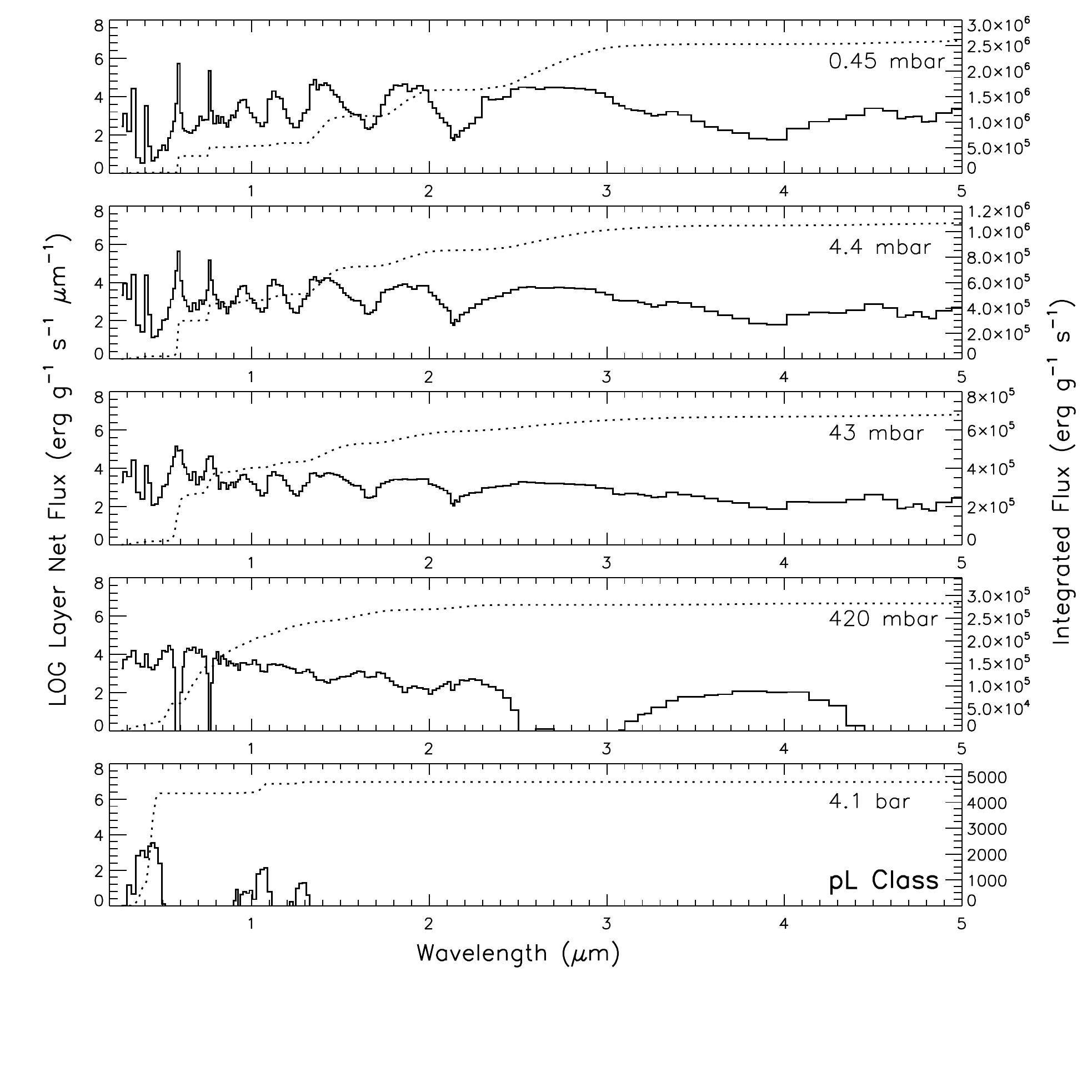} 
\vspace*{-15mm}
\caption{Incident net flux (erg g$^{-1}$ s$^{-1}$ $\mu$m$^{-1}$, solid line, left axis) in five model layers for a cloud-free hot Jupiter model, $g$=15 m s$^{-2}$, at 0.05 au from the Sun.  The dotted line (right axis) illustrates the integrated flux, evaluated from short to long wavelengths.  The layer \emph{integrated} flux is read at the intersection of the dotted line and the right axis.  Note the logarithmic scale on the left.  At 0.45 mbar, although the absorption due to neutral atomic alkalis are important in the optical, more flux is absorbed by water vapor in the infrared.  Heating due to alkali absorption becomes relatively more important as pressure increases.  By 420 mbar, there is no flux left in the alkali line cores, and by 4.1 bar, nearly all the incident stellar flux has been absorbed. Adapted from \ct{Fortney08a}, which also included a description of the hot Jupiter ``pL Class" noted on the figure.\label{fluxin}
}
\end{figure}

\begin{figure}[htp]
\includegraphics[clip,width=1.0\columnwidth]{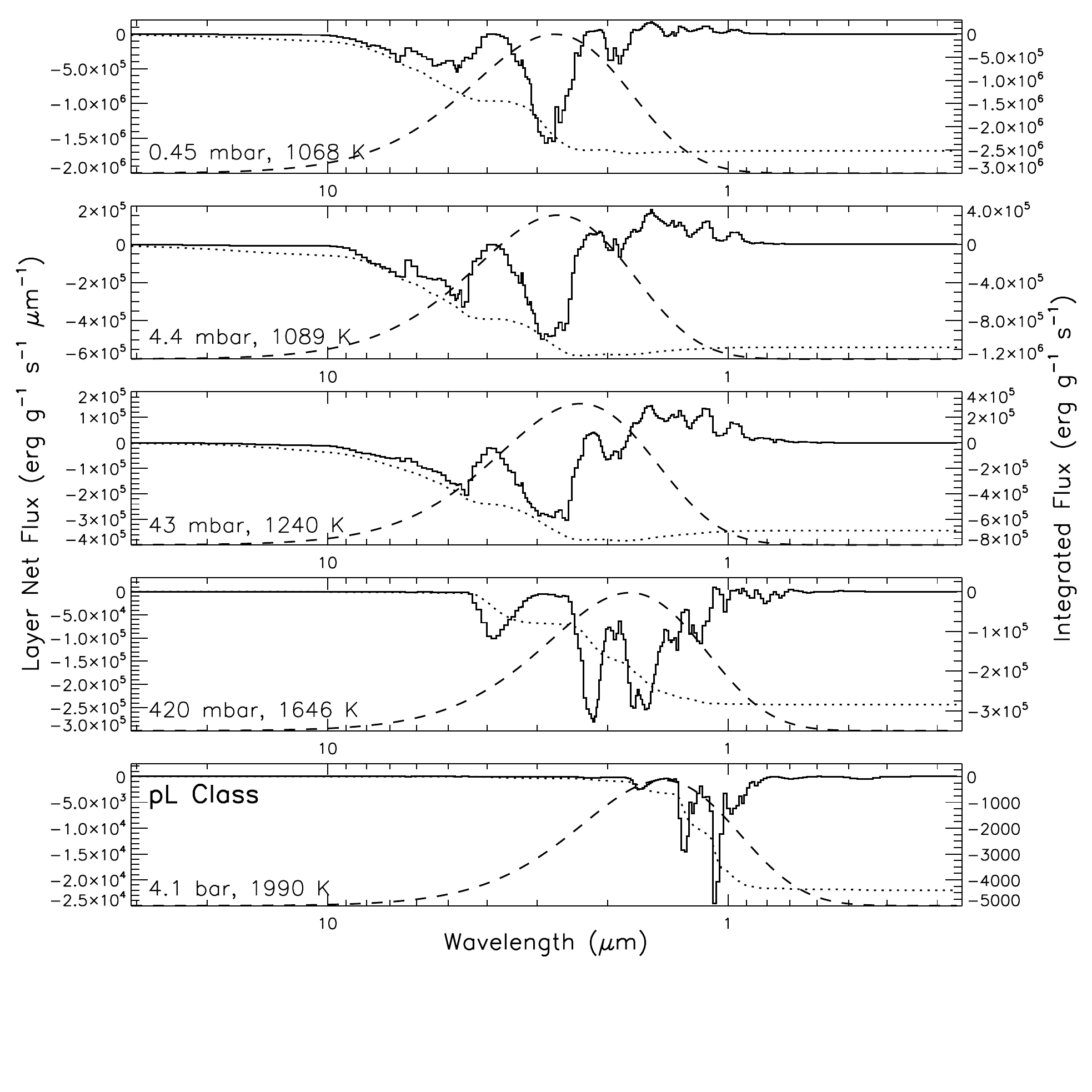}  
\vspace*{-15mm}
\caption{Similar to Figure~\ref{fluxin}, but for thermal emission, from 30 to 0.26 $\mu$m (long wavelengths to short).  Negative flux is emitted.  The dotted line is again integrated flux, evaluated from long to short wavelengths.  The negative value of a layer integrated flux in Figure~\ref{fluxin} equals the integrated flux here.  In addition, the scaled Planck function appropriate for each layer is shown as a dashed curve.  The temperature and pressure of each layer is labeled.  Cooling occurs mostly by way of water vapor, but also CO.  As the atmosphere cools with altitude progressively longer wavelength water bands dominate the layer thermal emission. Adapted from \ct{Fortney08a}.\label{fluxout}}
\end{figure}

\begin{figure}[htp]
\includegraphics[clip,width=0.75\columnwidth]{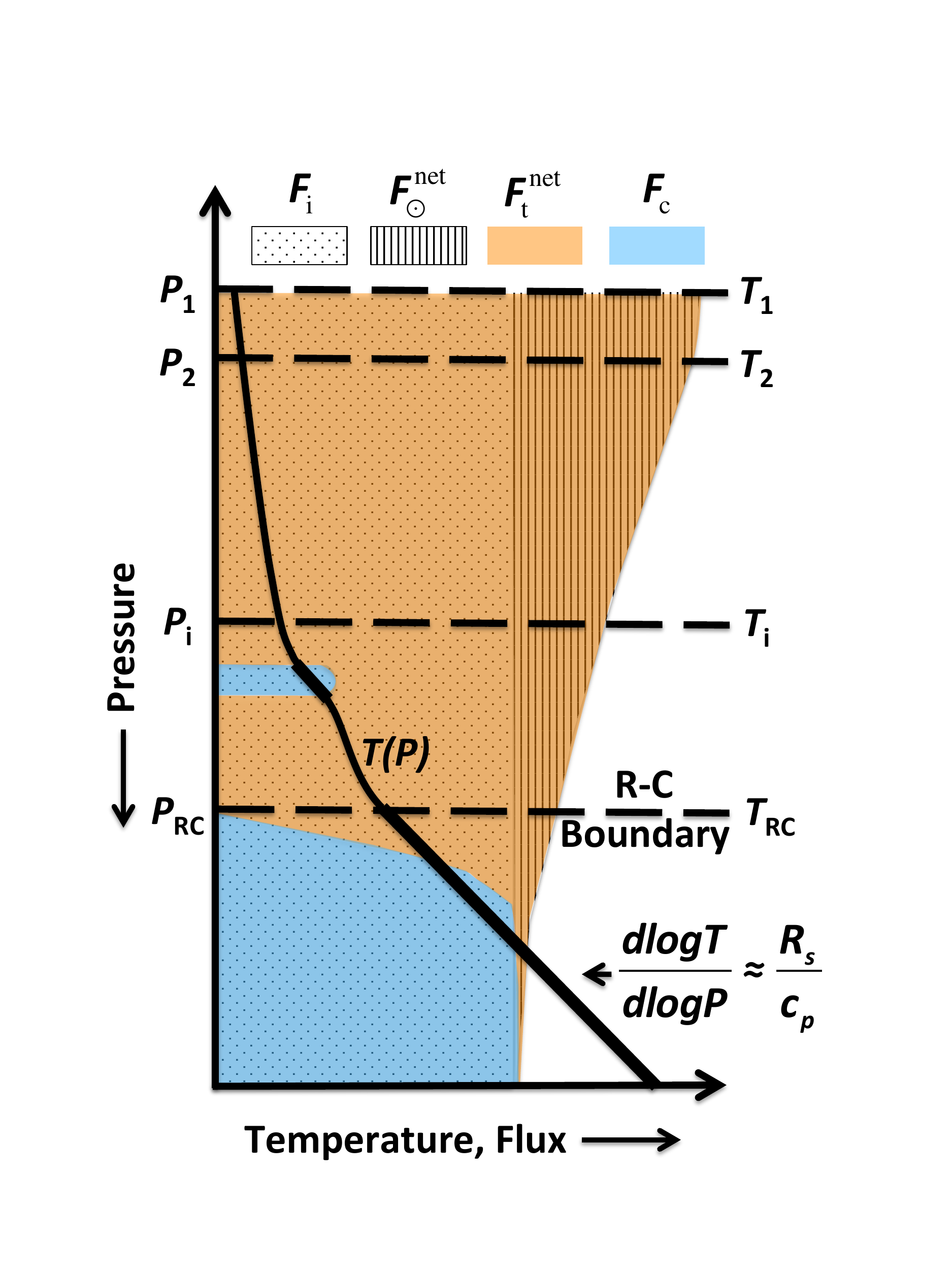} 
\vspace*{-8mm}
\caption{Schematic depiction of the temperature structure of a model atmosphere.  The y-axis is pressure, increasing downwards, and the x-axis shows temperature and energy flux.  Model levels are shown (horizontal dashed lines), and the solid line is the temperature structure profile, where bolded parts indicate a convective region. `RC' indicates the radiative-convective boundary.  In equilibrium, net thermal flux ($F^{\rm{net}}_{\rm{t}}$, orange) and the convective flux ($F_{\rm{c}}$, blue) must sum to the internal heat flux ($F_{\rm{i}}$, dotted, which is $\sigma T_{\rm int}^4$) and, for an irradiated object, the net absorbed stellar flux ($F^{\rm{net}}_{\odot}$, striped, which alone is $\sigma T_{\rm eq}^4$ ).  Note that the internal heat flux is constant throughout the atmosphere, whereas the schematic profile of net absorbed stellar flux decreases with increasing pressure, and eventually reaches zero in the deep atmosphere.  At depth, convection carries the vast majority of the summed internal and stellar fluxes, but is a smaller component in detached convective regions (upper blue region). Adapted from \ct{Marley15}.  \label{cartoon}}
\end{figure}

Recent reviews on radiative transfer in substellar and exoplanetary atmospheres can be found in \ct{Hubeny17} and \ct{Heng17b}.  Here will be merely show some illustrative plots of how a 1D radiative-convective atmosphere model operates in terms of the absorption of stellar flux (Figure \ref{fluxin}), the emission of thermal flux (Figure \ref{fluxout}), and the outgoing flux carried by an atmosphere in radiative-convective equilibrium (Figure \ref{cartoon}).

As we will see below, strongly irradiated planets are dominated by the absorbed incident stellar flux, rather than any intrinsic flux from the deep interior.  Atmospheric energy balance is satisfied by re-radiation of absorbed stellar flux.  For a generic gas giant planet at 0.05 au from the Sun, Figure \ref{fluxin} shows the wavelength-dependent absorption of stellar flux at five pressure levels within a 1D model.  At the top of the model, most flux is absorbed by the near infrared water bands (see Figure \ref{3abunds} for their exact locations in wavelength), while at deeper layers most absorption is via the pressure-broadened alkali (Na and K) doublets at 0.59 and 0.77 $\mu$m.  By 4 bars essentially all stellar flux has been absorbed.  The re-radiation of this energy to space, at these same pressure levels, occurs in the near and mid infrared, mostly longward of 3 $\mu$m in the top three panels shown in Figure \ref{fluxout}.  Deeper in the atmosphere, where temperatures are warmer, there is more overlap with shorter wavelength water bands.

Figure \ref{cartoon}, adapted from \ct{Marley15} shows the balance of several fluxes for a modestly irradiated planet, somewhat similar to Jupiter.  The planet's intrinsic flux is carried via radiation or convection, with convection dominating at the deepest levels where the atmosphere is dense and mostly opaque.  A second detached, convective zone forms in region of local high opacity, which carries some of the flux as well.  At depth, the profile in the convective region (thicker solid line) is that of an isentrope (constant specific entropy) with a temperature-pressure profile that is essentially adiabatic, as a only a minute super-adiabaticity is needed to transport flux via convection.  Absorbed stellar fluxes are shown as a an additional component that the atmosphere must also carry via radiation.  In practice, a 1D radiative-convective model needs to iterate to find a temperature structure that satisfies the constant flow of intrinsic energy through each layer (given that each layer both absorbs and emits flux) and the re-emission of absorbed stellar flux at each layer.  \ct{Marley15} features an in-depth discussion of the temperature corrections needed in each model layer, for each iteration, to converge to a model in radiative-convective equilibrium.

\section{Overview of Pressure-Temperature Profiles and Absorption Features}
\subsection{Pressure-Temperature Profiles}  \label{sec:3}
\begin{figure}[htp]
\includegraphics[clip,width=1.0\columnwidth]{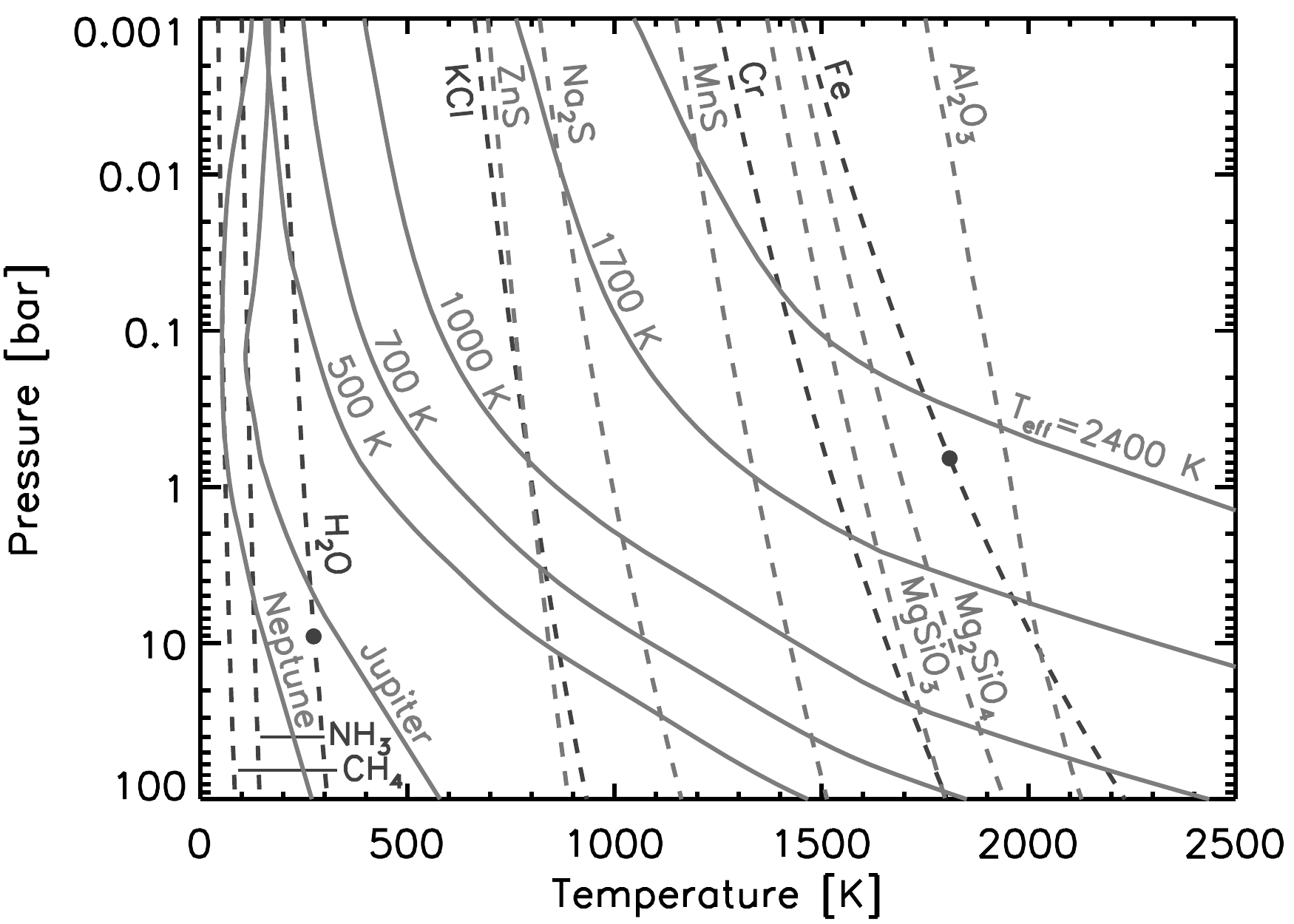}  
\caption{Pressure-temperature profiles from a warm brown dwarf (2400 K) to Neptune (50 K), showing the range of cool molecule-dominated H-He atmospheres.  Model atmospheric profiles are shown as solid curves.  Chemical condensation curves for cloud species are shown as dashed lines.  Figure adapted from \ct{Marley15}.\label{ptgrid1}}
\end{figure}

Much has been written over the years on the temperature structure of planetary atmospheres.  Just in the recent past, analytic models of atmospheric temperature structure have been published, mostly focusing on strongly irradiated planets, by \ct{Hansen08b}, \ct{Guillot10}, \ct{Robinson12}, and \ct{Parmentier14}.  These frameworks aim to understand the radiative (or radiative-convective) temperature structure as a function of the three temperatures outlined above, as well as the gaseous opacity relevant for flux incident upon the atmosphere (typically visible light) and the gaseous opacity relevant for emitted planetary fluxes (typically infrared light).  Figure \ref{ptgrid1} shows atmospheric pressure-temperature (\emph{T--P}) profiles from 2400 K down to 50 K, compared to relevant condensation curves for cloud-forming materials.

\section{Interpreting Spectra via Absorption Features}
While much of the physics of stellar atmospheres transfers over to our understanding of planetary atmospheres, the interpretation of spectra is not one such area.  While much of stellar atmospheres can be interpreted in terms of narrow atomic/ionic lines, caused by electronic transitions, on top of wavelength-independent ``continuum" opacity sources, the same is not true for planets.  In planets, few---if any---continuum opacity sources exist, and atmospheric opacities are dominated by the forest of rotational/vibration lines of dominant molecular absorbers like H$_2$O, CH$_4$, CO, CO$_2$, and NH$_3$, among other molecules.

\begin{figure}[htp]  
{%
  \includegraphics[clip,width=0.8\columnwidth]{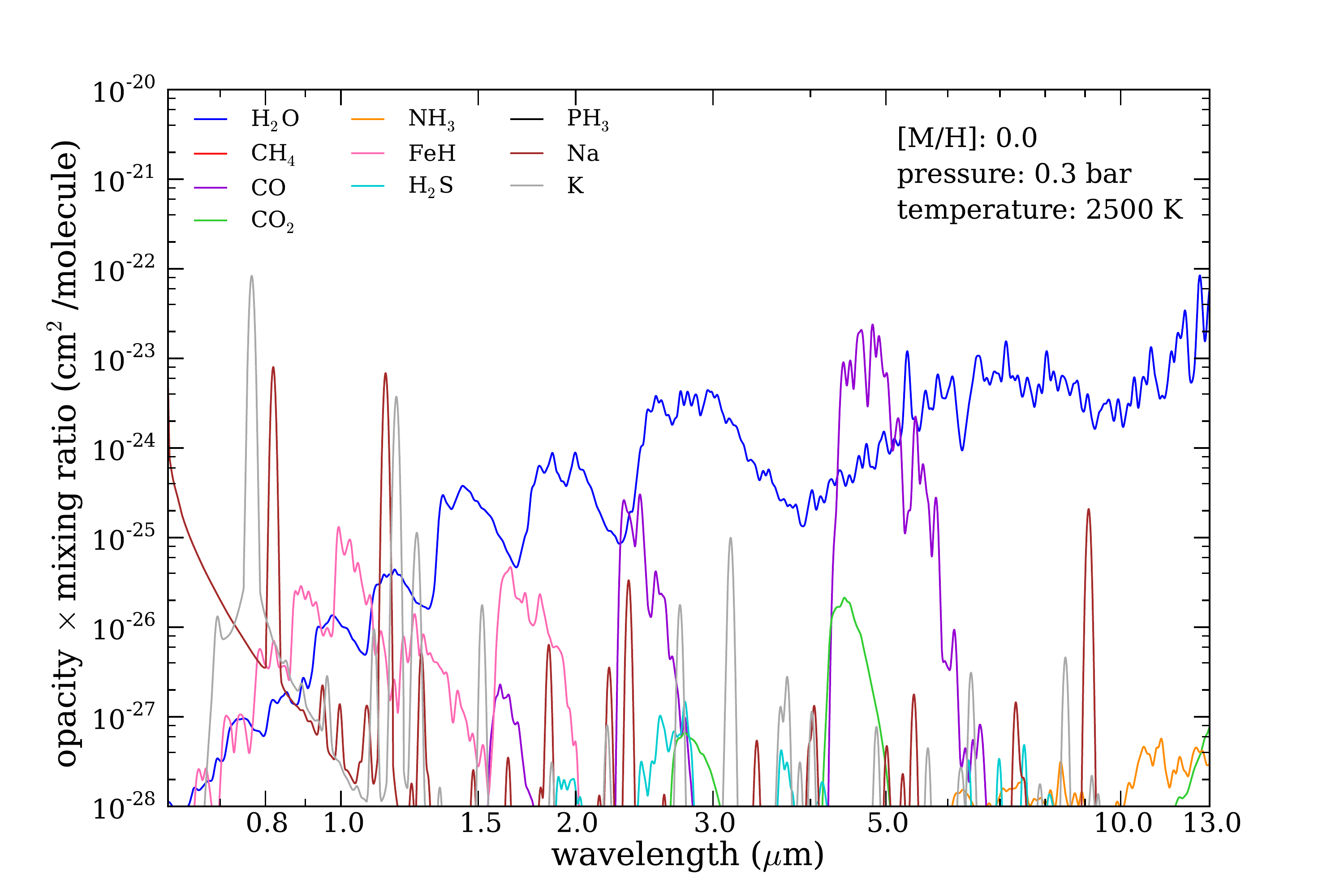}%
}
{%
  \includegraphics[clip,width=0.8\columnwidth]{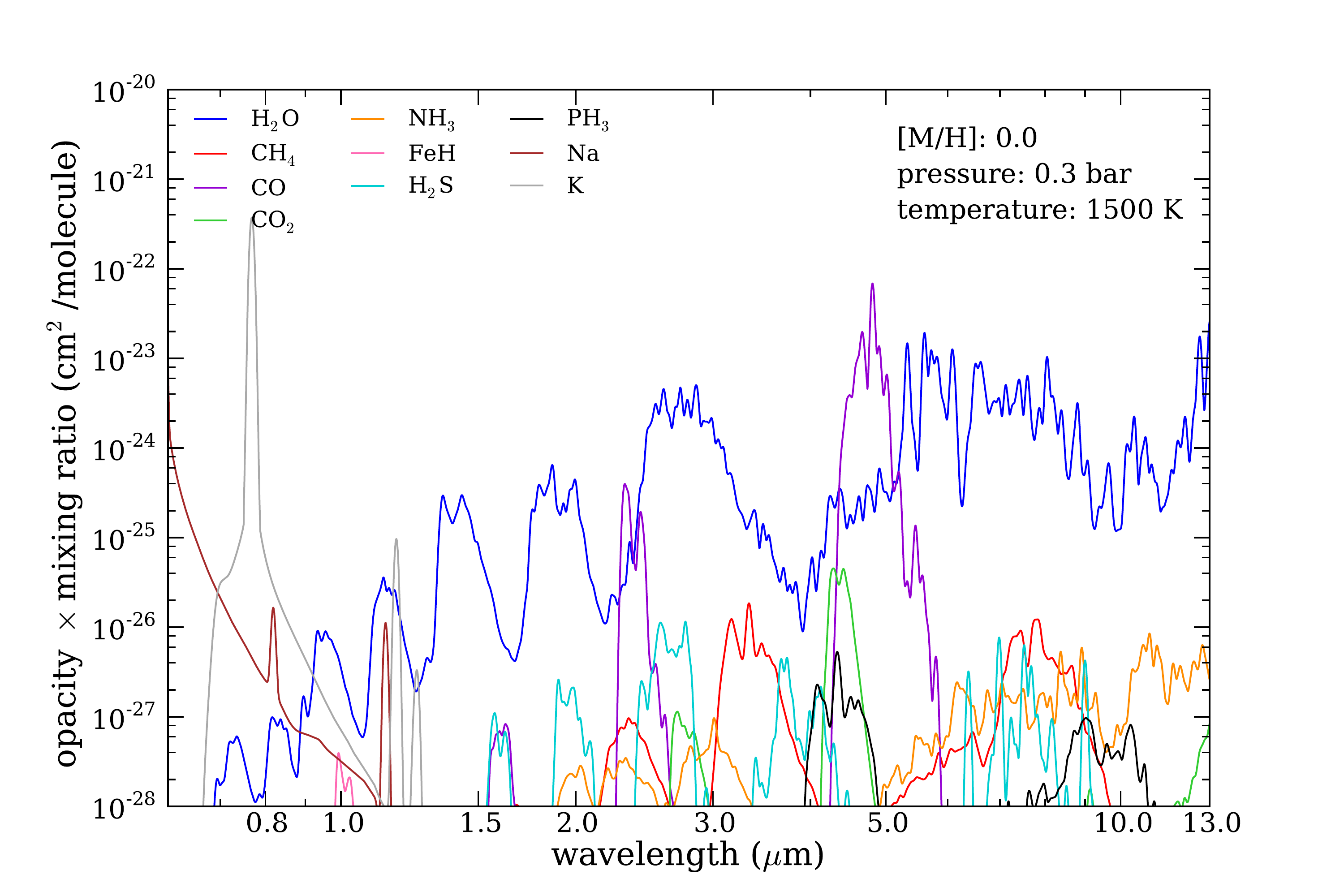}%
}
{%
  \includegraphics[clip,width=0.8\columnwidth]{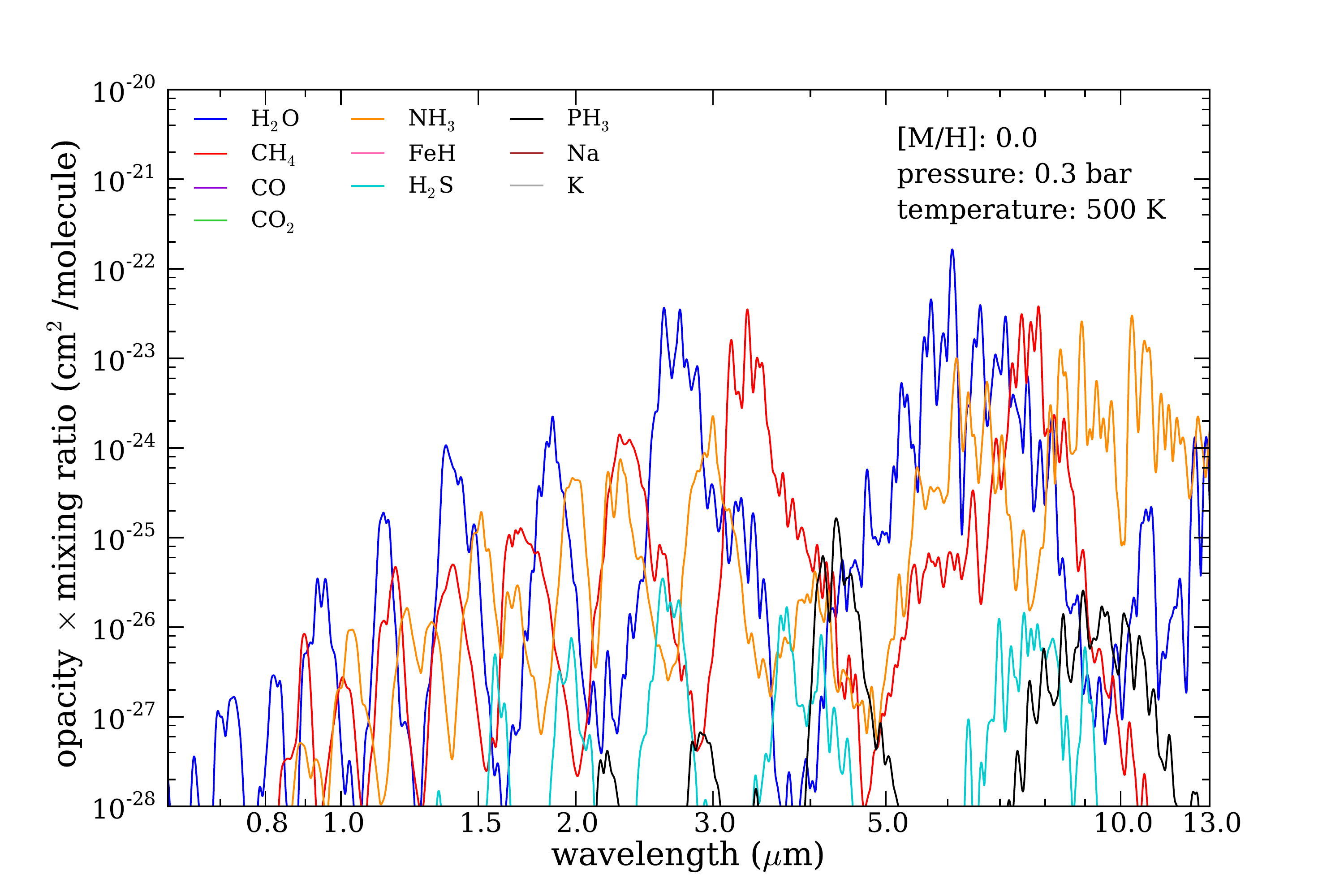}%
}
\caption{These three panels show the absorption cross-sections of molecules, weighted by their volume mixing ratios.  The x-axis is the wavelength range of interest for the \emph{James Webb Space Telescope}.  These calculations are for solar metallicity atmospheres at 0.3 bar, at 2500 K, 1500 K, and 500 K.  Water vapor is a dominant opacity source at all of these temperatures.  Figure courtesy of Caroline Morley. \label{3abunds}}
\end{figure}

The entire concept of a ``photosphere,'' the $\tau=2/3$ surface from which all photons are emitted, is nearly meaningless in a planetary atmosphere, where opacity can vary widely from wavelength to wavelength.  As an example, Figure \ref{3abunds} shows the absorption cross-section (cm$^2$ per molecule) for a solar mixture of gases at 0.3 bar and three separate temperatures (2500 K, 1500 K, and 500 K) which shows that at all temperatures the opacity is nowhere dominated by any continua but instead by the opacities of various molecules.

\begin{figure}[htp]
\includegraphics[clip,width=1.0\columnwidth]{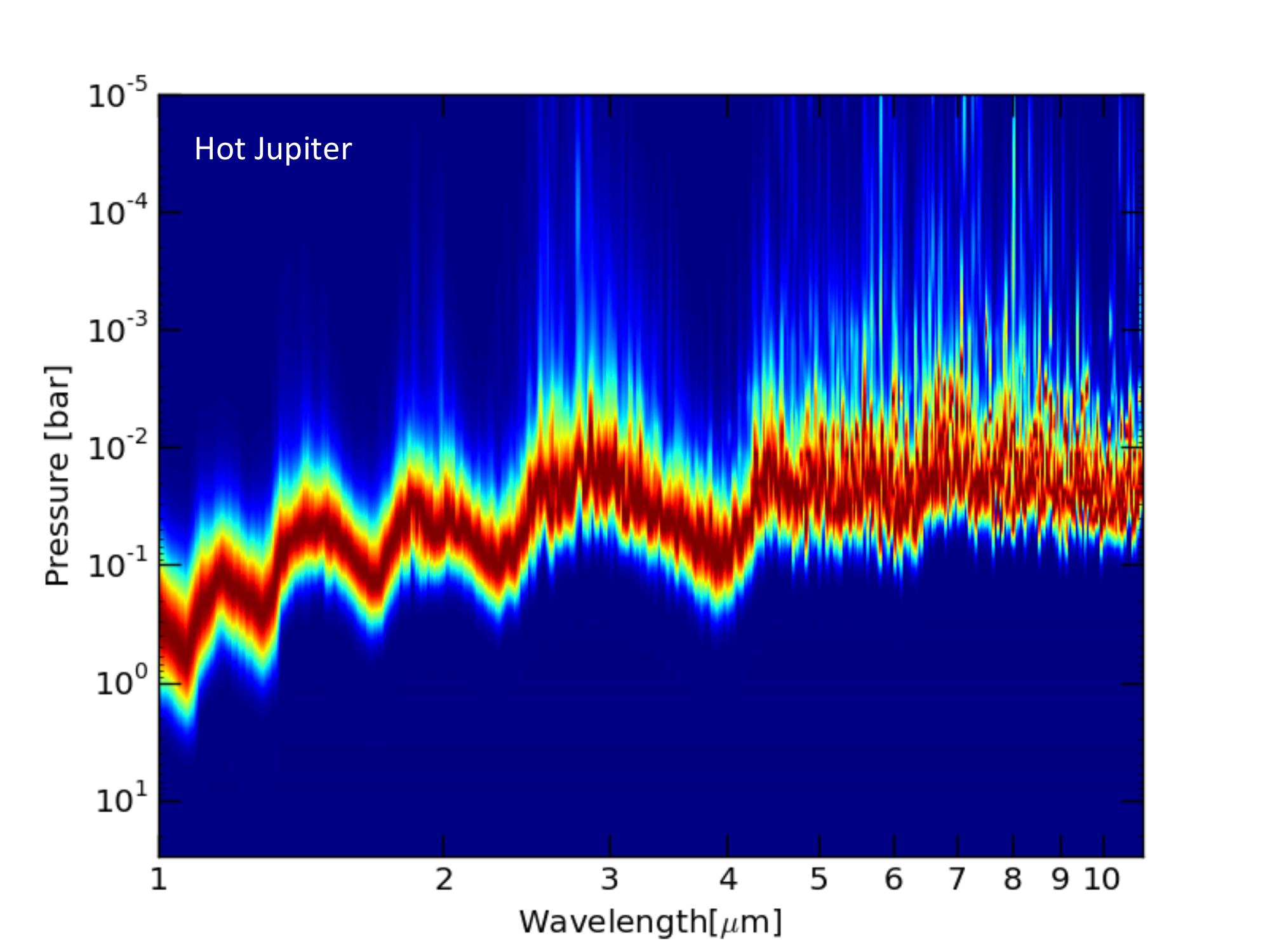}
\caption{Contribution function vs.~wavelength for an HD 209458b-like hot Jupiter.  Maximum contributions are shown in red.  Where opacity is highest, contributions come from the lowest pressures.  At wavelengths of lower opacity, flux emerges from higher pressures.  Figure courtesy of Mike Line. \label{cf}}
\end{figure}

The best way to interpret planetary emission spectra is with the concept of the brightness temperature, $T_{\mathrm B}$.  This is a wavelength-dependent quantity that is the temperature that a blackbody planet must have to emit the same amount of specific flux as the real planet, at that wavelength.  For an atmosphere in local thermodynamic equilibrium (LTE), this corresponds to the temperature of the real atmosphere where the optical depth reaches $2/3$ --- this is the level in the atmosphere that one ``sees'' down to.  This is an important way to think about spectra as it can show us, for instance, that difference levels of flux in different infrared bands could actually come from the same level in the atmosphere, if they have the same $T_{\mathrm B}$.  The spectra and \tb, which are \emph{both measured quantities}, can be turned into the \emph{pressure level} of that thermal emission by interpolating the \tb\ values on a model pressure-temperature profile.  One can then infer the wavelength-dependent pressure probed from an observed spectrum.

A more detailed analysis \cp[e.g.,][]{ChambHunt} of emergent emission spectra shows, and one could likely intuit, that thermal emission at a particular wavelength comes from a range of pressures, not from only one precisely defined pressure.  From this analysis emerges the definition of the ``contribution function,'' which shows the \emph{pressure range} from which thermal emission emerges.  The pressure that corresponds to \tb\ is then merely the location of the maximum of the contribution function.  The contribution function can be quantified, as by \ct{Knutson09}, as: 
\begin{equation}
    cf(P)=B(\lambda,T) \frac{d e^{- \tau}}{d \log(P)}
\end{equation}
A plot of the color-coded contribution function vs. wavelength for a hot Jupiter atmosphere model is shown in Figure \ref{cf}.

\section{Stepping Through Physical Effects}
In the following sections we will look at how a variety of physical and chemical processes affect the temperature-pressure profile and spectra of exoplanetary atmospheres.  We will do this through a series of model calculations starting with, and then deviating from, solar-composition H/He atmospheres.

The atmosphere code employed for calculating these models iteratively solves for radiative-convective equilibrium by adjusting the size of the convection zone until the lapse rate everywhere in the radiative region is sub-adiabatic. This code was originally developed for modeling Titan’s atmosphere \cp{Mckay89}, and has been extensively modified and applied to the study of brown dwarfs \cp{Marley96,Burrows97,Saumon08,Morley12} and solar and extrasolar giant planets \cp{Marley99,Marley12,Fortney05,Fortney08a,Fortney13,Morley15}.  The radiative transfer equations are computed using optimized algorithms described in \ct{Toon89}.

\subsection{Surface Gravity}
Typically, when one constructs a grid of model atmospheres, the first parameters of interest are the \teff\ and the surface gravity.  The surface gravity is often written as a log (in cgs units), such that ``log $g$ = 4.0'' means a gravity of $10^4$ cm s$^{-2}$, or 100 m s$^{-2}$.  For reference, Jupiter's surface gravity is around 25 m s$^{-2}$.  Surface gravity is a unit of choice because it is flexible.  We may not know the masses and radii of the objects that we are studying so these quantities can be swept into the gravity.

\begin{figure}[htp] 
\centering
\subfloat[]{
  \includegraphics[width=60mm]{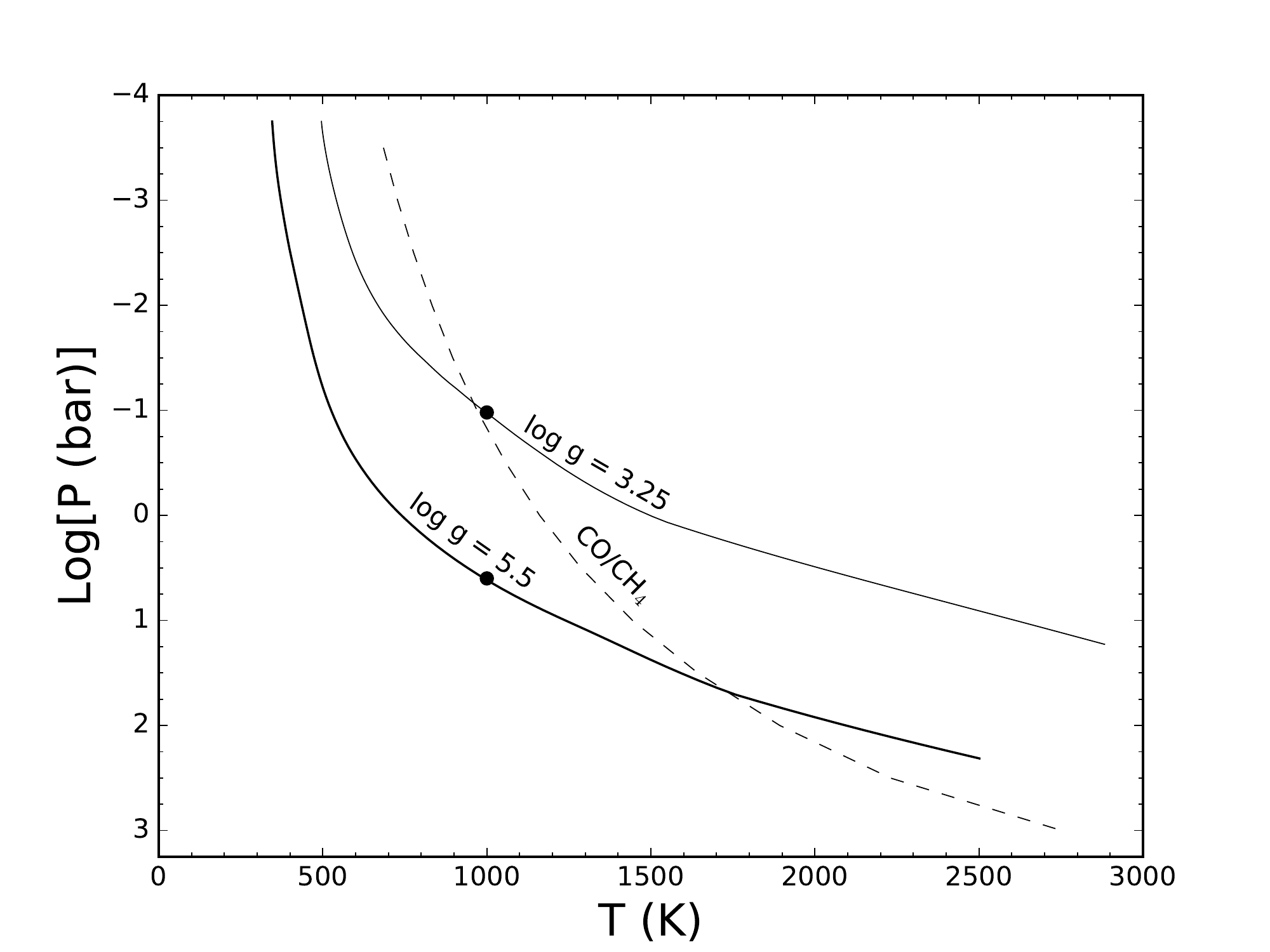}
}
\subfloat[]{
  \includegraphics[width=60mm]{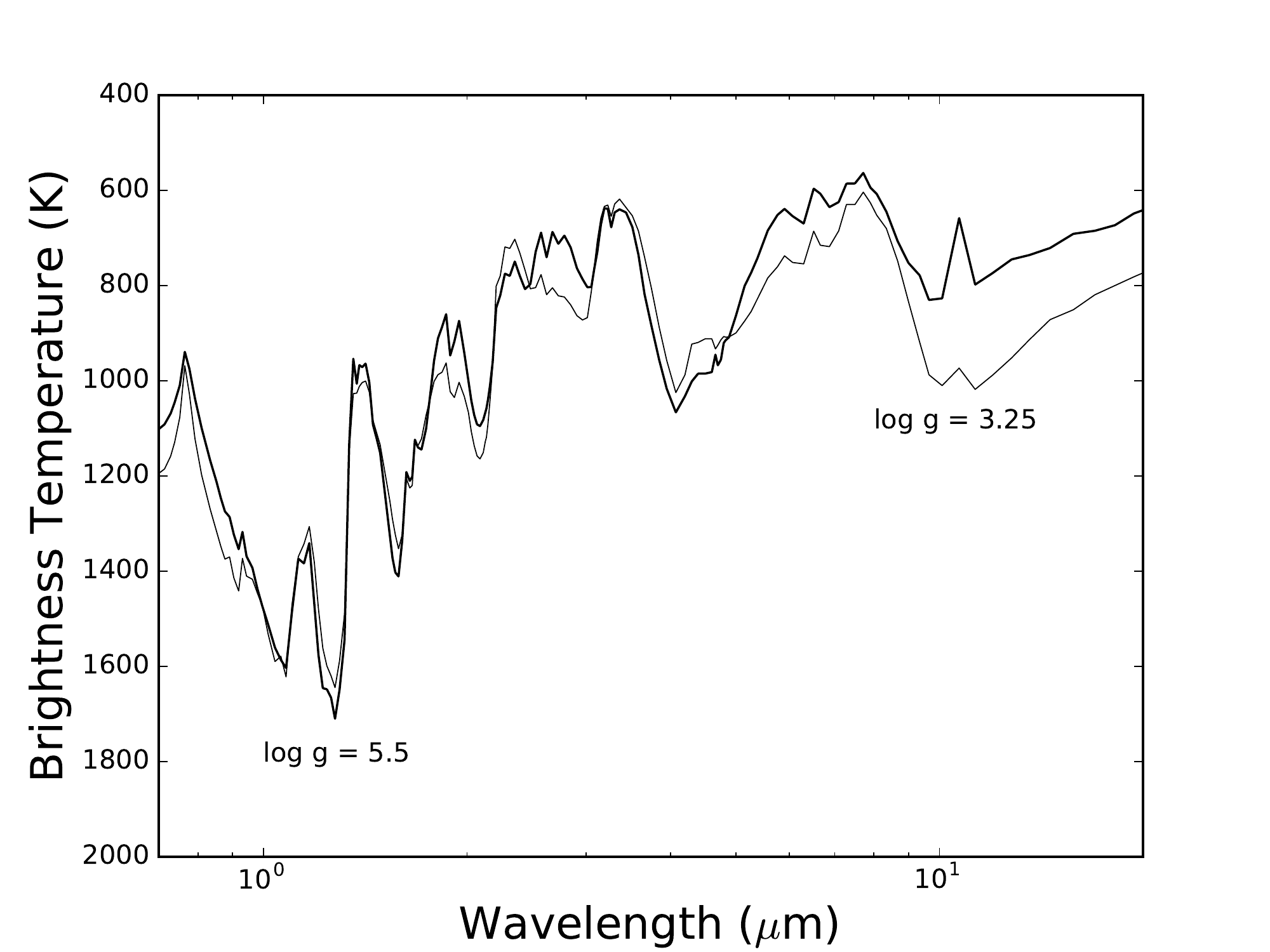}
}
\newline
\subfloat[]{
  \includegraphics[width=60mm]{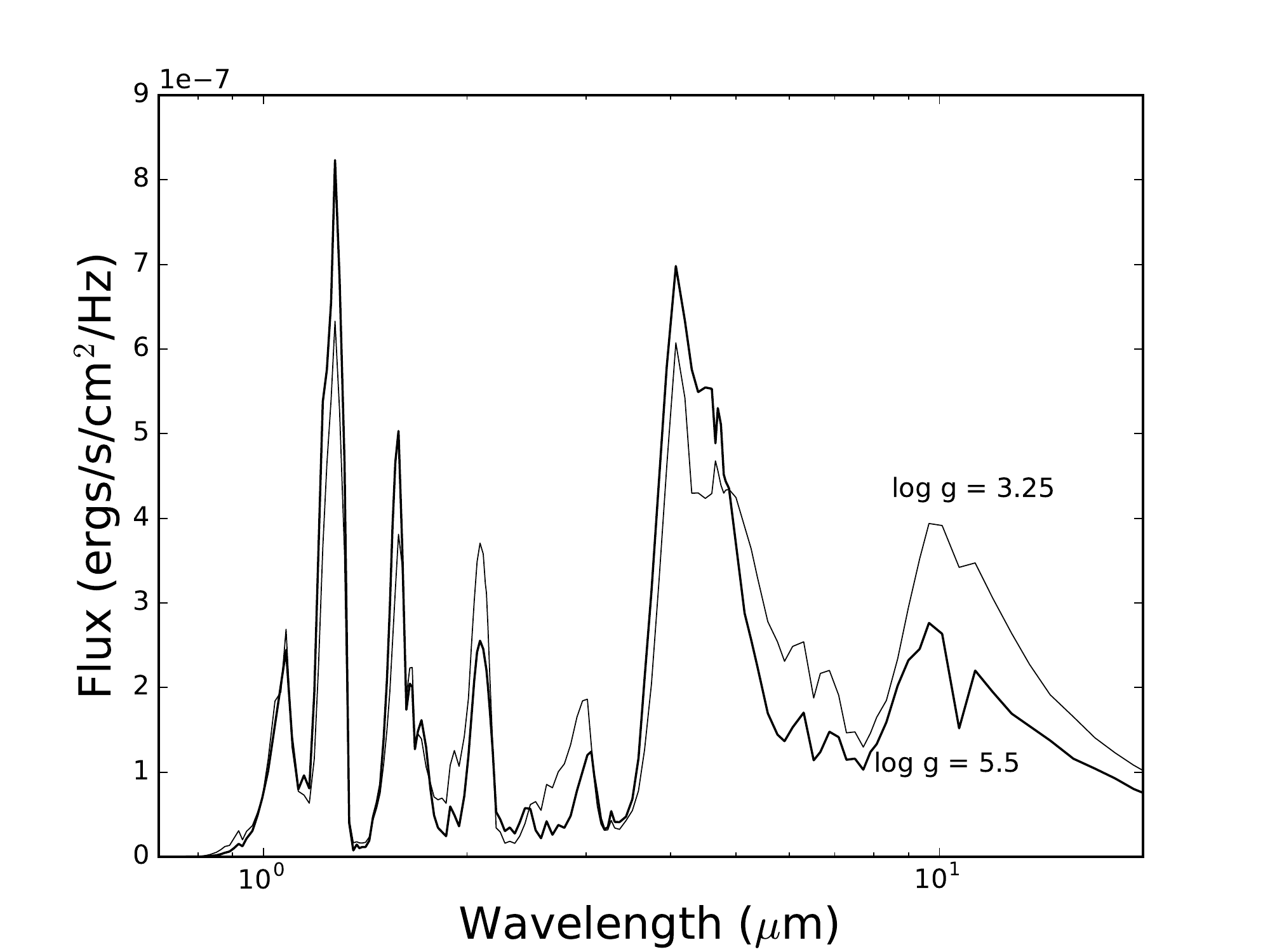}
}
\subfloat[]{
  \includegraphics[width=60mm]{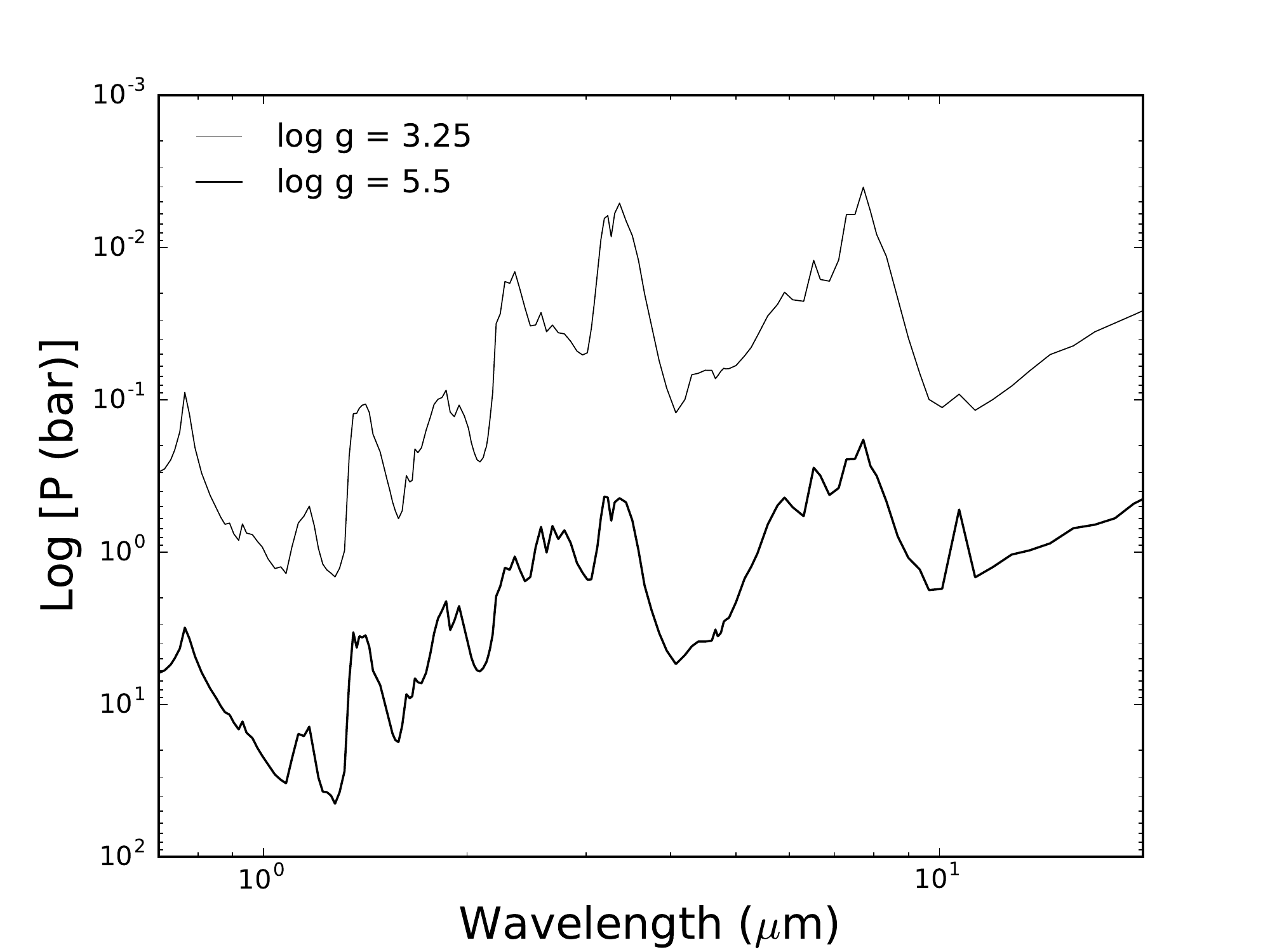}
}

\caption{These four panels demonstrate the influence of surface gravity.  Shown in each are two models with the same \teff\ of 1000 K, but surface gravities that differ by a factor of 178.  Panel (A) shows the pressure-temperature profiles.  Panel (B) shows the \tb\ for each wavelength.  Panel (C) shows the emitted spectrum for both models.  Panel (D) shows the pressure of the $\tau=2/3$ layer as a function of wavelength, which graphically shows that lower-gravity atmospheres have lower pressure photospheres (see text).  Pressure-dependent opacity like hydrogen collision-induced absorption (CIA) limits the depth one can see around 2 $\mu$m and longward of 10 $\mu$m in the higher gravity model.\label{gravity}}
\end{figure}

All things being equal, lower gravity objects have emission from lower atmospheric pressures.  This is clearly seen in Figure \ref{gravity}, which shows the \emph{T--P} profiles of two models with \teff\ of 1000 K.  The black dot on the profiles (top panel) at 1000 K show where the local temperature is equal to the model \teff.  This can be thought of as a kind of ``mean photospheric pressure,'' although keep in mind that the emission comes from a range of pressures.  Lower gravity leading to a lower photospheric pressure can be understood as the effect of gravity on the scale height, $H$, where $H = k T / \mu m_{\rm H} g$, and $k$ is Boltzman's constant, $T$ is the temperature, $\mu$ is the dimensionless mean molecular weight (2.35 for a solar abundance mix of gasses), $m_{\rm H}$ is the mass of the hydrogen atom, and $g$ is the surface gravity.  For an isothermal, constant gravity, and exponential atmosphere, the column density of molecules ($N$), from some reference location with local density $n_o$, vertically to infinity, is $N=n_o H$.  This value $N$ is directly proportional to the optical depth, $\tau$.  In either a low gravity or high gravity atmosphere, we see down to the $\tau=2/3$ level at a given wavelength.  If the gravity is lower at the same temperature, $H$ will be larger, meaning that $n_o$ will be smaller (found at a lower pressure) to reach the same $N$, or the $\tau=2/3$ level.

We should expect the spectra of these two profiles to differ given their quite different temperature structures, and that is what we see in the second and third panels.  A few things are worth noting:  In the higher gravity atmosphere, the potassium doublet at 0.77 $\mu$m (first seen in Figure \ref{3abunds} above) is much more pressure-broadened.  Also, the flux peaks in the J (1.2 $\mu$m), H (1.6 $\mu$m), and K (2.2 $\mu$m) differ significantly.  The high gravity model is much brighter in J and dimmer in K, compared to the low gravity model.  This is due to the opacity source known a hydrogen ``collision induced absorption'' (CIA), which goes as the square of the local density.  In higher pressure photospheres, this opacity source, which peaks in K band, is significantly more important.  Most of the rest of the spectral differences can be attributed the differing abundances of CO and CH$_4$.  The dashed curve shows where these molecules have an equal abundance in thermochemical equilibrium.  To the right of this curve, CO is dominant, and to the left, CH$_4$ is.  As one travels further from this curve, less and less of the ``unfavored'' species is found in the atmosphere.

\subsection{Metallicity}
\begin{figure}[htp]   
\centering
\subfloat[]{
  \includegraphics[width=60mm]{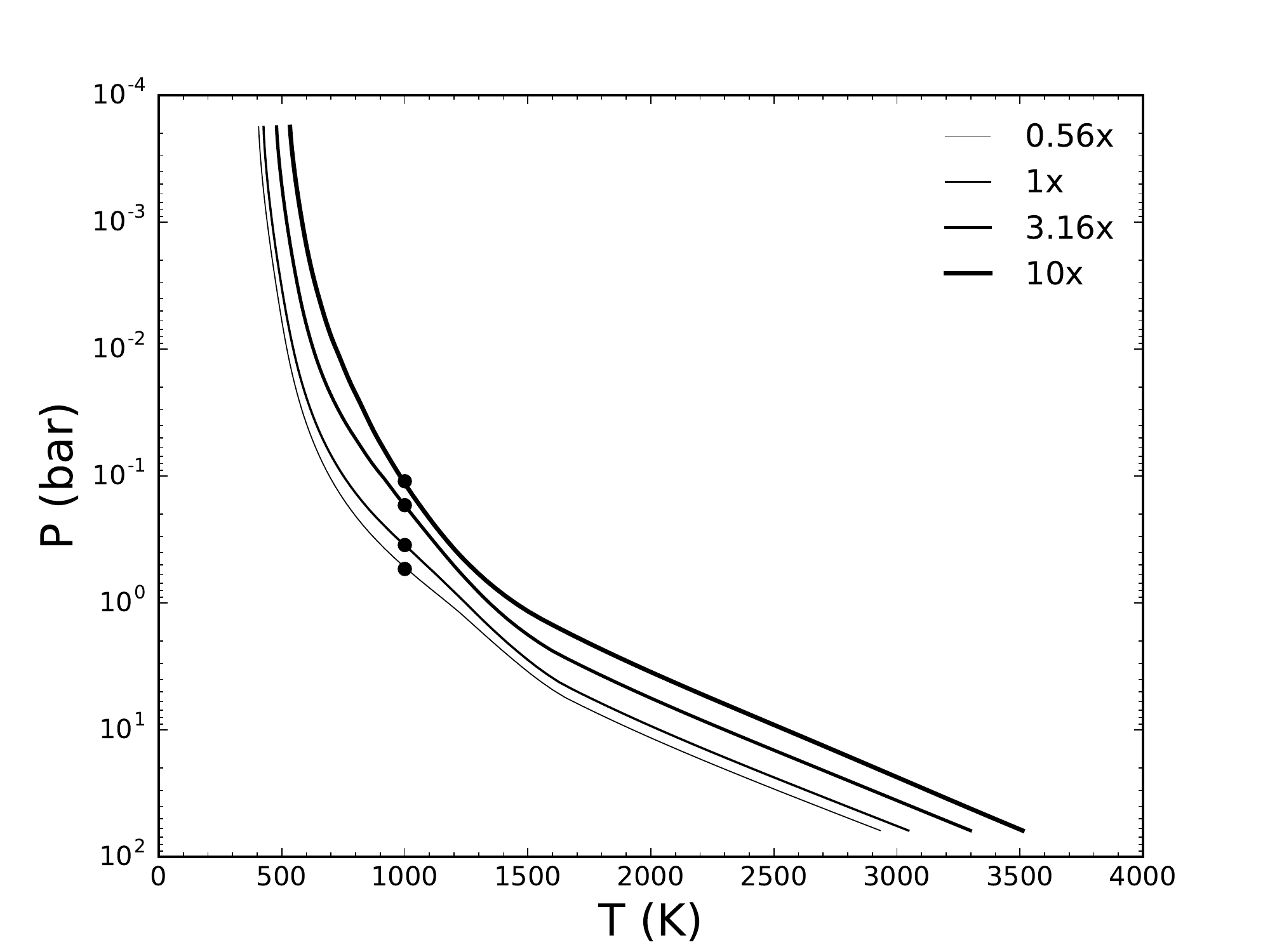}
}
\subfloat[]{
  \includegraphics[width=60mm]{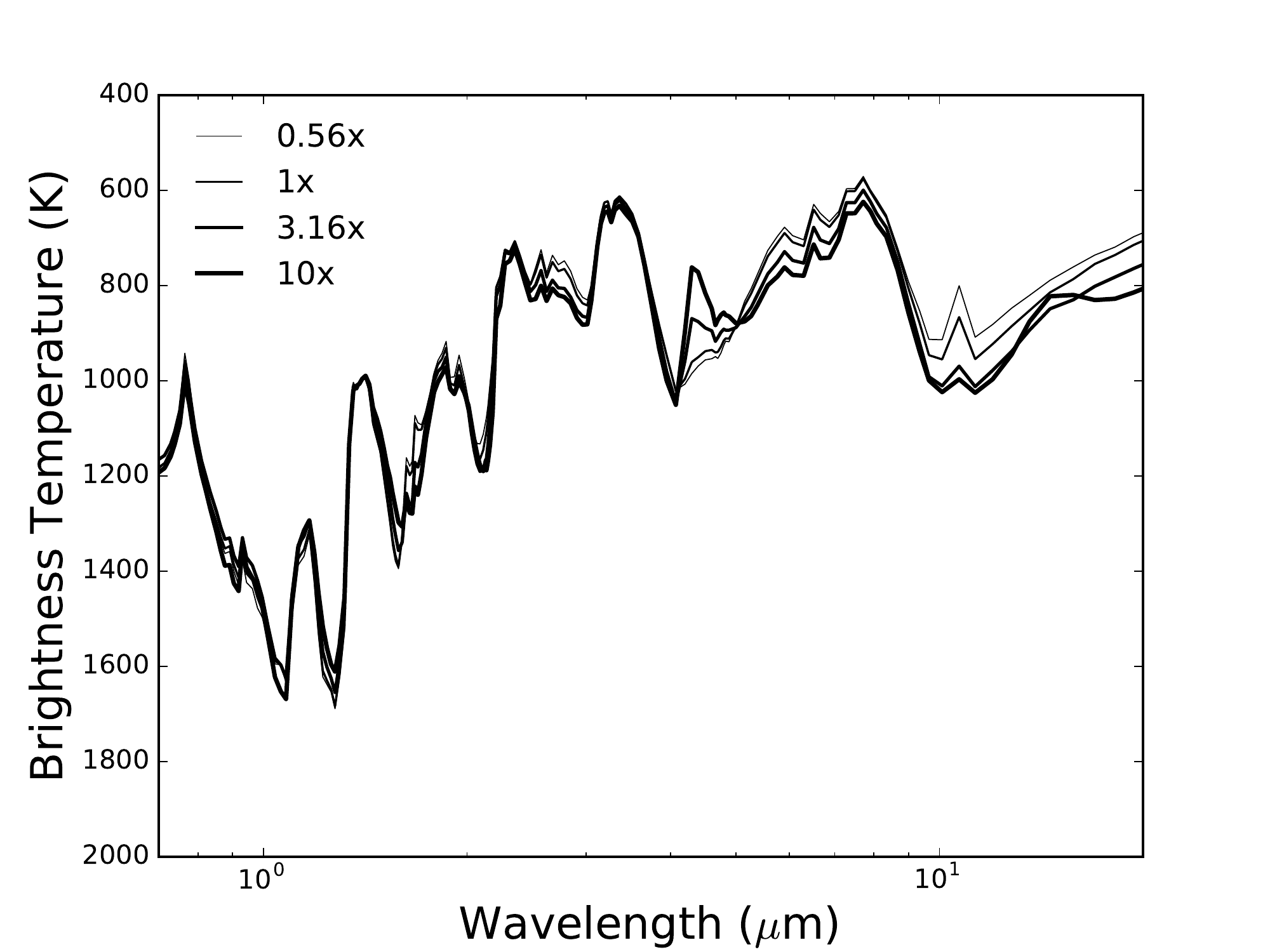}
}
\newline
\subfloat[]{
  \includegraphics[width=60mm]{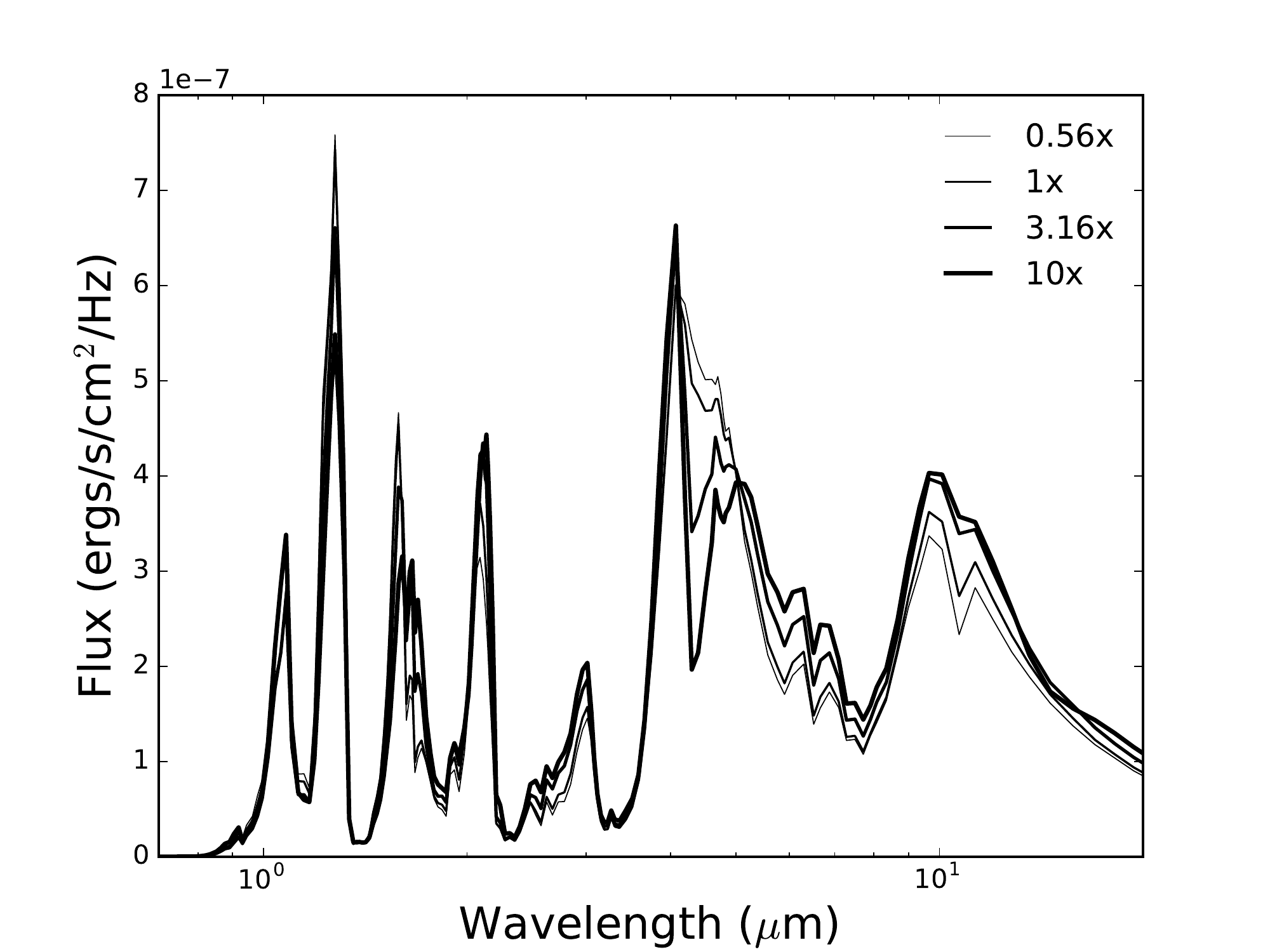}
}
\subfloat[]{
  \includegraphics[width=60mm]{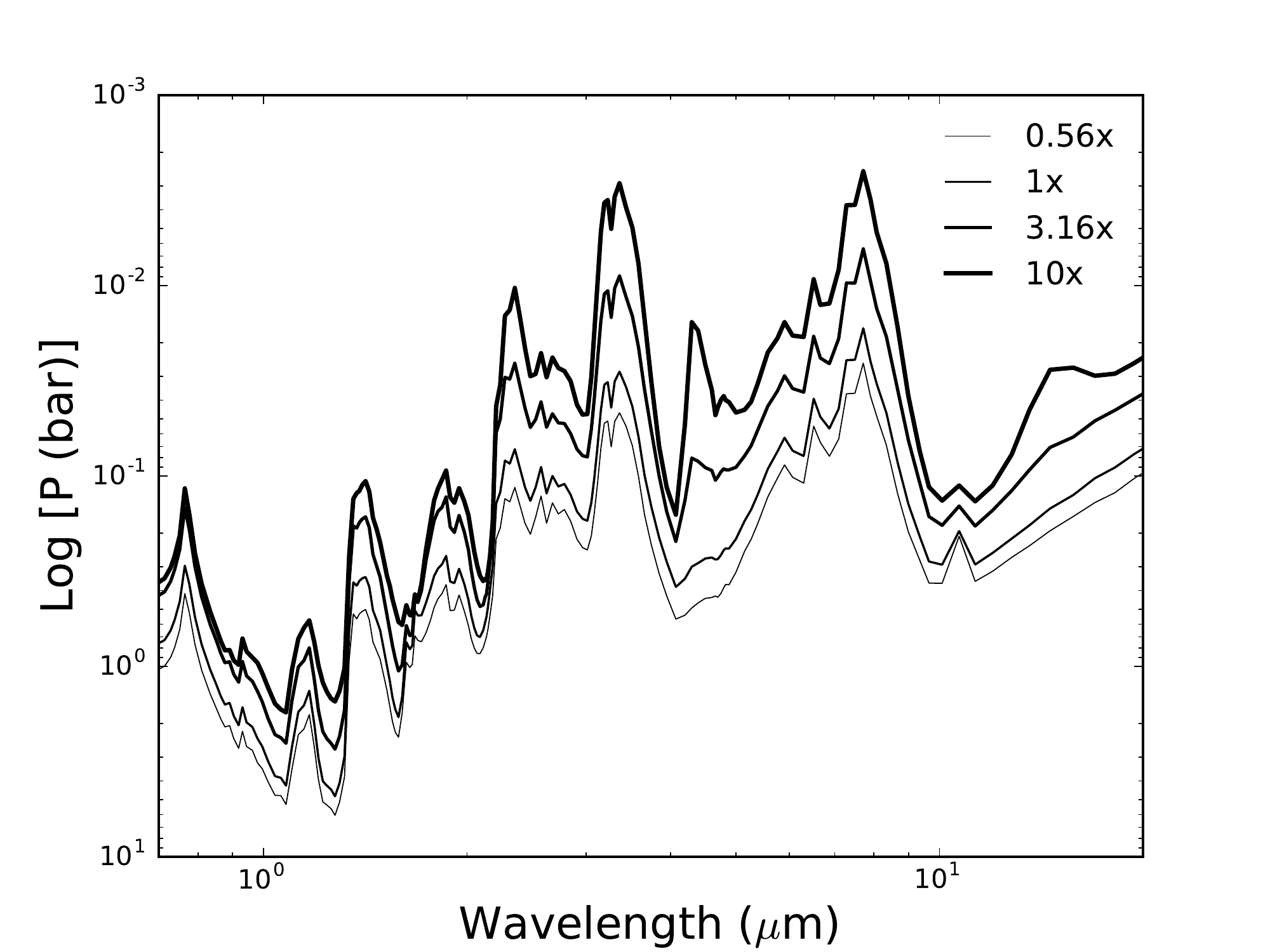}
}
\caption{Shown are four 1000 K log $g$ = 4 models at four different metallicities, $-0.25$, 0.0 (solar), +0.5, and +1.0.  The thickest line is the highest metallicity.  Panel (A) shows the \emph{P--T} profiles.  Panet (B) shows the \tb\ values for these models.  Panel (C) shows the emitted spectra.  Panel (D) shows the wavelength-dependent pressure level of the $\tau=2/3$ layer.  The spectra of the models look fairly similar, except for much larger absorption due to CO and CO$_2$ in the metal-rich models, as CO and CO$_2$ and increase linearly and quadratically with metallicity, respectively.  \label{metallicity}}
\end{figure}

The abundances of atoms and molecules in an atmosphere obviously also dictate the depth to which one can see at a given wavelength, and hence, the emitted spectra.  For a H/He dominated atmosphere, as the metallicity increases, the opacity increases, and the photospheric pressures decrease.  This can be seen in Figure \ref{metallicity}, which shows four models, all with \teff\ $=1000 $K, but with metallicity values of [M/H]$=-0.25$, 0.0, +0.5, and +1.0.  [M/H] is a log scale referenced to solar abundances, such that "0.0" is solar and "+1.0" is ten times solar.  The metal-rich models have lower pressure photospheres so their deep atmospheres end on a warmer adiabat.

The spectra of these atmospheres for the most part look fairly similar.  The main differences here are due to how metallicity influences chemical composition.  The abundances of CO and CO$_2$ scale linearly and quadratically with metallicity, respectively \cp{Lodders02}.  This is seen most clearly from 4-5 $\mu$m (again refer to Figure \ref{3abunds}, where the metal-rich models show significantly more absorption from CO/CO$_2$).

\subsection{Carbon-to-Oxygen Ratio}
In a solar metallicity gas, the carbon-to-oxygen (C/O) ratio is 0.54 \cp{Asplund09}.  In such a mixture, at higher temperatures, CO takes up just over half of the available oxygen, leaving the remainder of the oxygen to be found in H$_2$O.  At cooler temperatures, carbon is found in CH$_4$, leaving most of the oxygen free to be in H$_2$O.  This is readily seen in the spectral sequence of cool brown dwarfs, as seen in Figure \ref{cushing} and \ct{Kirkpatrick05}.  The C/O ratio also effects the condensation sequence of the elements which are lost into clouds (Figure \ref{ptgrid1}), as many of these refractory species take oxygen out of the gas phase and into the solid phase.

However, the details of particular abundances of molecules at a given $P$, $T$, and base elemental mixing ratios, quite sensitively depend on the abundances of C and O, as one might expect.  In particular, there is a dramatic change at C/O$\,>1$.  If C/O$\,>1$, then at hot temperatures nearly all oxygen will be tied up in CO, with little left for H$_2$O.  Extra carbon can then go into CH$_4$, which is never seen at high temperatures for ``normal'' C/O ratios.  The implications for spectra and the atmospheric structure of hot Jupiters have been examined in detail by \ct{Madhu11}, \ct{Madhu11b}, and \ct{Molliere15}.

\begin{figure}[htp] 
{%
  \includegraphics[clip,width=0.75\columnwidth]{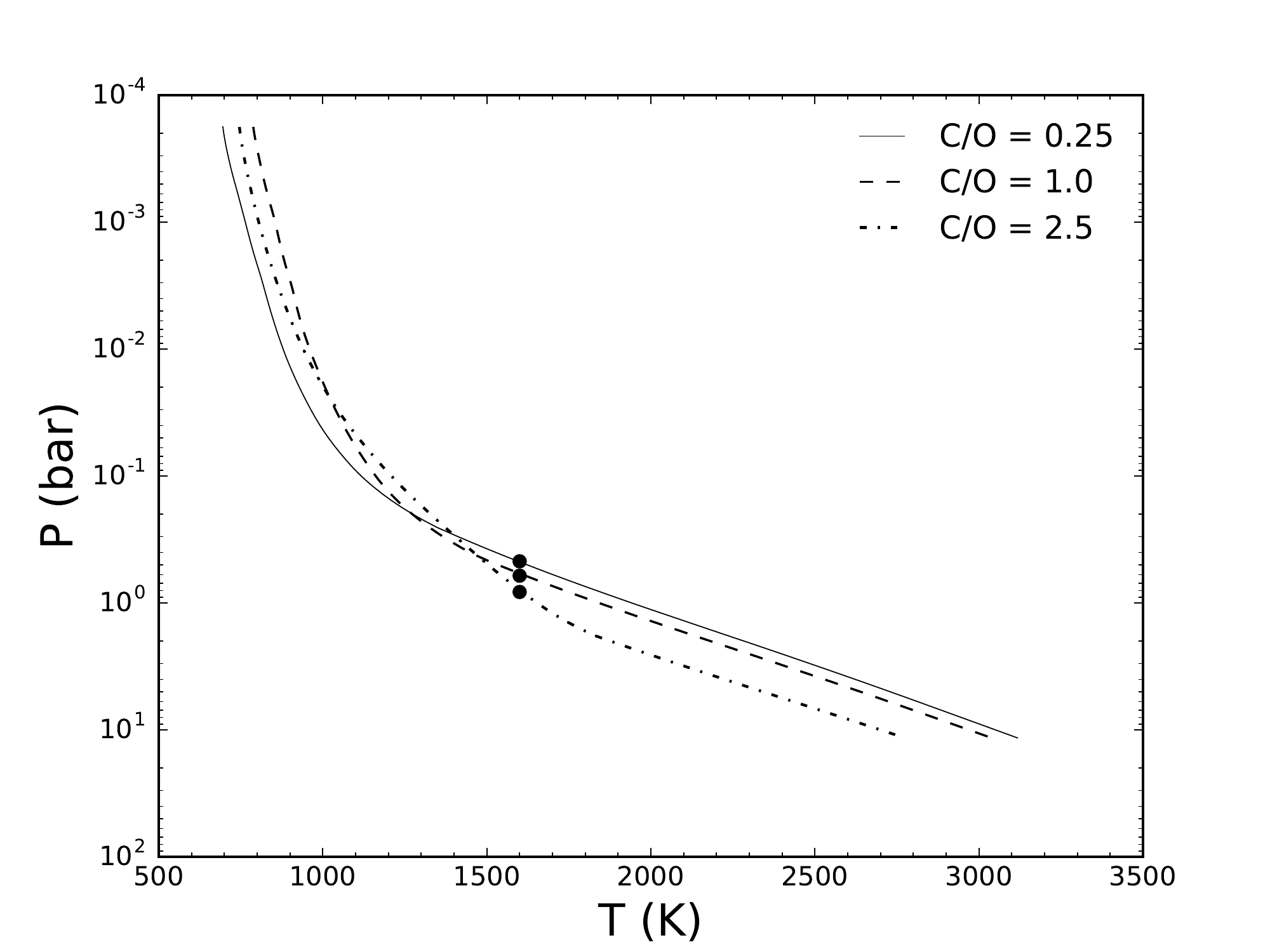}%
}
{%
  \includegraphics[clip,width=0.75\columnwidth]{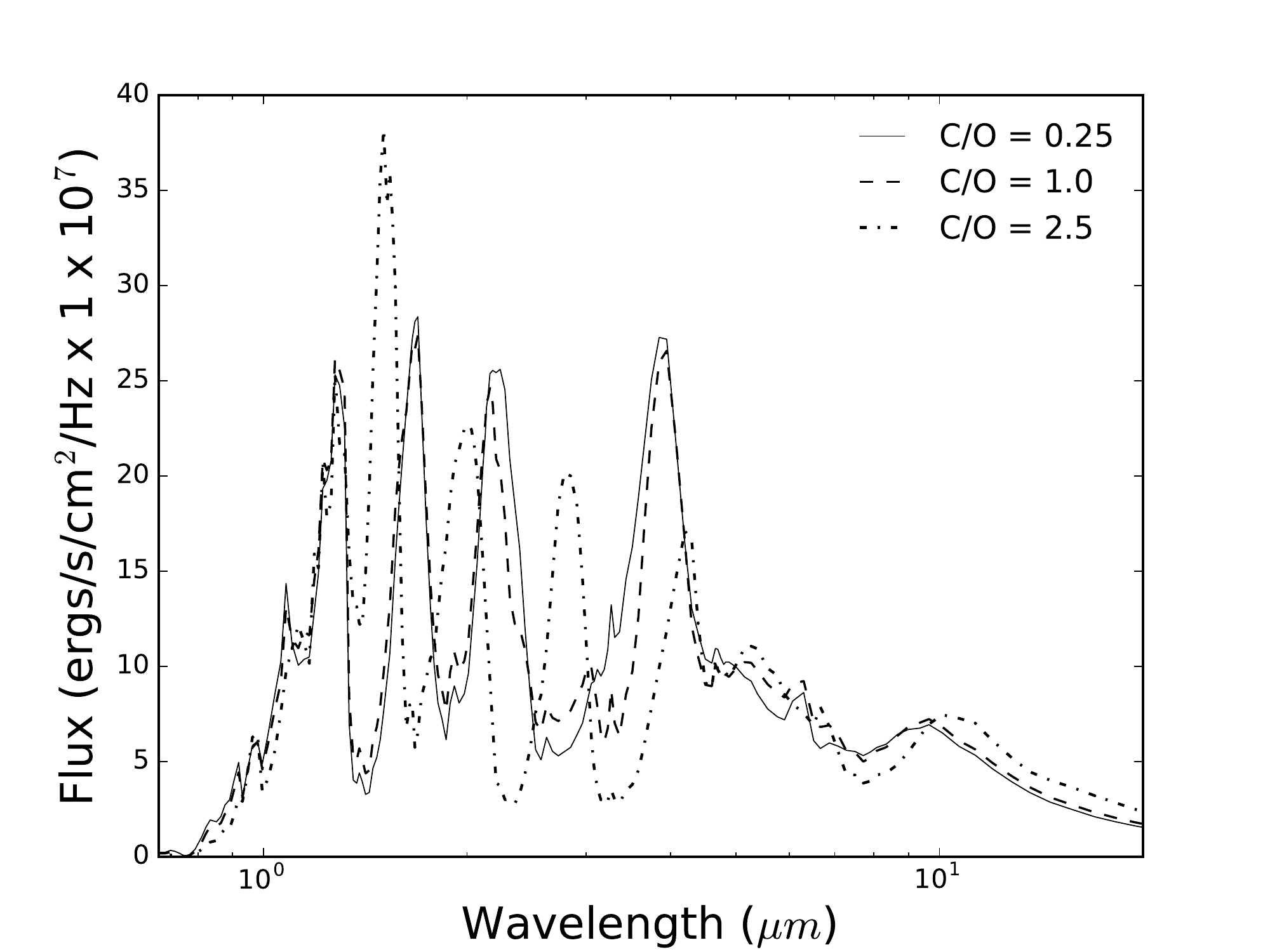}%
}
{%
  \includegraphics[clip,width=0.75\columnwidth]{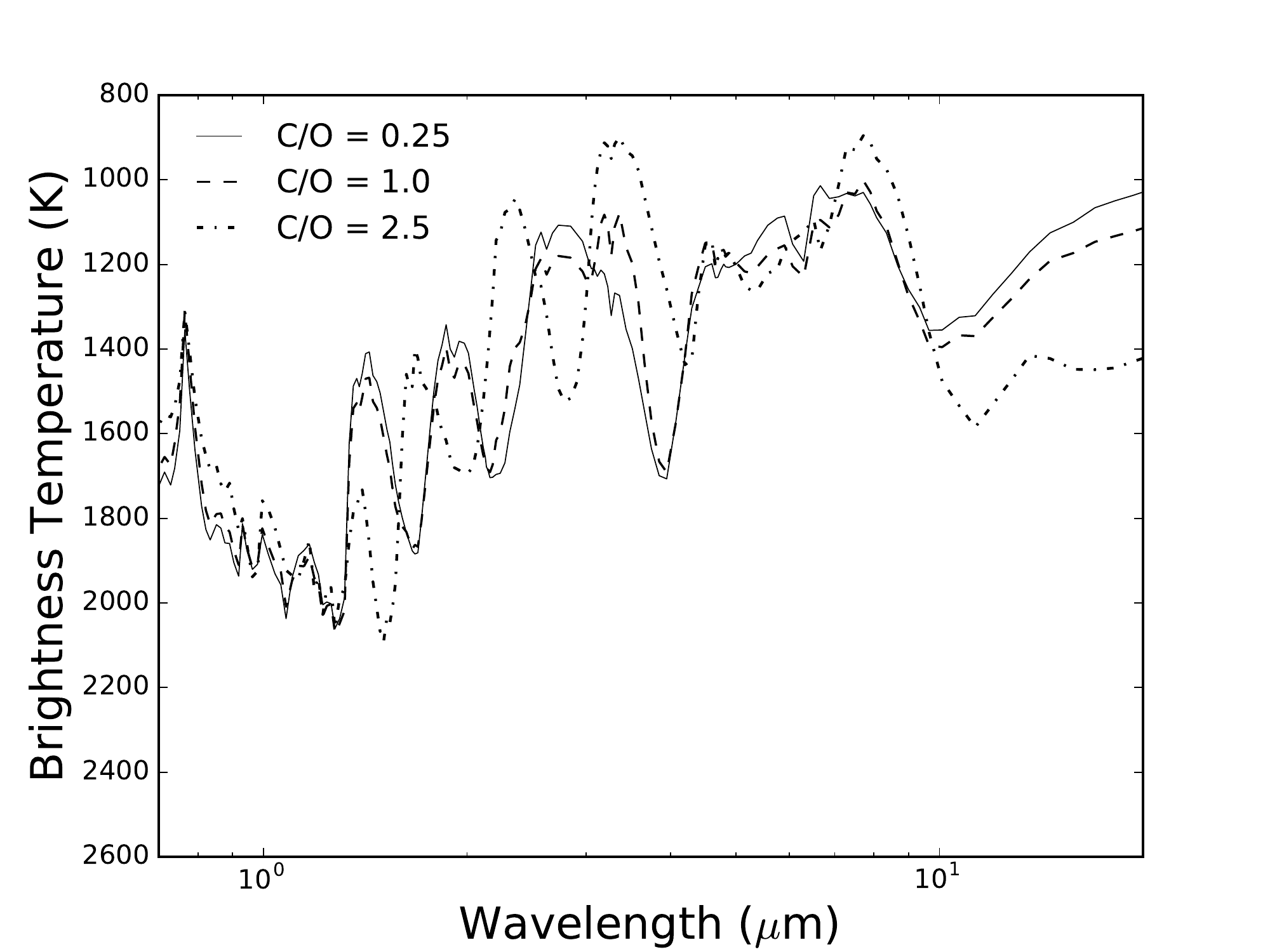}%
}

\caption{These three panels show the effect of the atmospheric C/O ratio for three models at \teff\ $=1600$ K and log $g$=4.  Differences are mostly subtle in the \emph{P--T} profile and are predominantly due to changes in the water vapor opacity, which is the dominant absorber. \label{co} }
\end{figure}

Here we will examine the changes in the \emph{P--T} profiles and spectra for three illustrative values of the C/O ratio in Figure \ref{co} for a 1600 K, log $g=4.0$ model.  These numbers are shown referenced to the solar value, such that C/O=1.0 is the solar value, \emph{0.25} is $1/4$ the solar value, and \emph{2.5} is $5/2$ of the solar value.  These 1600 K models are hot enough that CO is the dominant carbon carrier.  The models have relatively similar temperature structures, but the highest C/O value yields the highest pressure photosphere.  This is because this model has the most C and O tied up in CO gas, which is relatively transparent in the infrared compared to H$_2$O and CH$_4$ (see Figure \ref{3abunds}).  The spectra of the solar model and the \emph{0.25} model are relatively similar, as both are dominated by H$_2$O opacity, with some contributions from CO.  However, for the high C/O model, we can see that the H$_2$O bands essentially vanish and are replaced by strong CH$_4$ bands, which dramatically alters the emitted spectrum.  Such objects have not been seen in the collection of brown dwarfs, but it is possible there are formation pathways for such carbon-rich giant planets, as discussed in Section \ref{formation}.

\subsection{Incident Flux}
The incident flux from the parent star, often known as irradiation, insolation, or instellation, has a dramatic effect on the temperature structure of a planet.  Indeed, for a terrestrial planet, the incident flux is the planet's only important energy source.  For the Earth, most of the Sun's flux penetrates the atmosphere, which is optically thin in most of the optical, and is absorbed or scattered at the surface.  This provides the Earth's atmosphere with a warm bottom boundary.  Earth's atmosphere is optically thick at thermal infrared wavelengths near the surface, such that convection dominates in our troposphere.

\begin{figure}[htp]  
{%
  \includegraphics[clip,width=0.75\columnwidth]{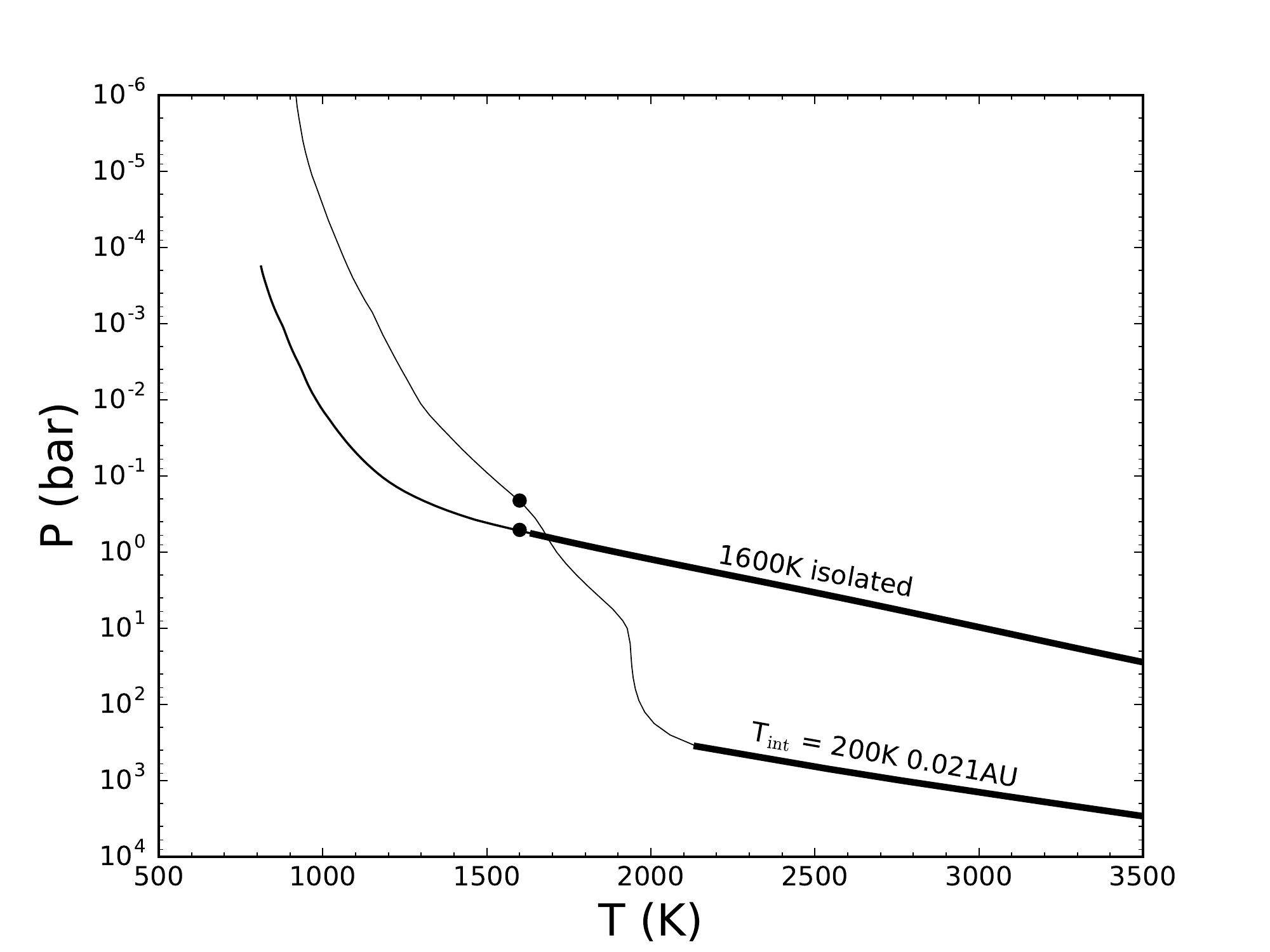}%
}
{%
  \includegraphics[clip,width=0.75\columnwidth]{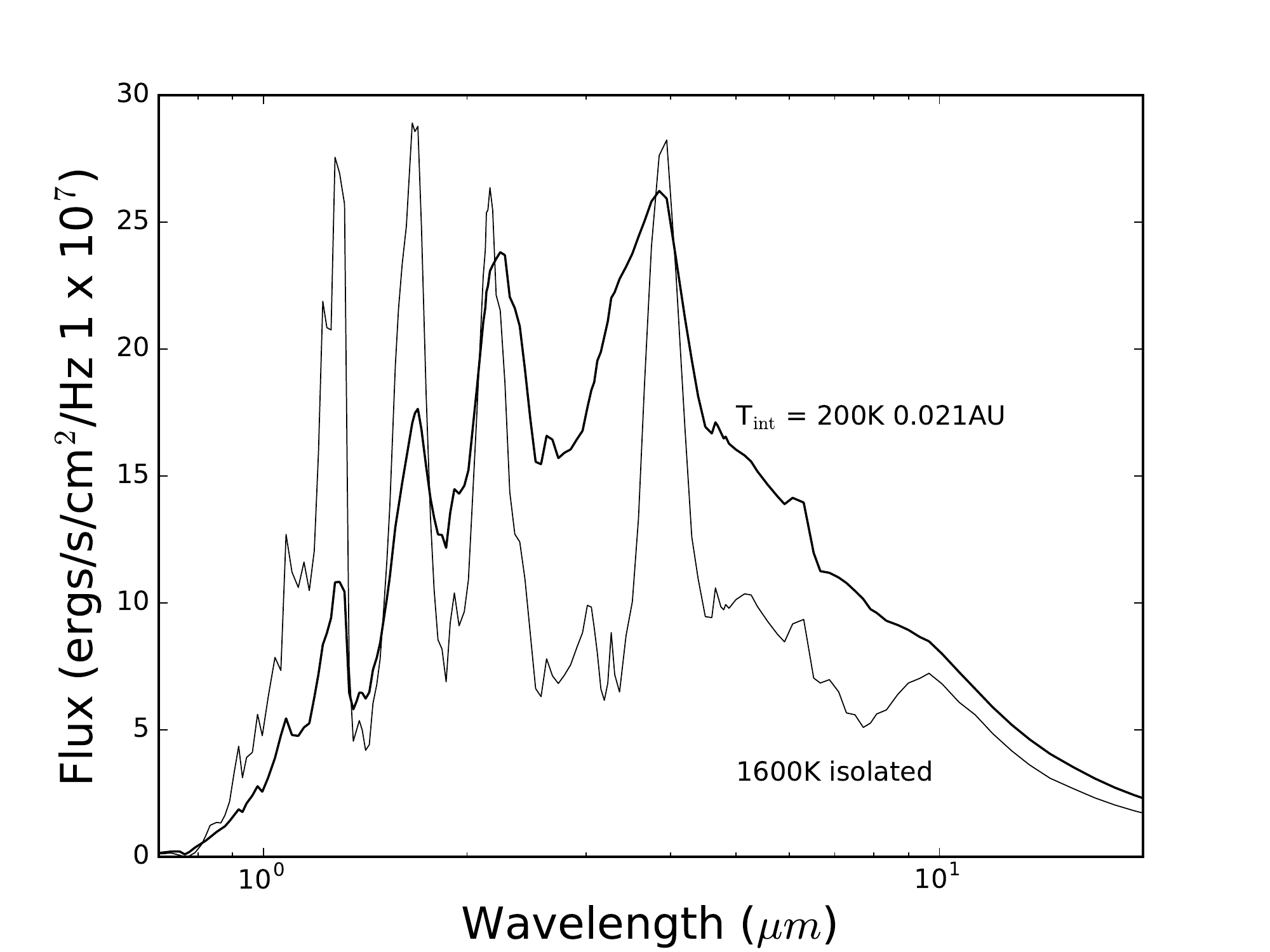}%
}
{%
  \includegraphics[clip,width=0.75\columnwidth]{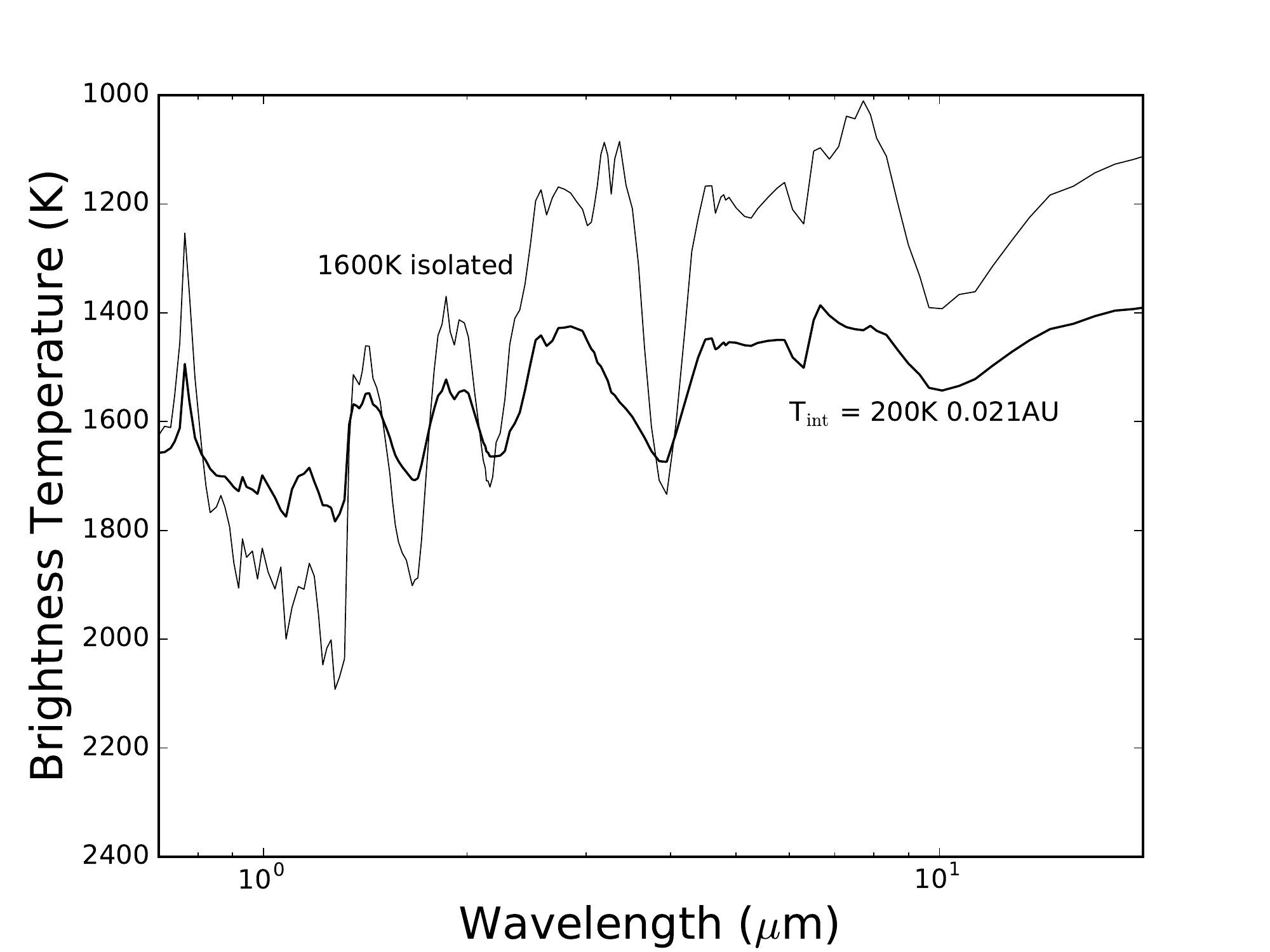}%
}
\caption{Shown are two models with the same \teff (1600 K) and surface gravity (log $g$ =4).  One is an isolated object and one is in a close-in orbit around a main sequence G0 parent star.  The irradiated planet model has a much shallower temperature gradient.  The radiative-convective boundary is pushed to much higher pressures in the irradiated model.  In the top panel, the convective part of the atmosphere is shown with a thicker line.  The more isothermal atmosphere yields a more muted (modestly more blackbody-like) emission spectrum.  \label{irrad}}
\end{figure}

For a strongly irradiated planet, such as a hot Jupiter at 0.03 au, or even a sub-Neptune at 0.2 au, the absorption and re-radiation of stellar flux carves the planetary \emph{T--P} profile shape.  An illustrative example of the difference between a H/He atmosphere heated from below (like a young giant planet on a wide orbit, or brown dwarf) or a planet heated from above by stellar flux, is show in Figure \ref{irrad}.  These two models have the same \teff\ of 1600 K and same surface gravity of (log $g =4$), but the temperature structures appear radically different.  In this figure, the convective parts of the atmosphere are shown in a thicker line, while the radiative parts are shown as a thinner line.  The irradiated model is forced to have a much hotter upper atmosphere.  This also forces the more isothermal, radiative part of the atmosphere to be relatively large in vertical extent, pushing the radiative-convective boundary down to nearly 1000 bar.  This is a generic finding for all strongly irradiated atmospheres and it is a significant difference between these atmospheres and those under modest stellar irradiation.  Weakly irradiated atmospheres can typically be convective up to the visible atmosphere.  Mixing processes in the radiative part of the atmosphere are probably slower than in the convective part of the atmosphere, but it is incorrect to think of these radiative regions as being quiescent.

\begin{figure}[htp]  
{%
  \includegraphics[clip,width=1.0\columnwidth]{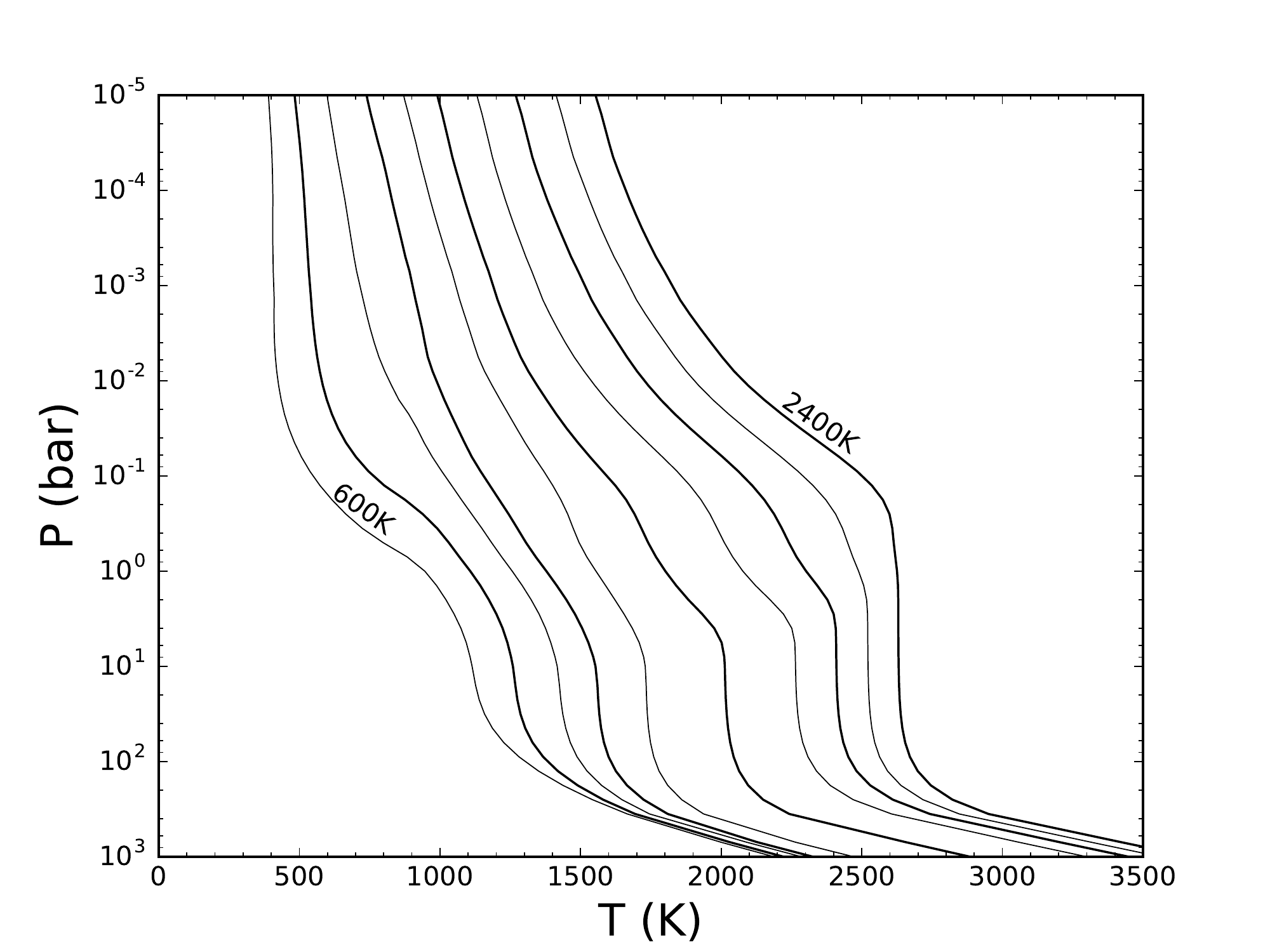}%
}
\caption{A sample calculation of the effects of stellar insolation, for planets at various distances from a 6000 K G0 main sequence star.  The models go from \teff\ of 2400 K to 600 K in 200 K increments.  All models have a \tint\ value of 200 K. \label{ptgrid}}
\end{figure}

The spectra of the two models (middle panel) look almost nothing alike, owing to the radically different temperature structures.  In the near infrared, where one sees most deeply into the atmosphere (see Figure \ref{3abunds}, middle panel), the isolated model is significantly hotter, which yields much higher near-infrared fluxes.  The bottom panel of Figure \ref{irrad} graphically shows that the more isothermal temperature structure directly leads to a smaller dynamic range in the temperatures probed, which leaves the spectrum more blackbody-like than the isolated object.  This suggests that while brown dwarfs and imaged planets will provide (and have been providing) essential lessons in terms of atmospheric abundances and molecular opacities, we should \emph{not} expect spectra of the strongly irradiated planets to necessarily follow the same sequence that is clear for the isolated objects at these same \teff\ values \cp[e.g.][]{Kirkpatrick05}.  Figure \ref{ptgrid} shows the result of a calculation of placing this same model planet at different distances from its parent star to yield \teff\ values from 2400 K to 600 K.  All models have a \tint\ value of 200 K.

\subsection{Outer Boundary Condition:  Parent Star Spectral Type}
As we have seen in Figures \ref{fluxin}, \ref{fluxout}, and the previous section, the pressure levels and wavelengths at which incident stellar flux is absorbed and then re-radiated back to space dictate the temperature structure, especially for strongly irradiated planets.   Not all parent stars have the same spectra energy distributions, as hotter A-type parent stars will peak in the blue or even UV, while cool M-star hosts will peak in the near-infrared, in accordance with Wien's law.

It is difficult to create a one-size-fits-all grid of model atmospheres for strongly irradiated planets because each particular planet has its own particular parent star.  Typically, when modeling a given exoplanet atmosphere, investigators will create a synthetic spectrum for the parent star by interpolating in a grid of stellar model atmospheres \cp[][]{Hauschildt99} for the fluxes incident upon the planet.  This effect of stellar spectral type on hot Jupiter atmospheres has been investigated in some detail by \ct{Molliere15}.

\begin{figure}[htp]  
{%
  \includegraphics[clip,width=0.75\columnwidth]{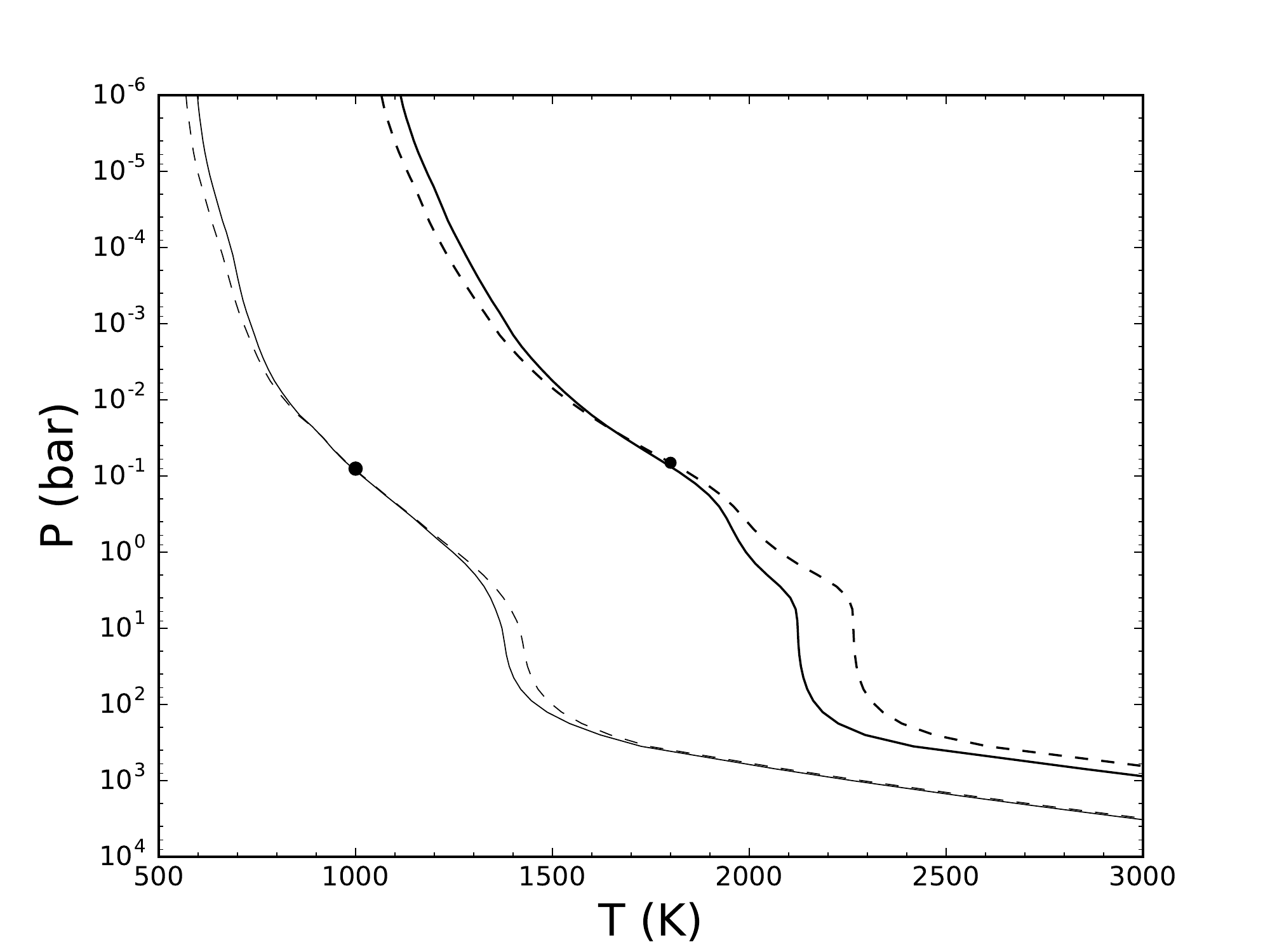}%
}
{%
  \includegraphics[clip,width=0.75\columnwidth]{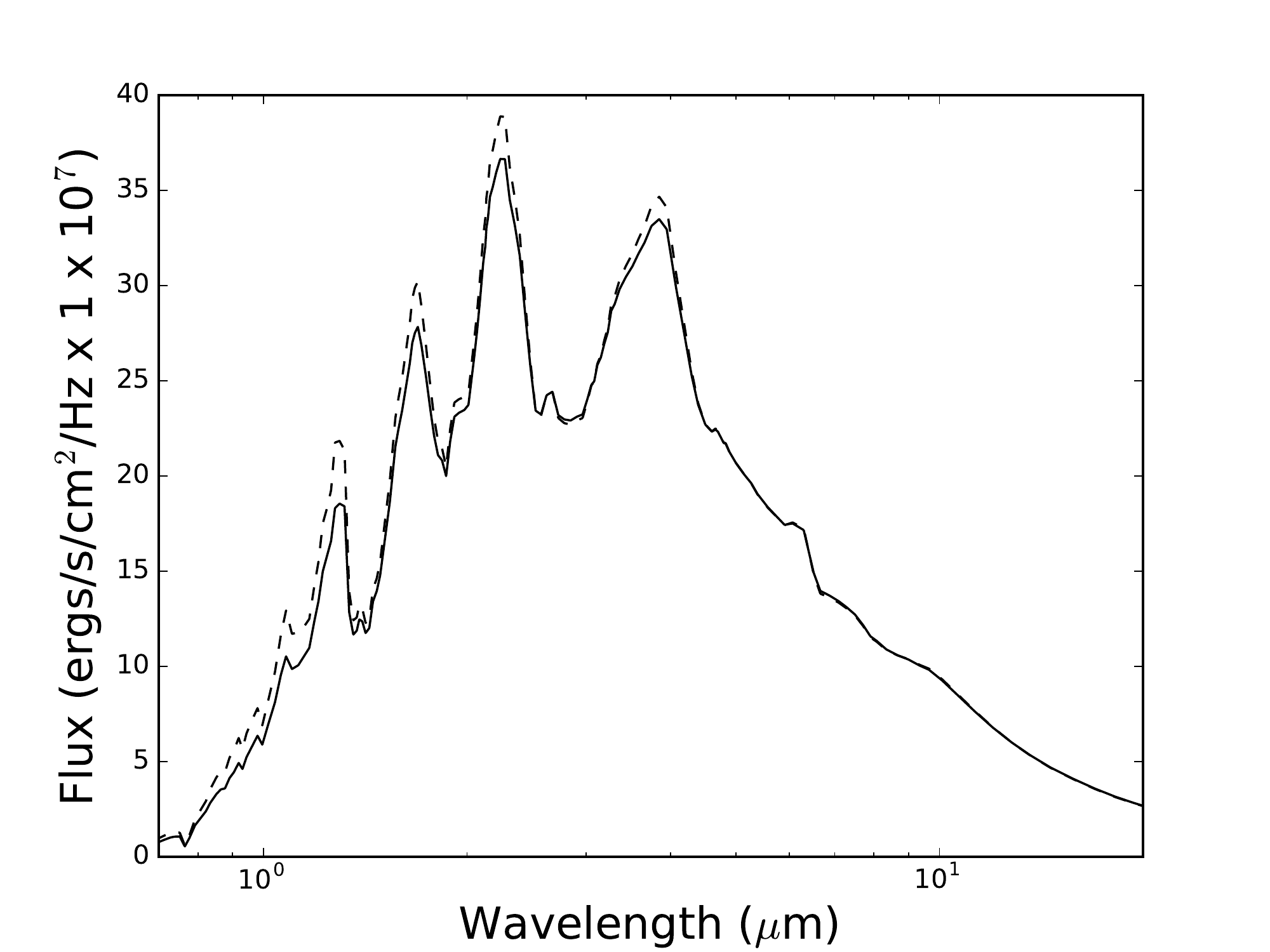}%
}
{%
  \includegraphics[clip,width=0.75\columnwidth]{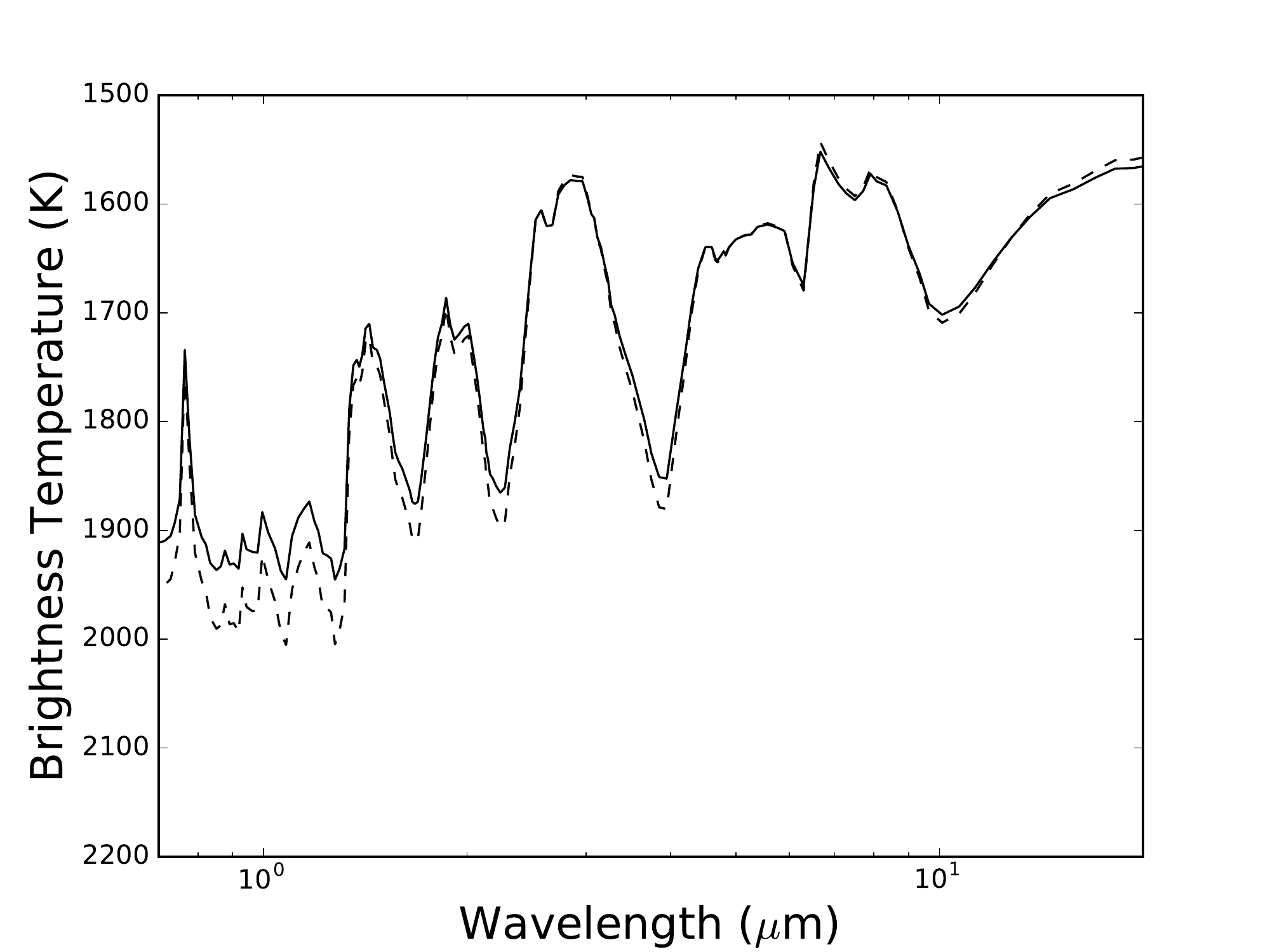}%
}
\caption{The effect of parent star spectral type on two hot Jupiter atmosphere models at \teff\ values of 1000 K and 1800 K.  The solid line is a main sequence K2 star, while the dashed line is G0.  The cool star puts out more flux in the infrared, which can be absorbed higher in the atmosphere by strong infrared bands leading to a slightly shallower \emph{P-T} profile, which mutes the spectrum.  The spectrum is only shown for the hotter model.\label{star}}
\end{figure}

Here in Figure \ref{star} we look at just a subset of models, for a 6000 K type G0 and 5000 K type K2 parent star.  These models are placed at distances such that the planet \teff\ are the same, yielding planetary \teff\ values of 1000 K and 1800 K.  The cooler parent star spectrum (solid curves), peaking at longer wavelengths, allows more incident energy to be absorbed higher in the atmosphere by the water bands (see Figure \ref{fluxin}), which warms the upper atmosphere and cools the lower atmosphere, relative to the hotter parent star (dotted curves).  The spectra from the more isothermal atmosphere are, as expected, slightly more muted, since the spectrum of a truly isothermal atmosphere would appear as a blackbody.

\subsection{Inner Boundary Condition:  Flux From the Interior}
Typically, one worries little about the inner boundary condition for a strongly irradiated planet.  Given the three temperatures discussed above, $T_{\rm eff}^4 = T_{\rm eq}^4 + T_{\rm int}^4$, the planetary \teff\ is typically dominated by absorbed stellar flux, with \tint\ contributing little to the planetary energy budget.  As one moves further from the parent star, or to younger planets which have interiors that have not yet cooled with age, the flux from the interior can matter considerably.  Obviously, for isolated brown dwarfs or young gas giants on wide separation orbits, this intrinsic flux is essentially all the flux that we see.

The inner boundary for strongly irradiated gas giants may be important in energy balance in some circumstances, however (Fortney et al., in prep).  As shown in \ct{Morley17a}, the emission spectrum of prototype warm Neptune GJ 436b is best matched by a model that has a high \tint\ value of $\sim$~350 K, much higher than the $\sim$~50 K one would expect from a Neptune-like evolution model \cp[e.g.][]{Fortney11}.  \ct{Morley17a} suggest the large intrinsic flux is due to ongoing tidal dissipation, as the planet is currently on an eccentric orbit.  With the aid of a tidal model these authors constrain the tidal $Q$ of the planet.  This was an interesting planetary science exercise where the planet's emission spectrum was tied to its interior structure and orbital evolution!
%Thorngren17 isn't in your references.bib file so the ct call is failing below.

\begin{figure}[htp]  
{%
  \includegraphics[clip,width=0.75\columnwidth]{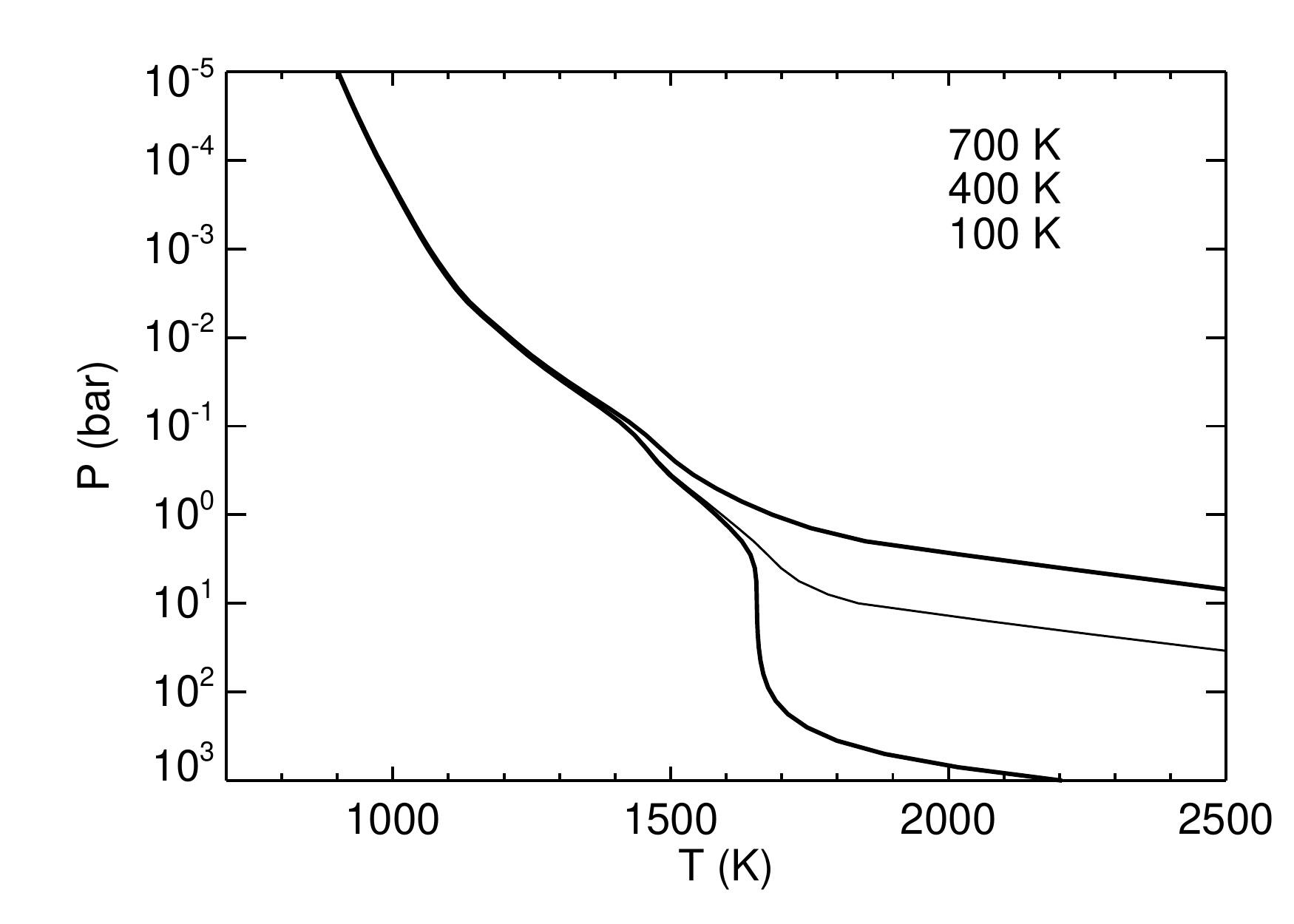}%
}
{%
  \includegraphics[clip,width=0.75\columnwidth]{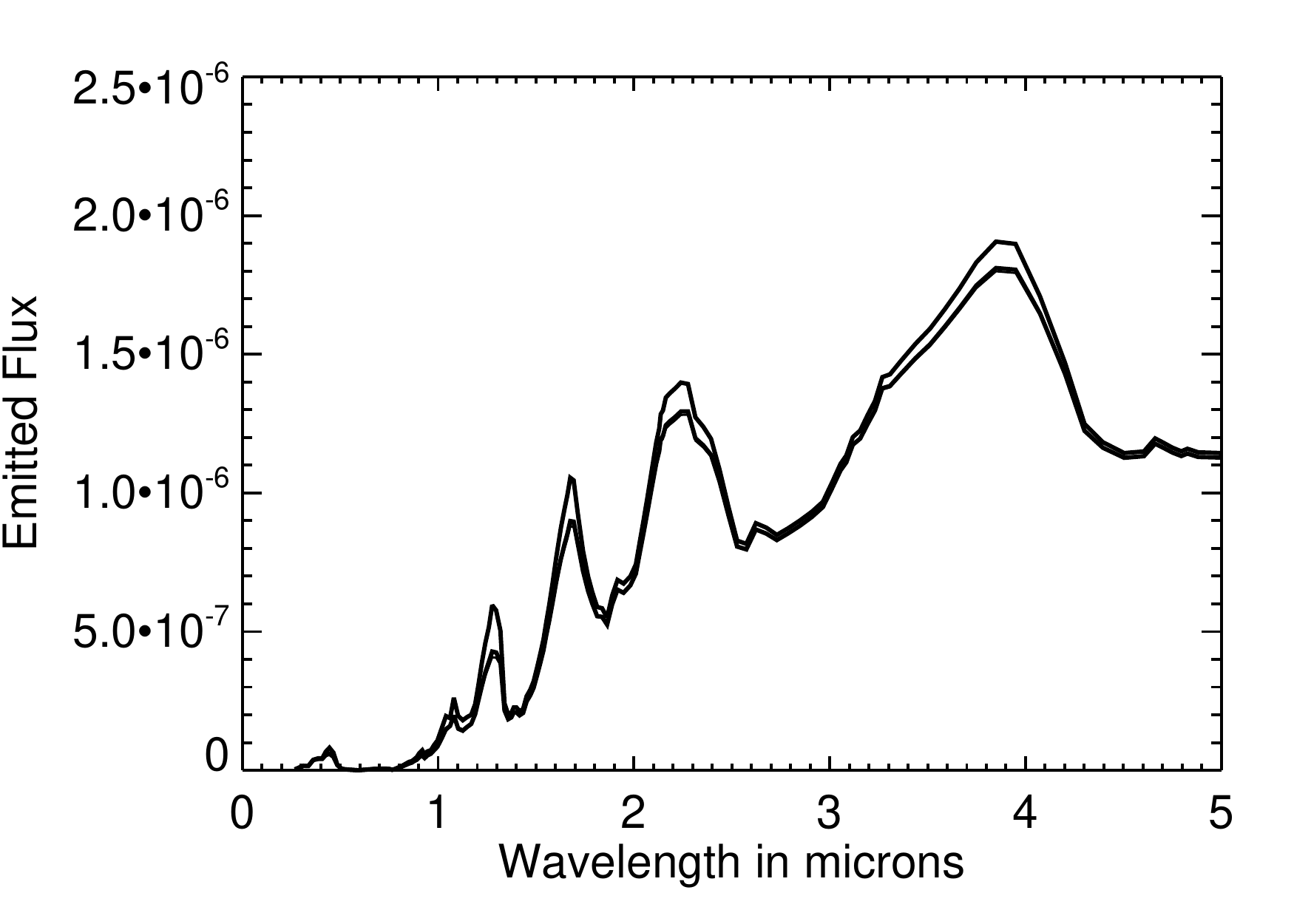}%
}
\vspace*{-5mm}
\caption{The effect of a hotter and cooler inner boundary on the \emph{T--P} profile and spectrum of a Saturn-like (gravity = 10 m s$^{-2}$) planet at 0.04 au from the Sun.  Values of \tint\ are 700, 400, and 100 K, with higher values leader to a hotter deep atmosphere and shallower radiative-convective boundary.  A hot inner boundary could be due to youth or additional energy sources in the planetary interiors.  The spectra from the 400 K and 100 K models are indistinguishable.  The differences in spectra are modest but likely detectable with \emph{JWST}.\label{inner}}
\end{figure}

Another case where the inner boundary may matter is for the largest-radius (``most inflated'') hot Jupiters.  This is because large radii imply hot interiors, which implies high fluxes from the interior \cp[e.g.][]{Fortney07a}.  \ct{Thorngren17} recently suggested that some hot Jupiter's may have \tint\ values of $\sim$~700 K, far in excess of Jupiter's value of 100 K.  Figure \ref{inner} shows an example calculation for a hot Jupiter at 0.03 au from the Sun, with inner boundaries \tint\ values from 100 to 700 K.  The flux enhancements are most prominent in the JHK near-infrared windows, which probe deepest in the planetary atmosphere.  Although small, these altered fluxes are likely detectable with \emph{JWST}.

\subsection{Role of Atmospheric Thickness}
For gas giant planets the atmospheric thickness takes up most of the radius of the planet.  Even for a sub-Neptune, we have little hope of seeing the bottom of the atmosphere.  For example, \ct{Lopez14} point out that for a 5 \me\ rocky planet with only 0.5\% of its mass in H/He (which would yield a radius of 2 \re), the surface pressure would be 20 kbar.

However, for terrestrial planets the atmospheric thickness is tremendously important, as it helps to set the surface pressure.  Surface temperature tends to scale with surface pressure as well, due to the greenhouse effect.  Our nearby example is Venus, which actually has a \emph{lower} \teq\ than the Earth, due to Venus's high Bond albedo. Venus's extremely high surface temperature is mostly due to its atmospheric pressure, which is 90 times larger than the Earth's.  Determining the surface pressure is not necessarily straightforward.  Perhaps the most straightforward way is if one could detect signatures from the ground in the planetary spectrum, in wavelengths where the atmospheric opacity is low, such that it could be optically thin to ground level.

\begin{figure}[htp]  
{%
  \includegraphics[clip,width=0.85\columnwidth]{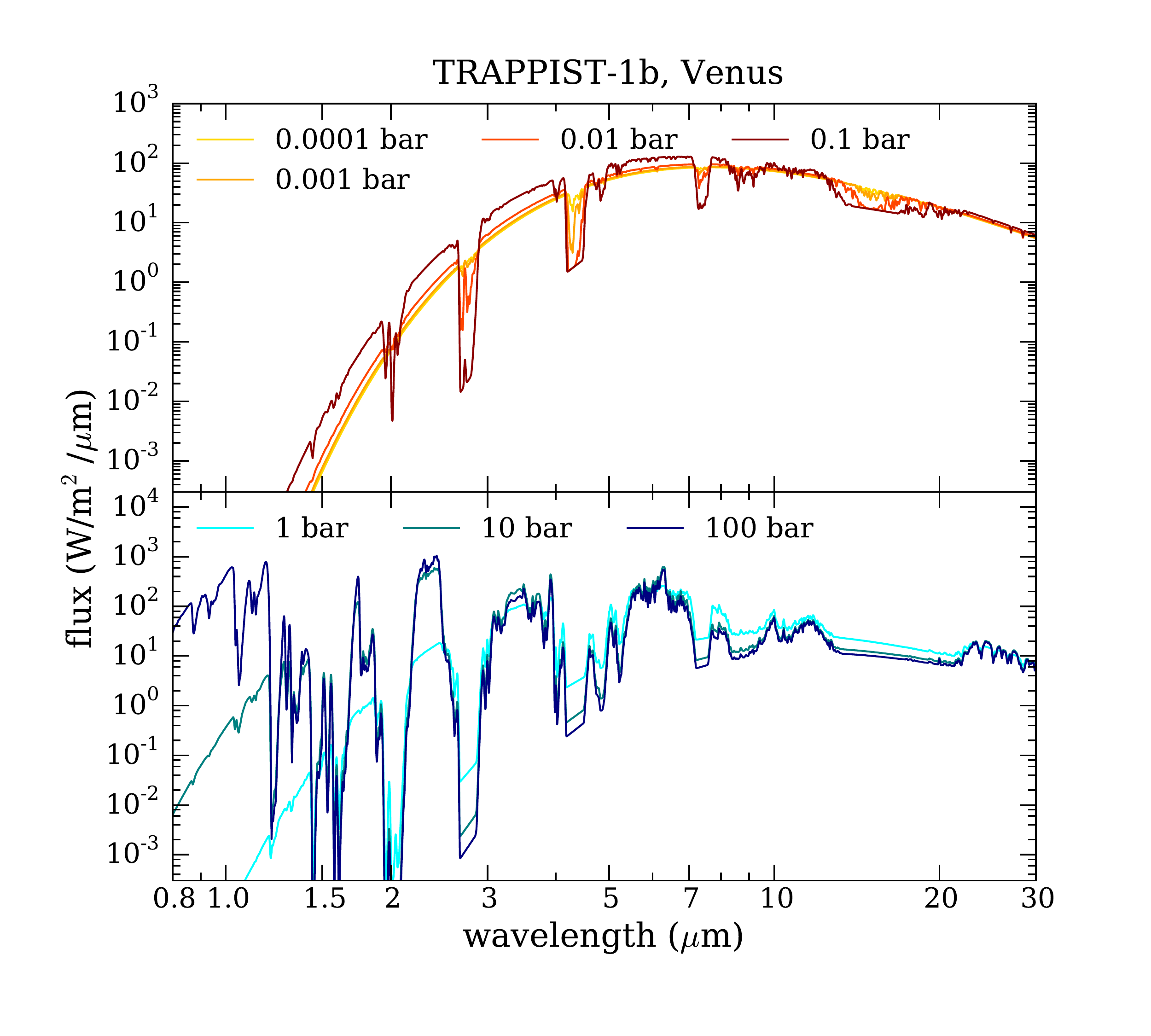}%
}
\vspace*{-5mm}
{%
  \includegraphics[clip,width=0.85\columnwidth]{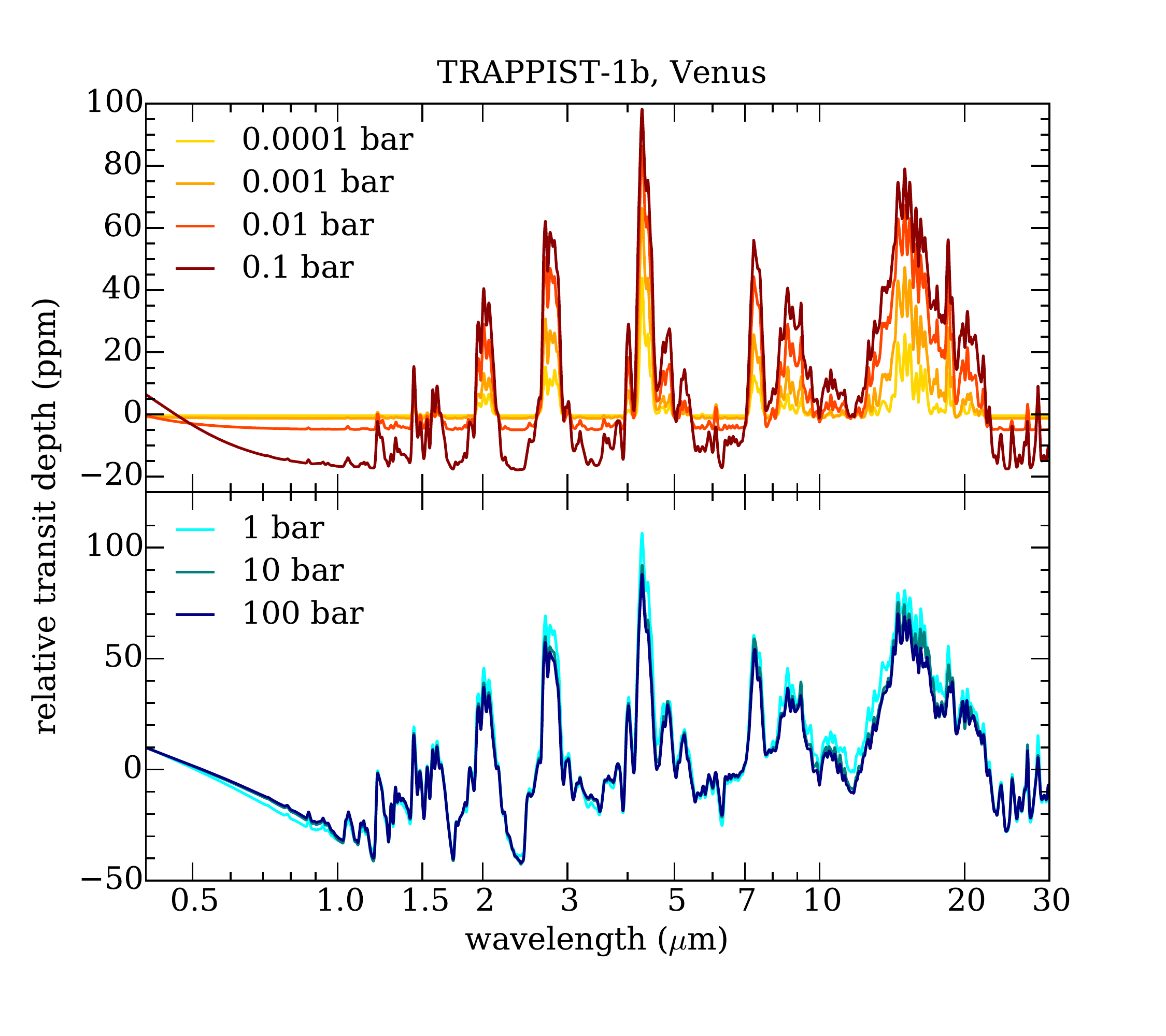}%
}
\vspace*{-2mm}
\caption{Emission (top) and transmission (bottom) spectra of Venus-like models of planet TRAPPIST-1b, courtesy of Caroline Morley, from \ct{Morley17b}.  Colors denote the surface pressure of the models.  These models have an adiabatic \emph{P--T} profile with depth that terminates at the planetary surface.\label{trappist}}
\end{figure}

Figure \ref{trappist} shows models from \ct{Morley17b} that examine the spectra of planet TRAPPIST-1b, the innermost planet in the TRAPPIST-1 system, around a very late M dwarf.  These plots examine the role of surface pressure on emission and transmission, with pressures from $10^{-4}$ to $10^2$ bar.  It should be noted that these models are cloud-free. In emission, in the top panel, the atmosphere is everywhere optically thin enough to see emission from the blackbody surface.  In transmission, which has a much longer atmospheric path length \cp[e.g.][]{Fortney05c}, one can no longer see the surface for pressures above $10^{-3}$ bars.  An interesting dichotomy between these plots is that for a wide range of thick atmospheres the transmission spectra are the same.  However, the emission spectra differ substantially.  This is because for transmission spectra, the atmosphere is basically a passive absorber of the stellar flux.  In emission, flux originates from a diverse range of pressure levels, meaning that the atmospheres appear distinct until the pressure is high enough that the atmosphere is optically thick at all pressures, which is nearly seen in the emission spectra models at 10 and 100 bars.

\subsection{Effects of Clouds}
The phase change of molecules from the gas phase to the liquid or solid phase is an unavoidable consequence of lower temperatures.  Within the solar system, clouds are ubiquitous in every planetary atmosphere.  As shown in Figure \ref{ptgrid1}, a wide range of species condense to form solids and liquids in H/He dominated atmospheres.  While sometimes known as ``dust'' in the astrophysical literature, here we will reference these condensates as ``clouds."  At the highest temperatures, the most refractory species such as Al$_2$O$_3$ (corundum) and Ca-Ti-O bearing species will condense.  Next are silicates (MgSiO$_3$ or Mg$_2$SiO$_4$) and iron.  A variety of sulfide species condense at moderate temperatures, later followed by water (H$_2$O) and ammonia (NH$_3$) at the coolest temperatures.  For reference, the cloud layers in Figure \ref{ptgrid1} should all occur in Jupiter at great depth, as one could mentally extrapolate the deep atmosphere's adiabat to higher pressures, past $\sim$~1 kbar or higher.

Many of the \emph{P--T} curves that indicate condensation can be readily derived from the Claussius-Clapeyron relation, as discussed in \ct{Seager10}.  A detailed look at the chemistry of condensation across $\sim$~500 to 1500 K can be found in \ct{Morley12} and \ct{Morley13}, with applications to cool brown dwarfs and warm sub-Neptune transiting exoplanets, respectively.  Cloud modeling is quite important because cloud opacity can be just as important as gaseous opacity.  However, while gaseous opacity can in principle readily be measured over a range of $P$ and $T$ via laboratory work, or via first-principles quantum mechanical simulations (see the contribution from Jonathan Tennyson, Chapter 3), cloud opacity is generally not amenable to this kind of analysis.

The complexity of clouds comes from a number of reasons.  Most importantly, there is a tremendous amount of complex and poorly understood \emph{microphysics}.  One must try to understand, at a given pressure level in the atmosphere, the mean or mode particle size, the (likely non-symmetric) distribution in sizes around this value, which could be bi-modal, how this mode and distribution change with height, the absorption and scattering properties, and scattering phase functions of these particles.  All of these quantities likely change with latitude and longitude, as Earth and Jupiter both have clear and cloud patches.  In addition, the coverage fraction of a given visible hemisphere likely changes with time.

There are several different ways investigators have aimed to understand the role of clouds in exoplanetary atmospheres.  A relatively simple cloud model is that of \ct{AM01} who aim to understand the 1D distribution of cloud particles by balancing the sedimentation of particles with the upward mixing of particles and condensible vapor.  All microphysics is essentially ignored and parameterized by a sedimentation efficiency parameter, $f_{\rm sed}$ (called $f_{\rm rain}$ in the original paper).  $f_{\rm sed}$ is tuned to fit observations of planets and brown dwarfs.  This methodology has been beneficial and has been applied across a wide range of \teff, from $\sim$~200 to 2500 K.  However, it lacks predictive power.  Another framework is that of Helling and collaborators \cp{Helling06,Helling08,Helling17} who use a sophisticated chemical network and follow the microphysics of tiny ``seed particles'' that fall down through the planetary atmosphere, from the atmosphere's lowest pressures.  Cloud particles sizes and distributions vs.~ height naturally emerge from these calculations, which typically yield ``dirty'' grains of mixed compositions.  The comparison of these models to observed planetary and brown dwarf spectra has not progressed quite as far at this time.

\begin{figure} 
\centering
\subfloat[]{
  \includegraphics[width=60mm]{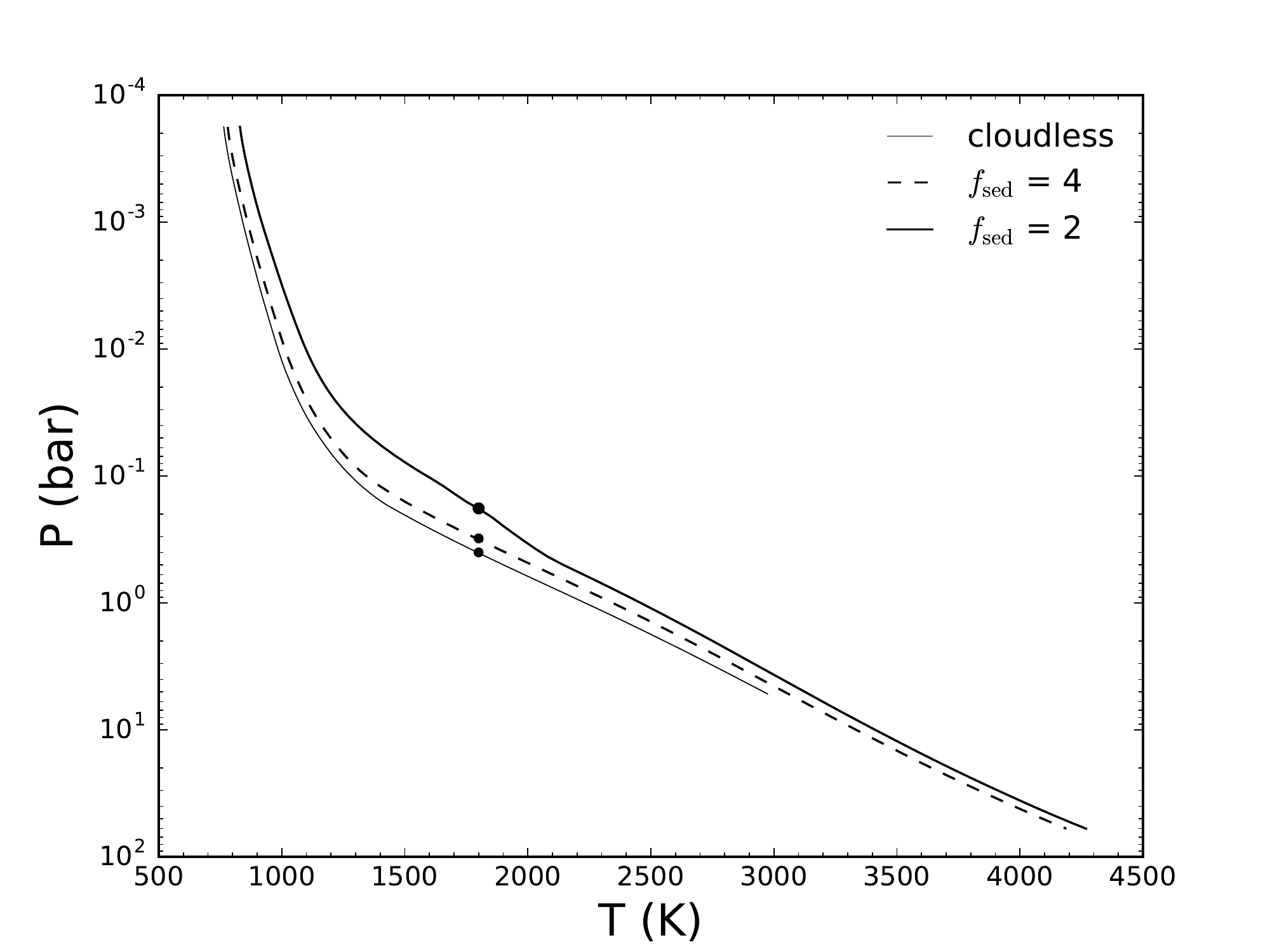}
}
\subfloat[]{
  \includegraphics[width=60mm]{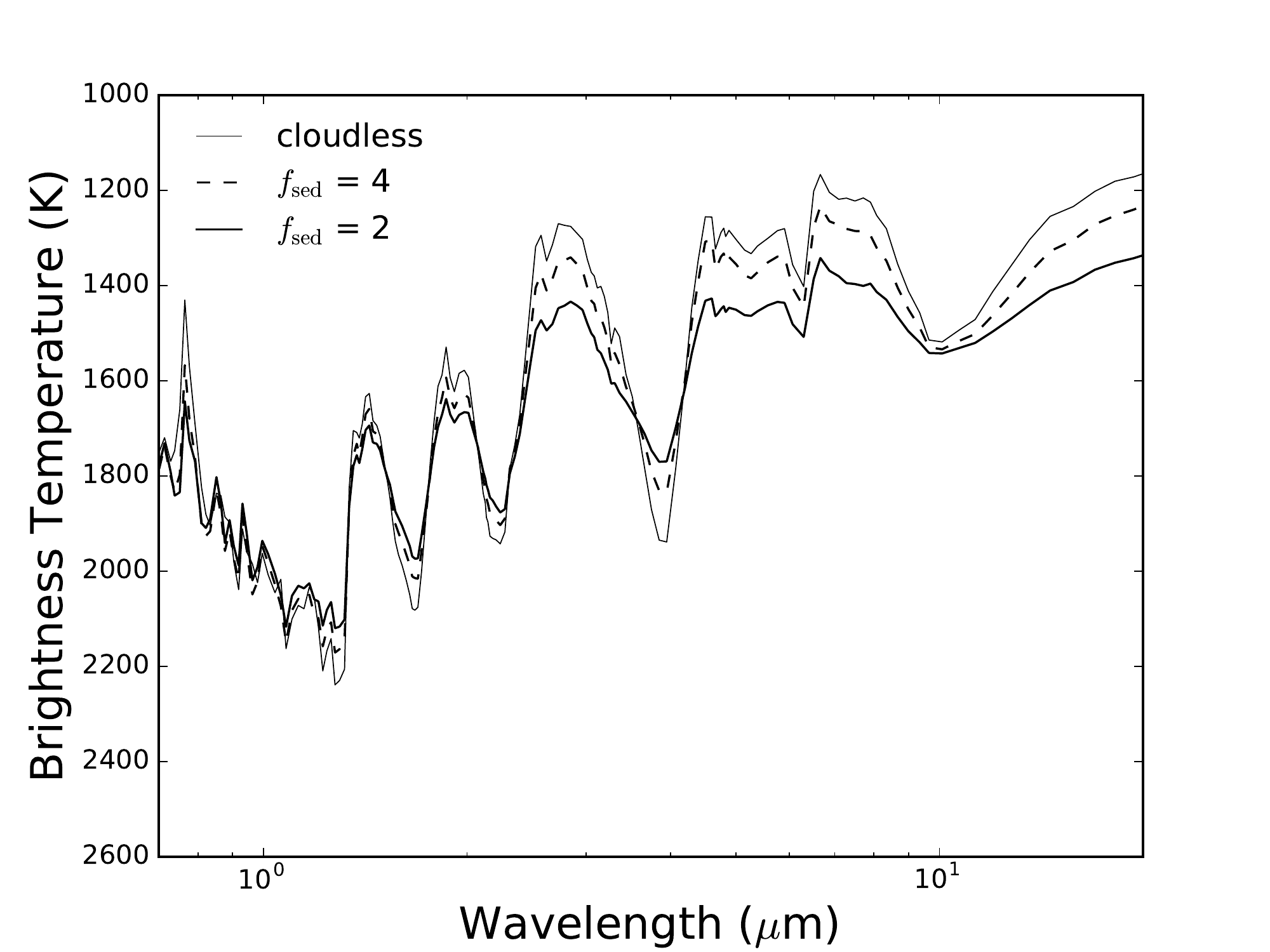}
}
\newline
\subfloat[]{
  \includegraphics[width=60mm]{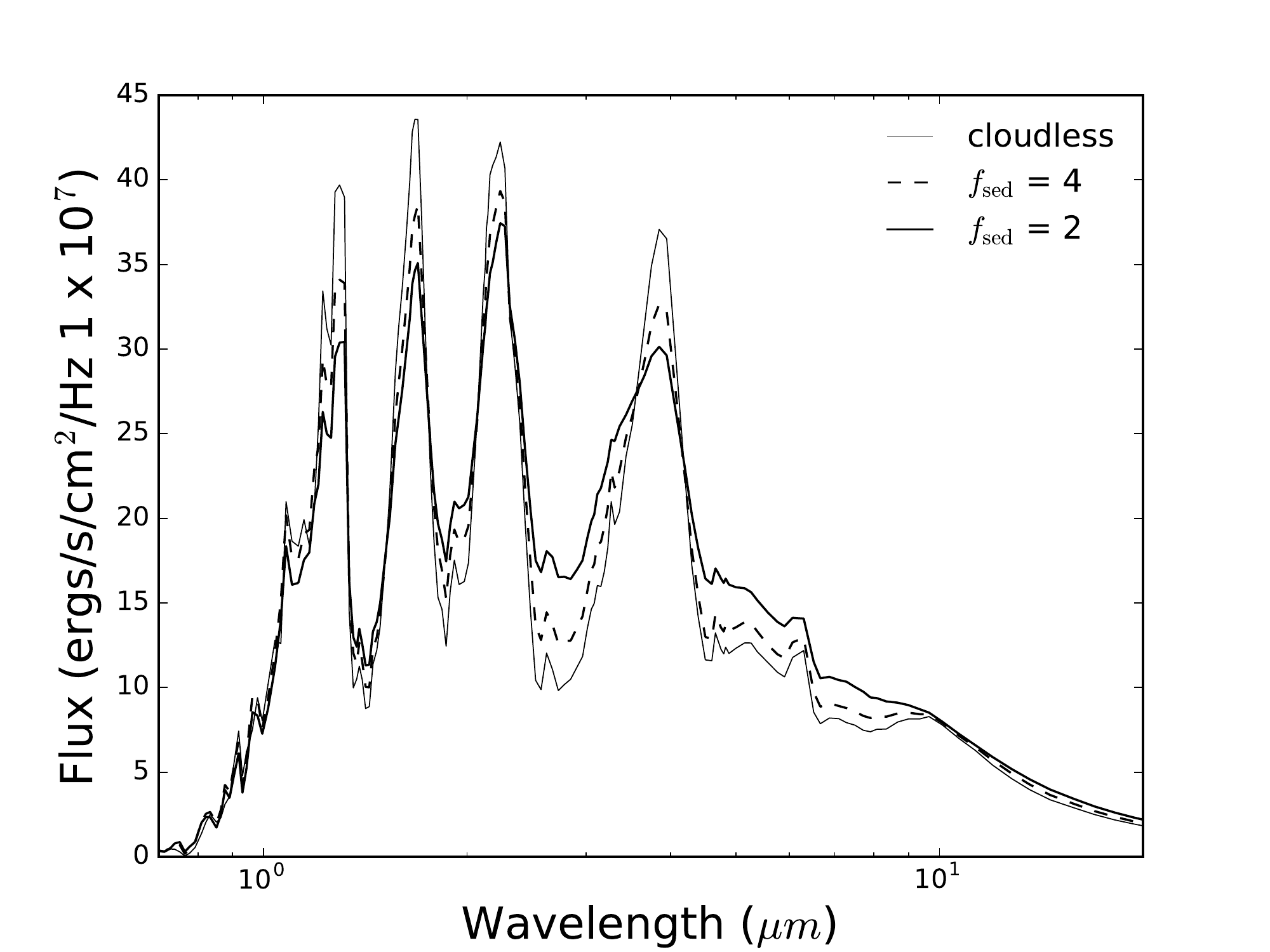}
}
\subfloat[]{
  \includegraphics[width=60mm]{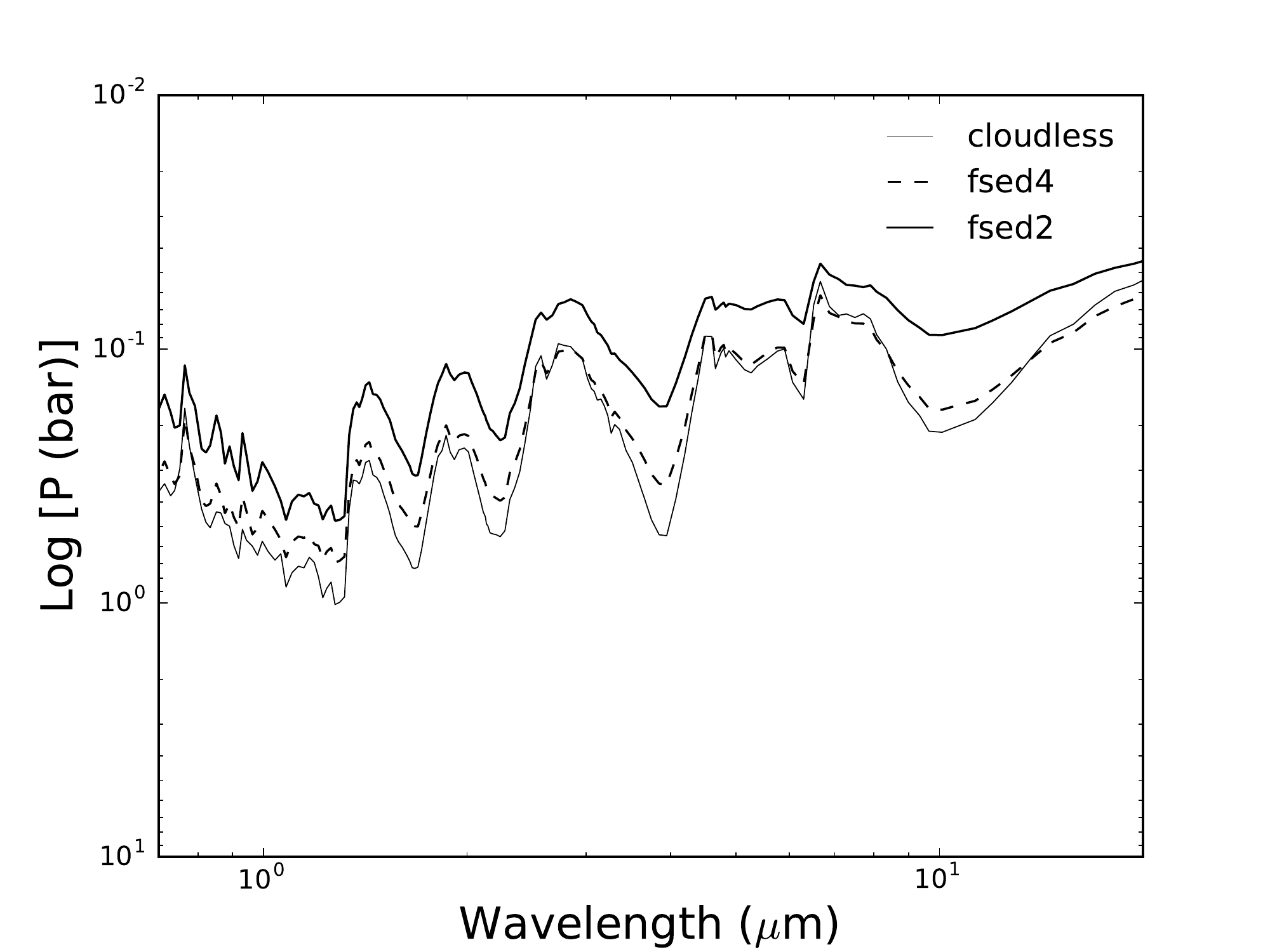}
}
\caption{The effect of clouds on a thermal emission spectrum of three models with the same \teff\ (1800 K) and gravity (log $g=4.0$).  Panel (A) shows the atmospheric \emph{T--P} profile, which is warmer with increased cloudy opacity, and has a lower pressure photosphere.  This thinner solid line is cloud-free.  Panel (B) shows the brightness temperatures one sees as a function of wavelength.  Note how the the effect of (gray) cloud opacity is to mute the ``hills and valleys," and limit the depth to high temperatures one might see.  This manifests in the spectrum in (C) where, with increasing cloud opacity, the spectrum becomes more muted and modestly more black-body like, and the JHK peaks are muted.  Panel (D) shows the pressures probed (the $\tau = 2/3$ level) at each wavelength.  Note the lower pressure photosphere as well as the smaller dynamic range in pressure as cloud thickness increases.\label{clouds}}
\end{figure}

The effects of clouds on the spectra of planets can be readily understood.  Cloud opacity is typically ``gray,'' meaning that there is little wavelength dependence to the opacity.  (While typical, this is not a rule.)  Compared to a cloud-free model, clouds are an additional opacity source, and limit the depth to which one can see.  Clouds typically then raise the planetary ``photosphere'' to lower pressures.  In Figure \ref{clouds} we can examine the effect of clouds on a 1400 K, log $g=4$ young giant planet.  These examples lack external irradiation.  These four panels show a cloud-free model (thin solid curve), an optically thinner cloud ($f_{\rm sed}=4$) cloud as a dashed curve, and a thicker cloud ($f_{\rm sed}=2$) in a thicker solid curve.

Figure \ref{clouds}\emph{A} shows the atmospheric \emph{P--T} profiles. The black dot shows where the local temperature matches the 1400 K \teff, which one can think of as the mean photosphere.  With larger cloud opacity, the \emph{P--T} profile is shifted to lower pressures at a given temperature.  \ref{clouds}\emph{B} shows the resulting brightness temperatures.  In the near infrared the clouds limit our ability to see deeply down in the JHK near-infrared opacity windows.  The inclusion of the gray opacity source limits the dynamic range in temperatures that are seen in the cloud-free model.  For each profile, the pressure that corresponds to each value of the brightness temperature, where the optical depth out of the atmosphere is 2/3, is shown just below in \ref{clouds}\emph{D}.  The higher pressures probed in the cloud-free model, and the higher dynamic range of pressures probed, is clearly seen.  The resulting emergent spectra are shown in \ref{clouds}\emph{C}.  Although all spectra are nothing like a blackbody, the cloudiest models have the most muted absorption features.  The J-band at $\sim$~1.3 $\mu$m, where water opacity is at its minimum, and one would normally see deepest into the atmosphere, is the wavelength range that is most affected by the cloud opacity.

\begin{figure}[htp]
\includegraphics[clip,width=1.0\columnwidth]{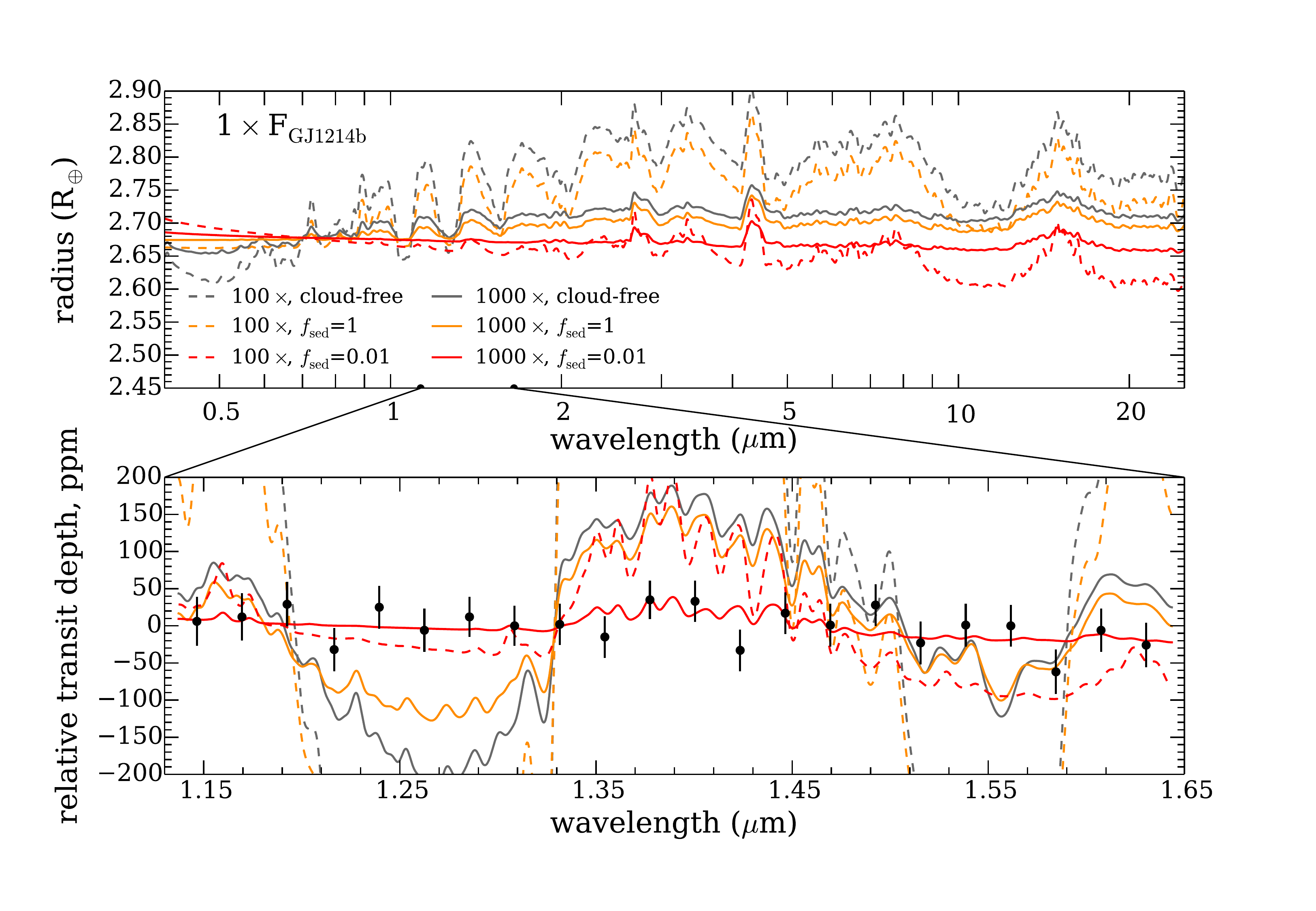}
\vspace*{-8mm}
\caption{Example transit transmission spectra of planet GJ 1214b at high metallicity, with and without clouds.  Models are at metallicity values of 100 and 1000$\times$ solar.  The top panel shows the optical and infrared transmission
spectra. The bottom panel shows the same spectra, zoomed in to focus on the featureless \ct{Kreidberg14a} data in the near-infrared from \emph{Hubble}. Cloud-free transmission spectra are
shown as dotted and solid gray lines and cloudy spectra are shown as colored lines. Note that the only model that fits the data is the 1000$\times$ solar model with $f_{\rm sed}= 0.01$ (very highly lofted) clouds.  Figure courtesy of Caroline Morley. \label{1214}}
\end{figure}

After the past 15 years of observing transit transmission spectra, we are fully aware of how clouds manifest themselves in transiting planet atmospheres.  Typically, weaker than expected absorption features are seen \cp[e.g.,][]{Sing16}.  There are numerous examples where cloud opacity blocks all molecular absorption features \cp{Kreidberg14a}.  An illustrative example of how clouds effect absorption features in transmission spectrum model is shown in Figure \ref{1214}.  These models already have enhanced metallicity atmospheres that drive up the mean molecular weight, $\mu$, which shrinks the scale height and size of absorption features.  The clouds provide additional (mostly gray) opacity that mutes the absorption features.

\section{Retrieval}
Over the past several years inverse methods have become an integral part of modeling exoplanetary atmospheres.  Within this framework one performs a wide exploration of a range of possible atmospheres models that can yield a best-fit to an observed spectrum.  Typically these methods, called ``retrieval,'' aim to find the combination of atmospheric abundances and atmospheric \emph{T--P} profile that yield a best fit to an observed spectrum.  Those models can find solutions outside the confines of self-consistent models.

\subsection{Forward Model}
The most important piece of any atmospheric retrieval algorithm is the \emph{forward model}.  The forward model takes a set of inputs, generally the parameters of interest, and then uses various physical assumptions to map these parameters onto the observable, e.g., the spectra. Depending on the situation of interest, there are three kinds of forward models one would implement, depending on if the observations were in thermal emission, transit transmission, or reflection.  First is an emission forward model, which computes the upwelling top-of-atmosphere flux, and would be used at secondary eclipse or for directly imaged planets.  Second is a transmission forward model, which computes the wavelength-dependent transit radius of the planet, and is used in defining the planet-to-star radius ratio.  Third is a reflection forward model, which sums the stellar flux scattered in any direction from an illuminated hemisphere, and would be used as a function of orbital phase for an imaged or a transiting planet phase curve.

These forward models differ in their geometry and in the atmospheric regions probed. The necessary inputs in all models are the temperatures at each atmospheric level--the thermal profile--and the abundance and type of opacity sources, whether they be molecular/atomic gases or cloud/grain opacities, gravity, host star properties, and basic instrumental parameters that convolve and bin the high-resolution model spectra to the data set in question.  These are the fundamental quantities that impact any emission, transmission, or reflection spectrum.

\subsection{Bayesian Estimator/Model Selection}
The Bayesian estimator is used to determine the allowed range of parameters (posterior), in the context of a given forward model, which can adequately describe the data.  Most investigators use a multi-pronged modeling approach \cp[e.g.,][]{Line13} that includes several MCMC samplers, including the powerful ensemble sampler EMCEE \cp{Foreman13}, implemented in \ct{Kreidberg15}, \ct{Greene16}, and multimodal nested sampling \cp{Feroz09}, PYMULTINEST \cp{Buchner14}, implemented in \ct{Line16L}. Using more than one inference method ensures robust results \cp[e.g.,][] {Lupu16}.  

An important aspect of Bayesian problems is that of model selection. Given two competing models, model selection aims to rigorously identify the simplest model that can adequately explain the data. This is done through the evaluation of the marginal likelihood, or Bayesian evidence \cp[e.g.,][]{Trotta08}. The validity and utility of these model selection-based approaches as applied to exoplanet atmospheres have been routinely demonstrated \cp{Benneke13, Kreidberg14b, Line16L}.  It is essential to use these evidence-based model selection methods to explore the hierarchy of model assumptions in order to diagnose the significance of the assumptions and/or missing model physics. 

\subsection{An Example: Cool T-Type Brown Dwarfs}
While exoplanet spectra typically have low signal-to-noise spectra and spectral coverage over a limited wavelength range, brown dwarfs have excellent spectra over a broad wavelength range, and have been studied in some detail since their first discovery in 1995.  An excellent review article on observations of these objects can be found in \ct{Kirkpatrick05}.  \ct{Line14,Line15,Line17} pioneered the use of retrieval methods for these objects.  T-type brown dwarfs are especially good targets because they typically lack the thick clouds in L-type brown dwarfs.  This means that their photospheres span a relatively large dynamic range in pressure, meaning that information on the atmosphere can be gleaned from many levels.

\begin{figure}[htp]  
{%
  \includegraphics[clip,width=1.0\columnwidth]{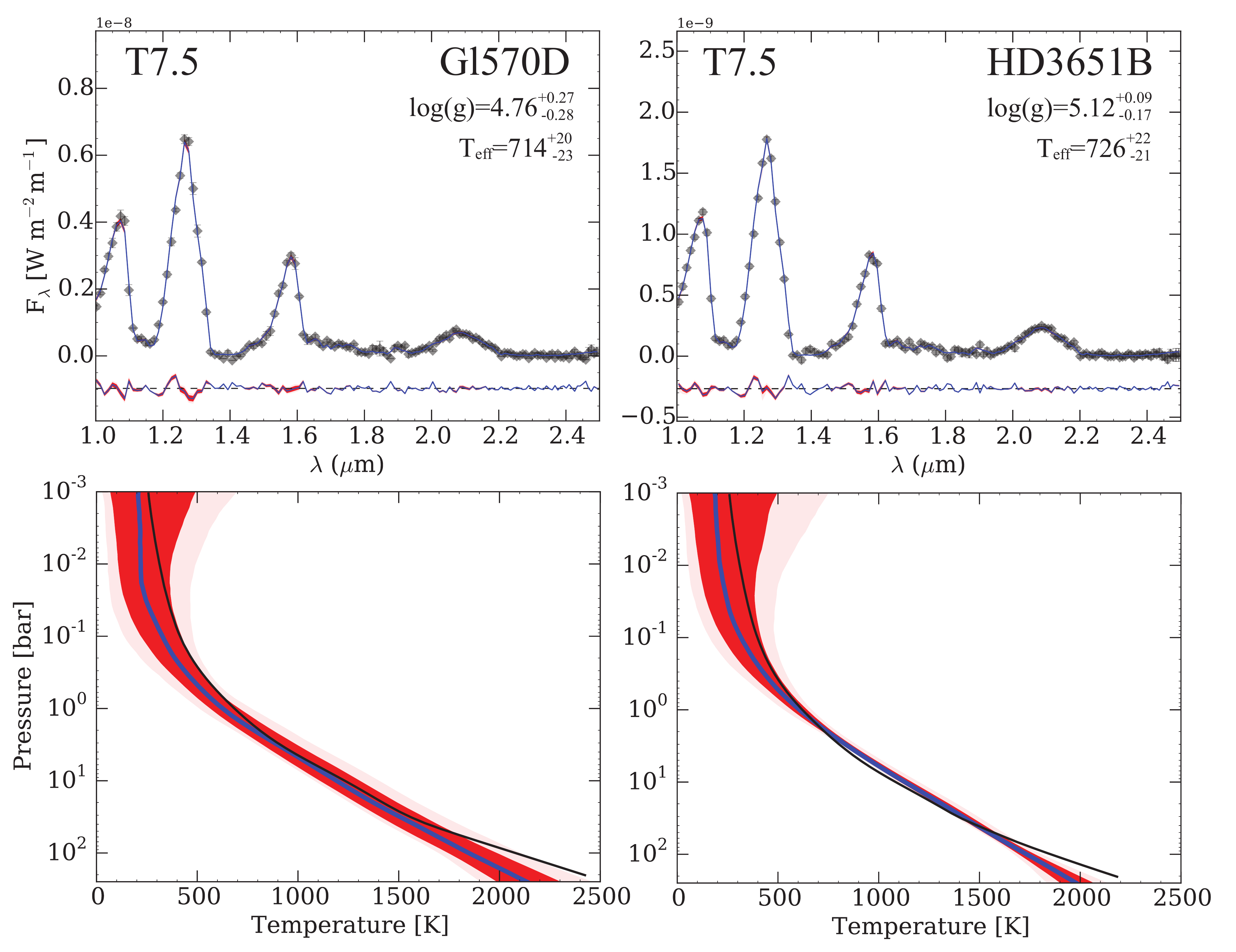}%
}
{%
  \includegraphics[clip,width=1.0\columnwidth]{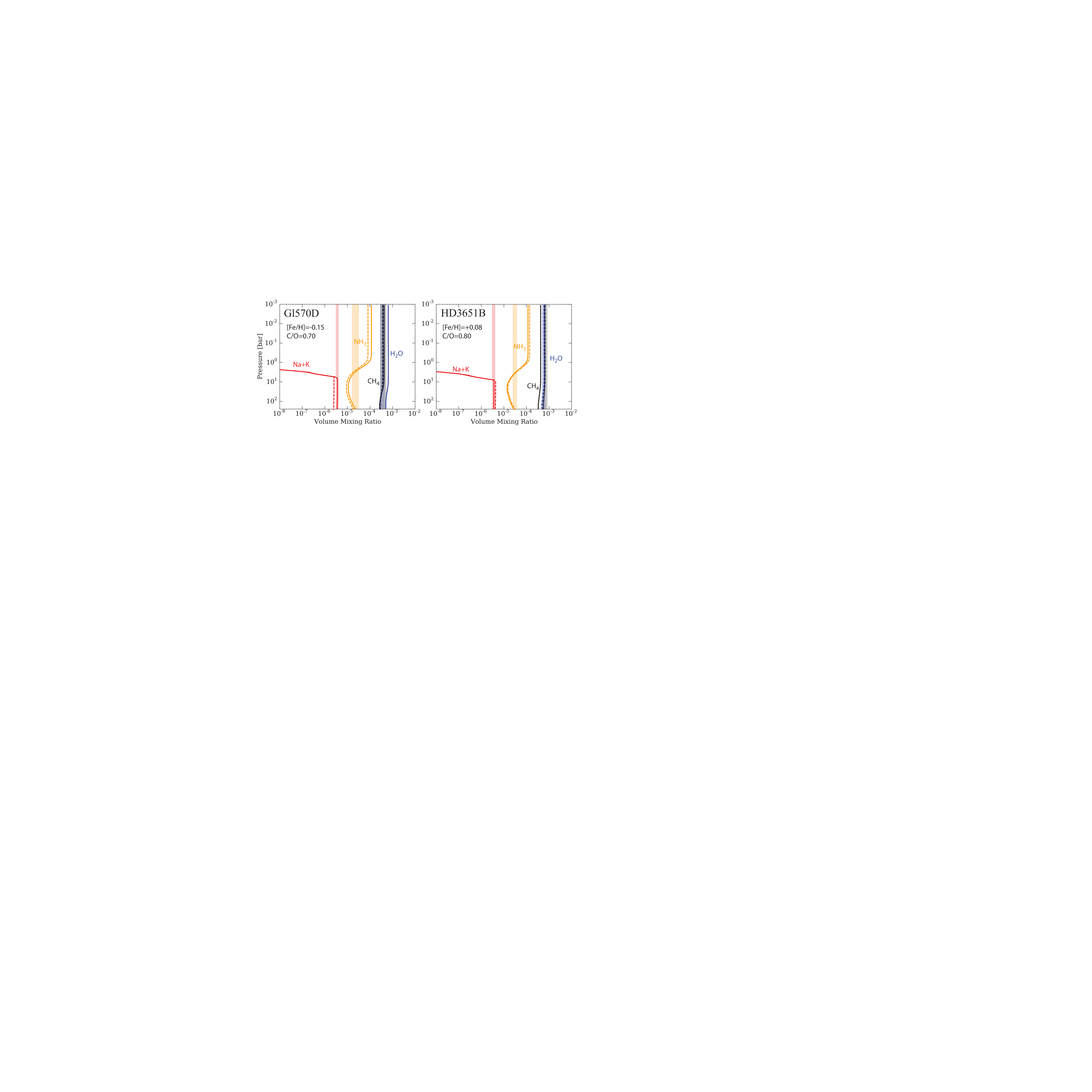}%
}
\caption{Adapted from \ct{Line15}, retrievals on two $\sim$700 K T-type brown dwarfs. Spectra (top row), retrieved temperature profiles (middle row), and retrieved atmospheric abundances (bottom row).  \emph{Top:} For the two objects we show the H-band calibrated SpeX spectra data as the diamonds with error bars, a summary of thousands of model spectra generated from the posterior and their residuals (median in blue, 1$\sigma$ spread in red), and their spectral type and bulk properties.  \emph{Middle:} This summarizes thousands of temperature profiles drawn from the posteriors for each object (median in blue, 1$\sigma$ spread in red, 2$\sigma$ spread in pink).  The black temperature profile shown for each object is a representative self-consistent grid model \cp{Saumon08} interpolated to the quoted $\log g$ and $T_{\rm eff}$ to demonstrate that the retrieved profiles are physical and are consistent with 1D radiative-convective equilibrium. \emph{Bottom:}  Comparison of the retrieved chemical abundances (shaded boxes) of the well constrained molecules with their expected thermochemical equilibrium abundances along the median temperature profile. The solid curves are the thermochemical equilibrium abundances for solar composition while the dashed curves are the thermochemical equilibrium abundances for the specified C/O and metallicity.  This shows that the retrieved abundances are thermochemically consistent. \label{lineBDs}}
\end{figure}

As a test-case in what we might expect for exoplanets in the future, when we have \emph{JWST}-quality (or better) data for cool atmospheres, \cite{Line15} studied two benchmark brown dwarfs, Gl 570D and HD 3651B, both of which are $\sim$~700 K brown dwarfs that orbit well-characterized Sunlike parent stars.  Figure \ref{lineBDs} shows the results of the retrievals.  The top panel shows the outstanding agreement between the model-derived spectra and the medium-resolution observations.  The work aimed to retrieve the atmospheric \emph{P--T} profile at 15 levels in the atmosphere.  The middle panel compares the derived profile to that of a best-fit radiative-convective self-consistent model atmosphere \cp{Saumon08}, and the agreement is striking.

Finally, the retrievals allow for the determination of chemical abundances, a first for a fit to brown dwarf spectra.  The bottom panel of Figure \ref{lineBDs} shows the derived abundances, within the 1$\sigma$ error bars, as shaded colors, with the expected chemical equilibrium abundances as solid and dashed curves.  The agreement is quite good.  (We note that the sharp drop in alkali metals Na and K is due to their condensation into solid cloud particles, and corresponding loss from the gas phase.)  The implied abundances for carbon and oxygen, and the C/O ratio, agree well with a detailed analysis of the a high-resolution spectrum of each object's parent star. Since the brown dwarf and the Sunlike star formed in a bound orbit, within the same giant molecular cloud, they should share the same abundances.  This gives confidence in the retrieval methodology.

\begin{figure}[htp]  
\includegraphics[clip,width=0.82\columnwidth]{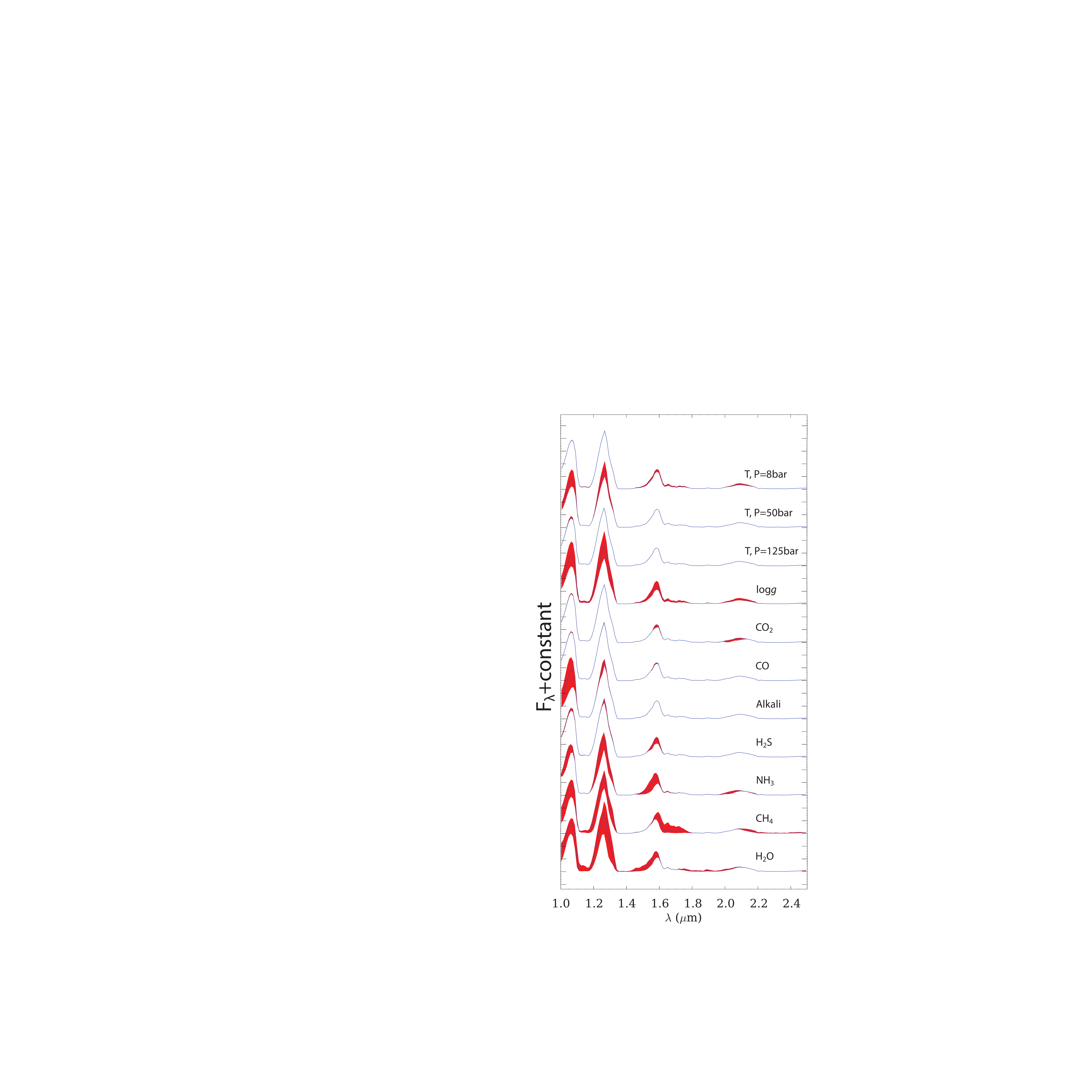}
\caption{Sensitivity of a cool brown dwarf spectrum to various parameters.  This is a synthetic spectrum with \teff\ of 700 K and a log $g$ of 5 and with purely thermochemical equilibrium composition.  The red regions represent a change in the spectrum due to a perturbation of each of the parameters.  For H$_2$O, CH$_4$,  NH$_3$, H$_2$S, alkali, this perturbation is $\pm$ 0.5 dex (where 1 dex is 1 order of magnitude) in number mixing ratio from the thermochemical abundance value. For CO the perturbation is +2 dex, CO$_2$, +6 dex, log$g$ $\pm$ 0.1 dex, and the temperatures are perturbed at each level by $\pm$50 K.   Figure courtesy of Michael Line.\label{sense}}
\end{figure}

Another nice feature that one can derive from retrieval models is the degree to which changes in chemical abundances, the temperature profile, or surface gravity affect the spectra at particular wavelengths.  This can give one an intuitive feel for which wavelengths are most sensitive to particular aspects of the model, and why some features are essentially insensitive to the spectrum itself.  This is nicely displayed in Figure \ref{sense} for a generic 700 K brown dwarf model.  One can first look at the three pressures, 8, 50, and 125 bars.  The spectrum gives very little sensitivity to the 8-bar temperature, because the atmosphere is optically thin at most wavelengths at this pressure.  However, at 50 bars, we are typically seeing down into the Y (1.0 $\mu$m), J (1.2 $\mu$m), and H (1.6 $\mu$m) band windows, so the spectrum gives us great leverage on the temperature there.  By 125 bars the atmosphere is nearly opaque at all wavelengths, so we have very little leverage on the temperature at such a high pressure.

For the atmospheric surface gravity all wavelengths contribute to our understanding, as this is a constant in the atmosphere.  For the molecular abundances, as one might expect, there is only ``power'' at wavelengths where the particular molecules are good absorbers \emph{and} if these molecules are abundant enough to create any spectral features.  At such a cool temperature (700 K), CO and CO$_2$ have very low mixing ratios, so we do not see them, and they have little impact on the spectrum.  The pressure-broadened alkali metals, Na and K, impact the spectrum via their pressure-broadened red wing that overlaps with the Y and J bands.  H$_2$S has little abundance so it impacts the spectrum modestly.  NH$_3$, CH$_4$, and H$_2$O all have multiple molecular bands throughout the near-infrared, so we should expect to be able to determine their abundances relatively robustly, in agreement with the bottom panel of Figure \ref{lineBDs}.

\section{Simplified Atmospheric Dynamics}  \label{dynamics}
There are a number of excellent reviews of atmospheric dynamics in the exoplanet context, including \ct{Showman08b} and \ct{Heng15} on hot Jupiters and \ct{Showman13} on terrestrial planets.  This is a huge subject and the interested reader can find a robust literature that connects the dynamics of solar system rocky planets to solar system gas giants to exoplanets.  Here we will suffice to discuss a few relevant timescales that help us understand the detected phase curves of (likely) tidally locked hot Jupiters, which are the planets whose atmospheres have been probed to understand dynamics.

\begin{figure}[htp]
\includegraphics[clip,width=1.0\columnwidth]{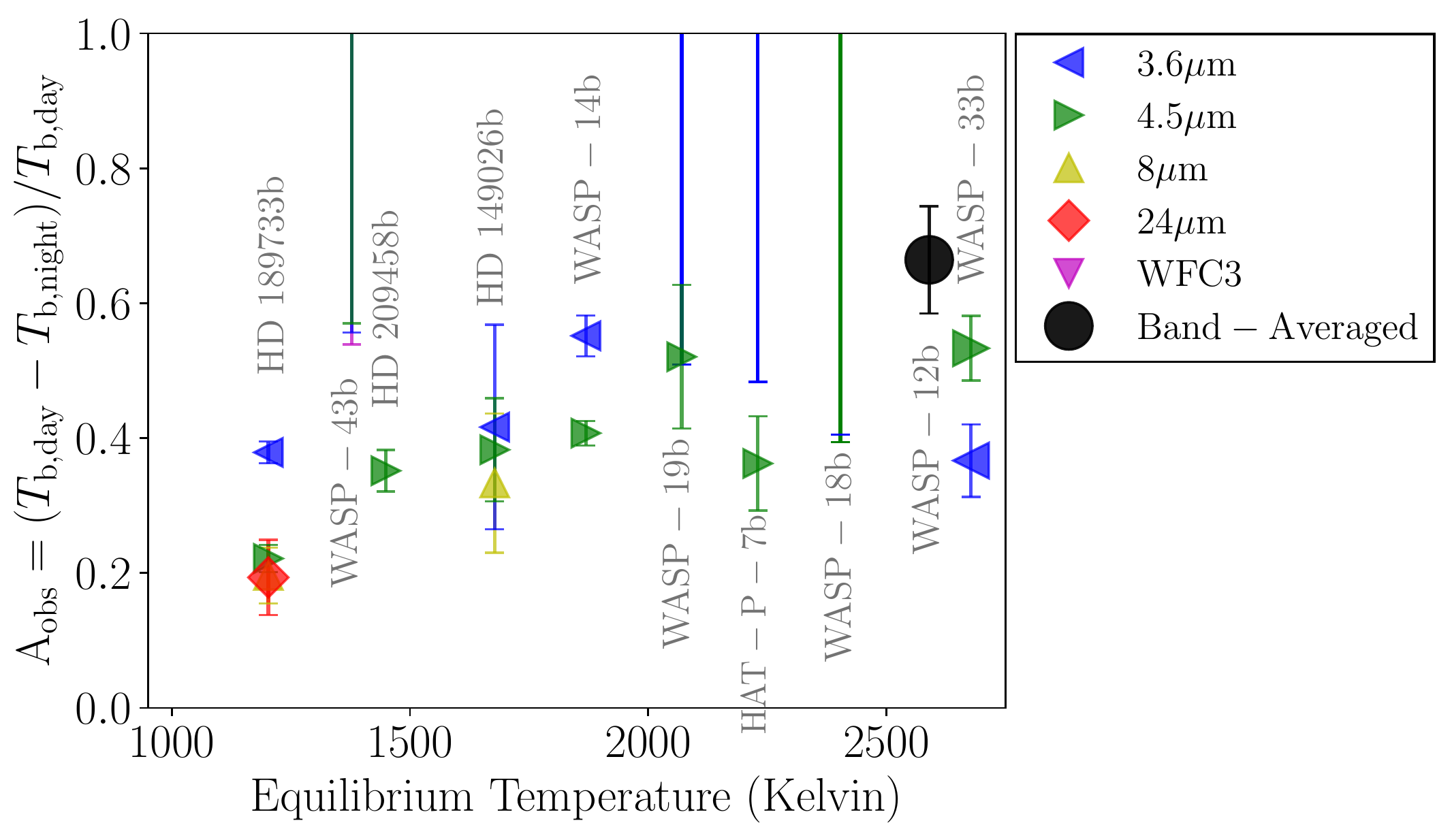}
\caption{This is a compilation of phase curve data, mostly from \emph{Spitzer}, that shows the day/night temperature contrast for transiting giant planets.  The y-axis is a measure of day/night temperature homogenization, with more homogenized planets plotting lower.  There is a weak general trend that the hottest planets have the largest day/night contrast, although these inferences depend on the wavelength observed (as they probe different depths in the atmosphere).  Figure courtesy of Tad Komacek.}

\end{figure}

Within the realm of hot Jupiters, state-of-the-art three dimensional circulation models include a treatment of the Navier-Stokes equations (or a simplified version of them that assume hydrostatic equilibrium, called the Primitive Equations), for the fluid dynamics.  This is combined with a treatment of the radiative transport in the atmosphere, which can be fully wavelength-dependent (termed ``non-gray'' in astronomical jargon) across a wide wavelength range from the blue optical to the far infrared, or can also averaged over in terms of weighted mean wavelength-independent visible (``shortwave'') and corresponding thermal (``longwave'') fluxes.  Such ``grey'' simplifications speed up codes dramatically, but have significant drawbacks when comparing to wavelength-dependent thermal infrared phase curves, as can be obtained from \emph{Spitzer} or \emph{Hubble} \cp{Knutson07b,Knutson09,Stevenson14}.

State of the art models including non-gray radiative transfer are described in \ct{Showman09,Lewis14,Mayne14}.  However, one should be very clear that \emph{all} levels of model sophistication are important within dynamics.  Within the dynamics literature one often finds a discussion of a ``hierarchy of models,'' from simple 1-layer models in 2D, to simplified 3D models, to 3D models with full radiative transfer.  It is only through an understanding of the physical mechanisms across multiple levels of complexity that one can begin to understand the diverse physical behavior of atmospheres.

As a basic introduction, one often discusses relevant physical timescales \cp{Showman02}.  These include the advective timescale, the timescale over which a parcel of gas is moved within an atmosphere.  This quantity is characterized by the atmospheric wind speed and a relevant planetary distance, on the order of a planetary radius.  Thus, the advective time is:
\begin{equation}
\tau_{\mathrm{advec}} = \frac{R_{\mathrm P}}{U} 
\end{equation}
where $R_{\mathrm P}$ is the planet radius and $U$ is the wind speed.  This can be compared to a radiative timescale, the time it takes for a parcel of gas to cool off via radiation to space: \begin{equation}
\tau_{\mathrm{rad}} = \frac{P}{g}\frac{c_{\mathrm P}}{4 \sigma T^3}.
\end{equation}
 
If $\tau_{\mathrm{advec}} << \tau_{\mathrm{rad}}$, we would expect temperature homogenization around the planet.  This is the case for Jupiter, as $\tau_{\mathrm{rad}}$ is quite large, owing to the cold atmospheric temperatures.  If $\tau_{\mathrm{advec}}>>\tau_{\mathrm{rad}}$, then we should expect large temperature contrasts on the planet.  All things being equal, for even hotter planets, $\tau_{\mathrm{rad}}$ will become smaller, given the strong temperature dependence.  With a relatively constant wind speed, this would imply larger temperature contrasts.  This could manifest itself in large day-night temperature differences observed for the hot Jupiters, which are expected to be tidally locked.  This is indeed what is observed.  In addition, other physical forces, such a Lorentz drag due to the thermal ionization of alkali metals, can slow advection as well \cp{Perna10}, also aiding large temperature differences.

\section{Connection With Formation Models}  \label{formation}
The atmospheres of all four of the solar system‚ giant planets are enhanced in ``metals'' (elements heavier than helium) compared to the Sun.  Spectroscopy of these atmospheres yields carbon abundances (via methane, CH$_4$) that show an enrichment of $\sim$~4, 10, 80, and 80, for Jupiter, Saturn, Uranus, and Neptune, respectively.  Spectra that determine other atmospheric abundances are challenging due to the very cold temperature of these atmospheres, which sequester most molecules in clouds far below the visible atmosphere.  The 1995 \emph{Galileo Entry Probe} into Jupiter measured in situ enrichments from factors of 2-5, for light elements (C, N, S, P) and the noble gases \cp[e.g.][]{Wong04}.  Taken as a whole, these measurements suggest that giant planet atmospheres are enhanced in metals compared to parent star values.

Before the dawn of exoplanetary science, these atmospheric metal enrichments were understood within the standard ``core-accretion'' model of giant planet formation \cp[e.g.][]{Pollack96}.  Within this theory, after a solid core of ice and rock attains a size of $\sim$~10 Earth masses, this core accretes massive amounts of H/He-dominated gas from the solar nebula, which can be several Earth masses to many hundreds of Earth masses.  This accretion also includes solid planetesimals that are abundant within the midplane of the solar nebula disk.  The accretion of solids and gas leads to an H/He envelope enriched in metals compared to parent star abundances.  Since giant planet H/He envelopes are mostly or fully convective, the metal enrichment of the envelope will include the visible atmosphere.

This planet formation framework was built upon a sample size of only 4 planets.  The promise of exoplanetary science is to understand metal enrichment, and its relevance to planet formation, over a vastly larger sample size. This planet \emph{mass-metallicity relation} needs to be understood in terms of the enrichment as a function of planet mass, but also the intrinsic dispersion in the enrichment at a given mass, since it appears that exoplanet populations are quite diverse \cp{Thorngren16}.   The metal enrichment observed today drives our understanding of the accretion of gas and solids in the planet formation era.

\begin{figure}[htp] 
\includegraphics[clip,width=1.0\columnwidth]{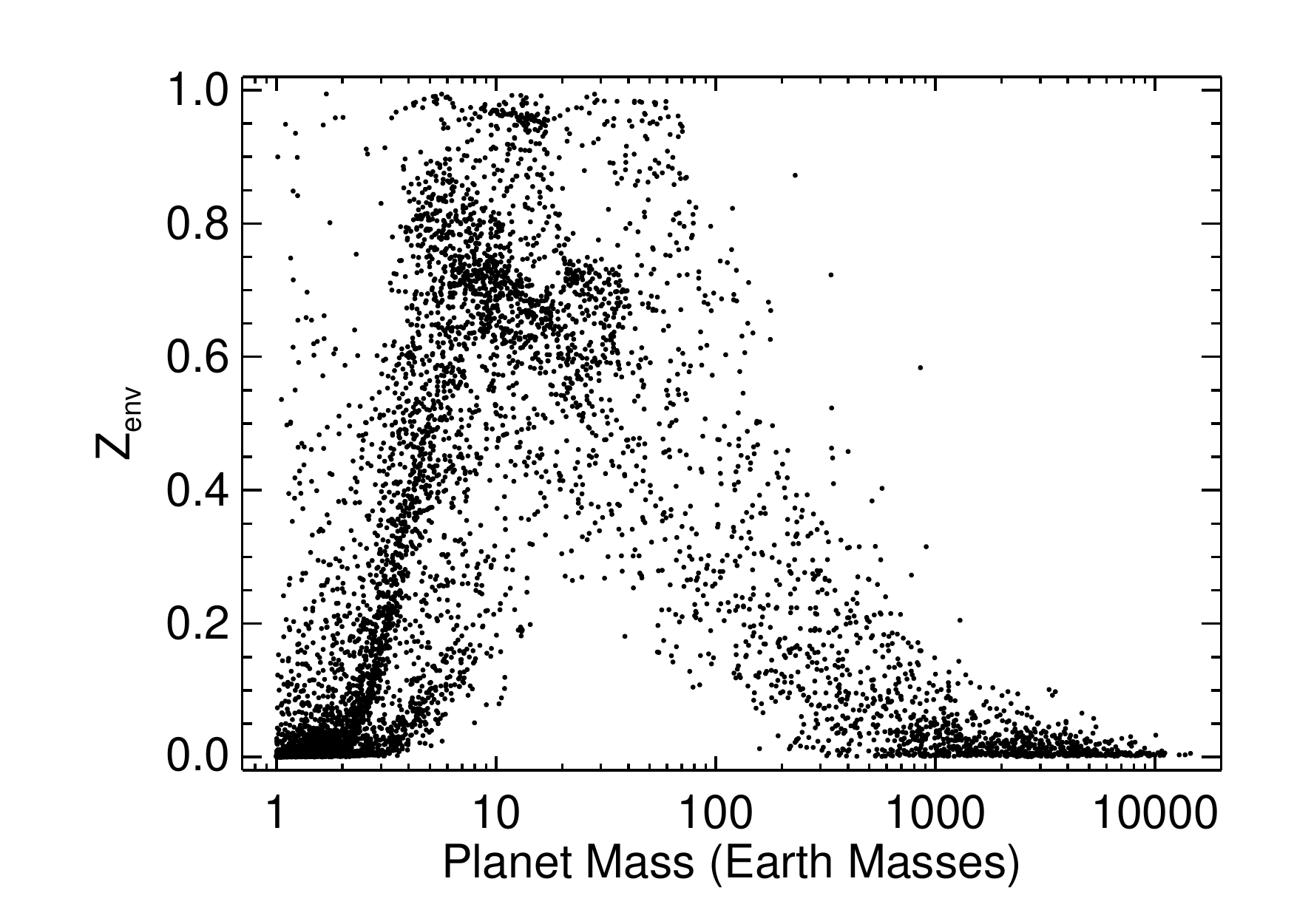}
\caption{Mass fraction of metals in the H/He envelope and visible atmosphere ($Z_{\rm env}$) as a function of
planet mass for the output of the population synthesis models of \ct{Fortney13}. The dots are individually formed planets and use 
100 km planetesimals. We make a simple
assumption of a uniform $Z_{\rm env}$ throughout the envelope. The turnover below 10 \me\ is due to planetesimals driving through the atmosphere and depositing their metals directly onto the core. The general trend of the mass / atmospheric metallicity trend of the solar system system \cp[e.g.,][]{Fortney13,Kreidberg14b} is reproduced by these models\label{pop}}.
\end{figure}

Recently, population synthesis formation models have aimed to understand the metal enrichment of atmospheres from massive gas giants down to sub-Neptunes.  An example from \ct{Fortney13} is shown in Figure \ref{pop}.  These models follow the accretion of gas and solids and aim to calculate the amount of solid matter accreted by the planet as well as the fraction that ablates into the atmosphere.  This is quite uncertain, as the size (or size distribution) of planetesimals is unknown, and the physics of ablation in these atmospheres is still not well understood theoretically.  Nonetheless, the general trend of the models agrees well with that seen in the solar system.

\begin{figure}[htp]
\includegraphics[clip,width=1.0\columnwidth]{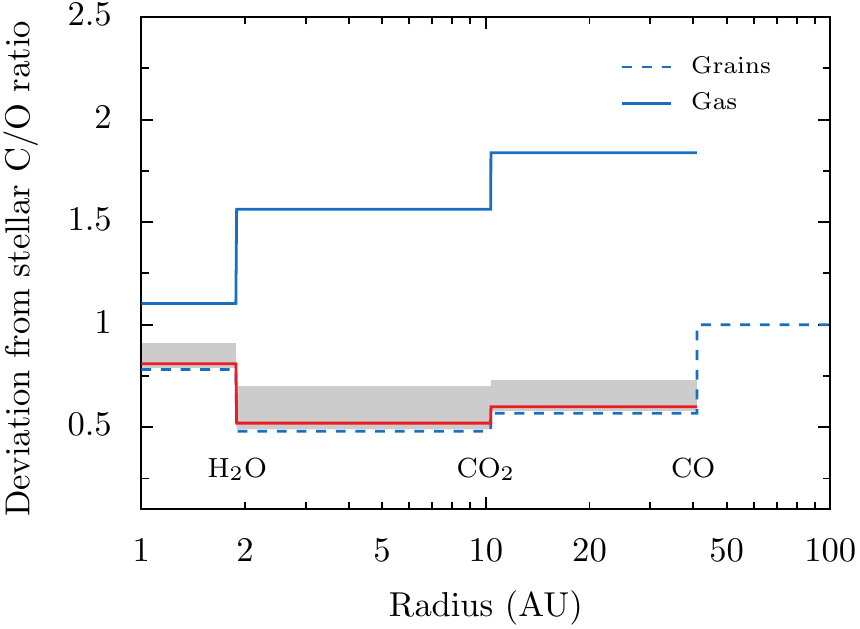}
\caption{For a typical model of the solar nebula, two values of the C/O ratio are plotted as a function of orbital separation, referenced to that of the parent star.  The solid line shows the C/O ratio in the nebular gas.  The dashed line shows the C/O ratios of grains, meaning the solids.  Inside of 2 au, condensation of oxygen-bearing rocks removes oxygen from the gas phase, increasing the C/O of the gas.  At the water snow line at 2 au, water converts to solid form, dramatically decreasing the amount of oxygen in gaseous form, therefore enhancing the C/O ratio of the gas while lowering the ratio in the solids as the amount of oxygen in solid form increases.  Since giant planets accrete massive amounts of gas and solids, the final C/O ratio of the planet traces its formation location as well as its relative accretion of solids and gas.  Figure courtesy of Nestor Espinoza, adapted from \ct{Espinoza17}. \label{ratio}}
\end{figure}

A recent update to the classic \ct{Pollack96} picture is the role of condensation (``snow lines'') in controlling the local composition of planet-forming disks like the solar nebula.  For instance, \ct{Oberg11} suggest that the condensation of water into solid form beyond the water snow line will drive up the C/O ratio of the local gas (since O is lost into solids) but will drive down the C/O ratio of the solids (due to the incorporation of O into solid water).  This is shown graphically in Figure \ref{ratio} for a standard solar nebula model.  The accretion of the H/He envelope of the planet, with a mix of gas and solids, could be a fingerprint of the local conditions in the disk.  This has been an essential new idea in connecting atmospheric observations to planet formation.

This upshot is that giant planet atmospheres are likely not enriched in individual metals by uniform amounts, which is consistent with the Galileo Probe data for Jupiter.  The original \ct{Oberg11} model has since been expanded by other groups to understand the roles of additional physical processes.  These processes include the accretion of solids and gas during disk-driven orbital migration, accretion via pebbles, and disk evolution/cooling over time \cp{Madhu14, Madhu17, Mordasini16, Espinoza17}.  These processes all affect the location of snow lines and the composition of accretions solids.  Importantly, these various theories make a wide range of predictions for the final C/O ratios of giant planets.  There is broad agreement within the planet formation field that giant planets will typically not take on the same carbon and oxygen abundances as their parent star.   Therefore, a derivation of the population-wide C/O ratio of the sample would be a new and unique constraint on planet formation.

\section{Perspective}
The study of exoplanetary atmospheres is still an extremely young field.  There is much room for adventurous investigators in observations, theory, and modeling.  We have barely scratched the surface of what there is to learn.  Given the complexity of planetary atmospheres, we should not expect them to readily fall into tidy spectral classes like main sequence stars. Planets have a diversity of initial abundances, formation locations, durations of formation, and subsequent evolution in isolation or in packed planetary systems, with incident stellar fluxes of both high and low energy across a broad parent star spectral range.  This rich diversity will make for a rewarding study in the near term, as \emph{JWST} transforms the data quality that we have for studying atmospheres.  In the long term, given the large phase space, the field likely will need a space telescope dedicated to obtaining high-quality spectra for a statistically large sample of atmospheres.

We should develop a hierarchy of modeling tools to confront these data sets, some simple to complex, from 1D to 3D.  Within this hierarchy we should endeavor for a better understanding of molecular opacities, cloud microphysics, radiative transport, and fluid dynamics.

\begin{acknowledgement}
I would like thank the organizers of the 2nd Advanced School of Exoplanetary Sciences (ASES2) for the opportunity to give lectures at the school on the beautiful Amalfi Coast.  It was a fantastic experience and I benefit from great interactions with the attendees.  The lectures and this chapter benefited significantly from many discussions with Mark Marley going back to 2004.  Christopher Seay aided the chapter greatly by generating many of the figures.  Mike Line provided figures and gave significant input on the atmospheric retrieval section.  Caroline Morley helped in ways large and small.  Callied Hood, Kat Feng, and Maggie Thompson provided essential comments.
\end{acknowledgement}

%\input{referenc}

%$%$\bibliographystyle{apj}
%$%$\bibliography{references}
\end{document}